\documentclass[aps,pra,reprint,superscriptaddress,showpacs,showkeys,floatfix]{revtex4-2}

\usepackage{graphicx}
\usepackage{lipsum}
\usepackage{color}
\usepackage{xcolor}
\usepackage{amssymb}
\usepackage{amsmath}
\usepackage{amsthm}
\usepackage{mathtools}
\usepackage{epsfig}
\usepackage{bm}
\usepackage[american]{babel}
\usepackage{hyperref}
\usepackage{braket}
\usepackage{tabularx}
\usepackage{wasysym}
\usepackage{subfig}
\usepackage{sidecap}
\usepackage[capitalize,nameinlink]{cleveref}
\usepackage[T1]{fontenc} 
\usepackage{txfonts} 
\usepackage{comment}
\usepackage{caption}
\newcommand{\dw}{$\mathcal{DW}$}
\newcommand{\xz}{$\mathrm{X}^3\mathrm{Z}^3$}
\newcommand{\id}{I}

\DeclareGraphicsExtensions{.pdf,.png,.jpg}
\hypersetup{colorlinks=true,urlcolor=blue,breaklinks=true,citecolor=blue,linkcolor=blue,pdfstartview=FitH,pdfpagemode=UseNone}

\begin{document}

\title{The domain wall color code}

\author{Konstantin Tiurev} 
\email{konstantin.tiurev@quantumsimulations.de}
\affiliation{HQS Quantum Simulations GmbH, Rintheimer Strasse 23, 76131 Karlsruhe, Germany}
\author{Arthur Pesah}
\affiliation{Department of Physics and Astronomy, University College London, London, WC1E 6BT, UK}
\author{Peter-Jan H. S. Derks}
\affiliation{Dahlem Center for Complex Quantum Systems, Freie Universit\"at Berlin, 14195 Berlin, Germany}
\author{Joschka Roffe}
\affiliation{Dahlem Center for Complex Quantum Systems, Freie Universit\"at Berlin, 14195 Berlin, Germany}
\author{Jens Eisert}
\affiliation{Dahlem Center for Complex Quantum Systems, Freie Universit\"at Berlin, 14195 Berlin, Germany}
\affiliation{Helmholtz-Zentrum Berlin f{\"u}r Materialien und Energie, 14109 Berlin, Germany}
\author{Markus S. Kesselring}
\affiliation{Dahlem Center for Complex Quantum Systems, Freie Universit\"at Berlin, 14195 Berlin, Germany}
\author{Jan-Michael Reiner}
\affiliation{HQS Quantum Simulations GmbH, Rintheimer Strasse 23, 76131 Karlsruhe, Germany}

\date{\today}

\begin{abstract}
We introduce the domain wall color code, a new variant of the quantum error-correcting color code that exhibits exceptionally high code-capacity error thresholds for qubits subject to biased noise. In the infinite bias regime, a two-dimensional color code decouples into a series of repetition codes, resulting in an error-correcting threshold of 50\%. Interestingly, at finite bias, our color code demonstrates thresholds identical to those of the noise-tailored XZZX surface code for all single-qubit Pauli noise channels. The design principle of the code is that it introduces domain walls which permute the code's excitations upon domain crossing. For practical implementation, we supplement the domain wall code with a scalable restriction decoder based on a matching algorithm. The proposed code is identified as a comparably resource-efficient quantum error-correcting code highly suitable for realistic noise. 
\end{abstract}

\maketitle

Quantum computers hold the promise to solve certain classes of computational problems with exponential speedups over the best known classical algorithms~\cite{9781107002173}. To enable large-scale quantum computations, information must be stored and processed in a nearly noiseless fashion. However, all components of the quantum computer, including physical qubits, gate operations, and measurements, are inevitably prone to errors. Fragile quantum information can be protected by countering errors with \emph{quantum error-correcting codes}~(QECCs), albeit at substantial resource overheads \cite{RevModPhys.87.307, QEC3, Devitt_2013, doi:10.1080/00107514.2019.1667078}. Such codes turn a collection of noisy, physical qubits into a more robust logical qubit by redundantly encoding information in a non-local way. 
Provided the physical qubit noise is below a certain threshold, the logical error rate can be made arbitrarily small by increasing the number of physical qubits in the code~\cite{doi:10.1126/science.279.5349.342,10.5555/2011665.2011666}. The challenge of practical error correction is to design codes that admit sufficiently high error thresholds and use a reasonable number of physical qubits 
to achieve the desired logical failure rates. Optimizing these figures of merit has been the core subject of research in the field of quantum error correction. 

Among the diverse range of QECCs, \emph{topological surface}~\cite{Kitaev-AnnPhys-2003,TopologicalQuantumMemory,PhysRevLett.108.180501,PhysRevA.86.032324} and \emph{color codes}~\cite{PhysRevLett.97.180501,LandahlColor,Kubica2018,Bombin_2015} are of special interest for practical purposes since they require only geometrically local operations on a two-dimensional qubit layout and exhibit remarkable abilities to protect quantum information. The color code is especially appealing as it supports the transversal implementation of the full Clifford gate set~\cite{PhysRevLett.97.180501,PhysRevX.7.031048,PhysRevA.91.032330}, such that single-qubit errors do not propagate to the remaining qubits of the code when logical gates are executed~\cite{PhysRevLett.102.110502,6006592}. Furthermore, the three-dimensional color code supports transversal non-Clifford gates~\cite{PhysRevLett.98.160502,Bombin_2015,PhysRevA.91.032330}, which, in conjunction with a technique called \emph{code 
switching}~\cite{PhysRevA.91.032330,PRXQuantum.2.020341,PhysRevLett.113.080501}, paves the way to a universal set of fault-tolerant gates. Finally---and importantly---the color code requires smaller resource overhead to encode a logical qubit compared to the surface code of the same distance~\cite{PhysRevA.76.012305,KesselringNew,Kesselring18,PhysRevA.98.052319,https://doi.org/10.48550/arxiv.1407.5103,PhysRevA.83.042310}.

The ability of a code to detect and correct errors strongly depends on the structure of noise affecting qubits on the physical level. 
Practical error correction strategies have to be \emph{tailored} to properties of realistic noise, such as the common situation where noise is biased towards a diagonal Pauli channel, e.g., dephasing. Such noise regimes are common in many hardware architectures, including superconducting qubits~\cite{Aliferis_2009,Brito_2008}, trapped ions~\cite{doi:10.1126/science.1253742}, and quantum dots~\cite{doi:10.1126/science.1217692}. Certain qubit architectures are purposely designed to exhibit strong bias in their noise characteristics~\cite{PhysRevX.9.041053,PRXQuantum.2.030345}. 
The capability of the standard surface and color codes to correct errors quickly deteriorates for such asymmetric biased-noise channels. The same asymmetry can be exploited to fit the error correction strategy~\cite{XZZX,PhysRevX.9.041031,PhysRevLett.120.050505,PhysRevLett.124.130501,PRXQuantum.2.030345,BiasedNonIID}, resulting in remarkable gains in the code's efficiency. The surface code in its so-called~\emph{XZZX configuration}~\cite{XZZX}
yields exceptionally high biased-noise thresholds which can match and even exceed the random-code hashing bound. In contrast, the existing noise-tailored version of the color code---the \emph{XYZ code}---demonstrates much more modest improvements, albeit admitting interesting features such as local decoding~\cite{miguel2022cellular}. Finding a high-threshold color code that can compete on par with the \emph{XZZX surface code} at realistic noise bias would be a significant milestone in the design of fault-tolerant architectures.

\begin{figure*}[t]
\includegraphics[width=0.9\textwidth]{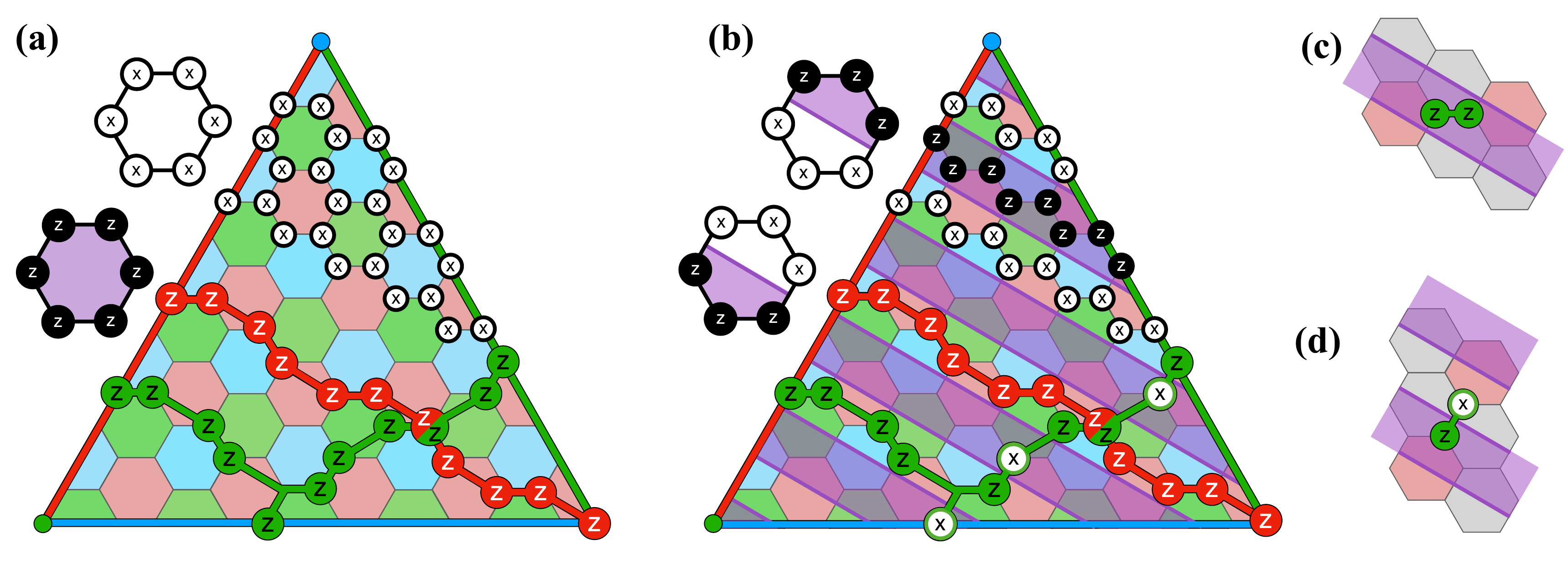}
\caption{\label{fig:1} A distance-11 color code in its (a)~CSS and (b)~\xz{} configurations. (a)~Qubits lie on the vertices of the hexagonal lattice. Each tile corresponds to a primal and a dual stabilizers of Eq.~\eqref{eq:CSS_stabilizers}. Logical Pauli operators are tensor products of physical Pauli operators acting on qubits supported on non-trivial strings that commute with all stabilizers of the code. Logical operators can be deformed by multiplying them with stabilizers. Two realizations of the logical $\bar{Z}$ operator are shown with the red and green strings of Pauli $Z$s. Similarly, logical $\bar{X}$ is a product of Pauli $X$s supported on the same qubits. Due to the CSS structure of the stabilizers, there exist many possible logical operators composed of only $X$ or only $Z$ Paulis. (b)~The \xz{} color code. For convenience, we show primal stabilizers along even diagonals and dual stabilizers along odd diagonals. The stabilizers form domains, where single-qubit Paulis of one type are measured by stabilizers. The domains are separated by DWs~(thin purple lines). Anyons of the same color can be paired by single-qubit Pauli errors of one type if they lie within the same domain, as in panel (c). In contrast, anyons located in different domains can only be connected by chains of errors that change their type when cross a DW, as in panel~(d). Hence, for the \xz{} code there is only one logical $\bar{X}$ consisting of Pauli-$X$ operators only, as shown with the red string, and one logical $\bar{Z}$ dual to it. Any other logical operator will be composed of Paulis of both types, as exemplified with the green logical operator.
}
\end{figure*}

In this letter, we introduce a family of color codes tailored for the efficient correction of biased noise, which we collectively refer to as \emph{domain wall}~(DW) \emph{color codes}. They are obtained by introducing domain walls which permute the code's excitations upon DW crossing~\cite{Kesselring18}. Certain instances of these new codes demonstrate strikingly high code-capacity thresholds, matching those achievable with the noise-tailored XZZX surface code and exceeding the random-code hashing bound~\cite{PhysRevA.54.3824} in the regime of experimentally relevant noise parameters. 
In our work, we extensively explore these superior thresholds. The proposed scheme provides a means to high-threshold logical qubits and serves as a potential test-bed for the experimental demonstration of the \emph{super-additivity of the coherent information}~\cite{9781107002173,PhysRevA.78.062335,doi:10.1137/20M1337375,doi:10.1126/science.1162242}. Lastly, we show that various noise-tailored topological QECC studied so far---such as the XZZX surface~\cite{XZZX} and the XYZ color~\cite{miguel2022cellular} codes---can be formulated as instances of DW codes. Our approach hence unifies previous findings within a single framework.

To introduce the DW code, we first describe the conventional color code, which, in this letter, we refer to as the \emph{Calderbank--Shor--Steane}~(CSS) \emph{color code}, along with an error correction procedure and a noise model. The DW code is then derived from the CSS color code by applying local transformations to the code stabilizers from the Clifford group. As we will see, this seemingly minor and innocent modification will have a dramatic effect on the error-correcting capabilities of the code under biased noise. To be specific, the \emph{color code} is a topological QECC formed by a stabilizer group $\mathcal{S}$ acting on physical qubits placed at the vertices of a trivalent, three-colorable lattice. The 6.6.6 color code is defined on a hexagonal lattice with triangular boundary conditions, as shown in Fig.~\ref{fig:1}~(a). The \emph{stabilizer group} $\mathcal{S}$ of the 
code is generated by operators associated with the faces of the lattice. Particularly, in the conventional, or the \emph{Calderbank--Shor--Steane}~(CSS),  color code each face supports two stabilizer generators,
\begin{equation}\label{eq:CSS_stabilizers}
S_{\textrm{p},f} = \prod_{v \in \partial f} X_v
\quad
\textrm{and}
\quad
S_{\textrm{d},f} = \prod_{v \in \partial f} Z_v,
\end{equation}
which we call \emph{primal} and \emph{dual stabilizers}, respectively. Here $v \in \partial f$ denotes all qubits in the support of face $f$ and $X_v$, $Z_v$ are the corresponding Pauli operators acting on qubit $v$. The code subspace is the $+1$ eigenvalue eigenspace of all elements of $\mathcal{S}$. One can verify from Fig.~\ref{fig:1}~(a) that the code contains one more physical qubit than there are independent stabilizer generators. This remaining degree of freedom constitutes a non-locally encoded logical qubit.

\emph{Logical operators} on the color-code qubit are non-trivial strings of single-qubit Paulis which commute with code stabilizers. Identifying a string of  single-qubit $X$ operators with a logical $\bar{X}$ operator, a logical $\bar{Z}$ is derived from it by replacing $X$ operators of the string with $Z$s. The correct anti-commutation relation $\{ \bar{X}, \bar{Z} \} = 0$ of logical operators is guaranteed due to an odd number of qubits in their support. Alternative configurations of logical operators can be constructed by multiplying these strings with code stabilizers and, due to the CSS structure of the code, any logical $\bar{X}$~($\bar{Z}$) operator can be implemented as a tensor product of single-qubit $X$~($Z$) operators. As an example, the red and green strings of Fig.~\ref{fig:1}~(a) are equivalent up to multiplication by the primal stabilizers of the code. 

\emph{Error correction} is achieved by measuring stabilizers throughout the computation. 
Since string-like logical operators 
commute with all stabilizers, such measurements 
do not perturb the encoded information. By combining the information from stabilizer measurements, collectively called a \emph{syndrome}, a correction can be proposed by a \emph{decoding algorithm}.
The performance of a QECC under an error model relies on its ability to correctly identify 
errors that have occurred on physical qubits and produced an observed syndrome. 

\emph{The noise model} we consider is independent and identically distributed across qubits. We characterize the noise model by a \emph{single-qubit error probability} $p$ and the \emph{bias} $\eta = p_Z/(p_X + p_Y)$, with $p_X = p_Y$ and $p_i$ for $i=X,Y,Z$ being the probability of the corresponding error channel. With this definition, $\eta$ takes values between $\eta_{\textrm{d}} = 0.5$~(depolarizing) and $\eta_{\textrm{ph}} = \infty$~(pure dephasing). The standard surface and color codes of the CSS type achieve their peak performance for depolarizing noise and become less efficient when dephasing prevails. To see why, consider pure dephasing noise. Since all the dual stabilizers of Eq.~\eqref{eq:CSS_stabilizers} 
commute with a phase-flip event, the dual syndrome indicates no information about the occurred error. Consequently, the ability of the standard surface and color codes to correct errors degrades as bias $\eta$ increases. 

\begin{figure}[t]	
\includegraphics[width=0.9\columnwidth]{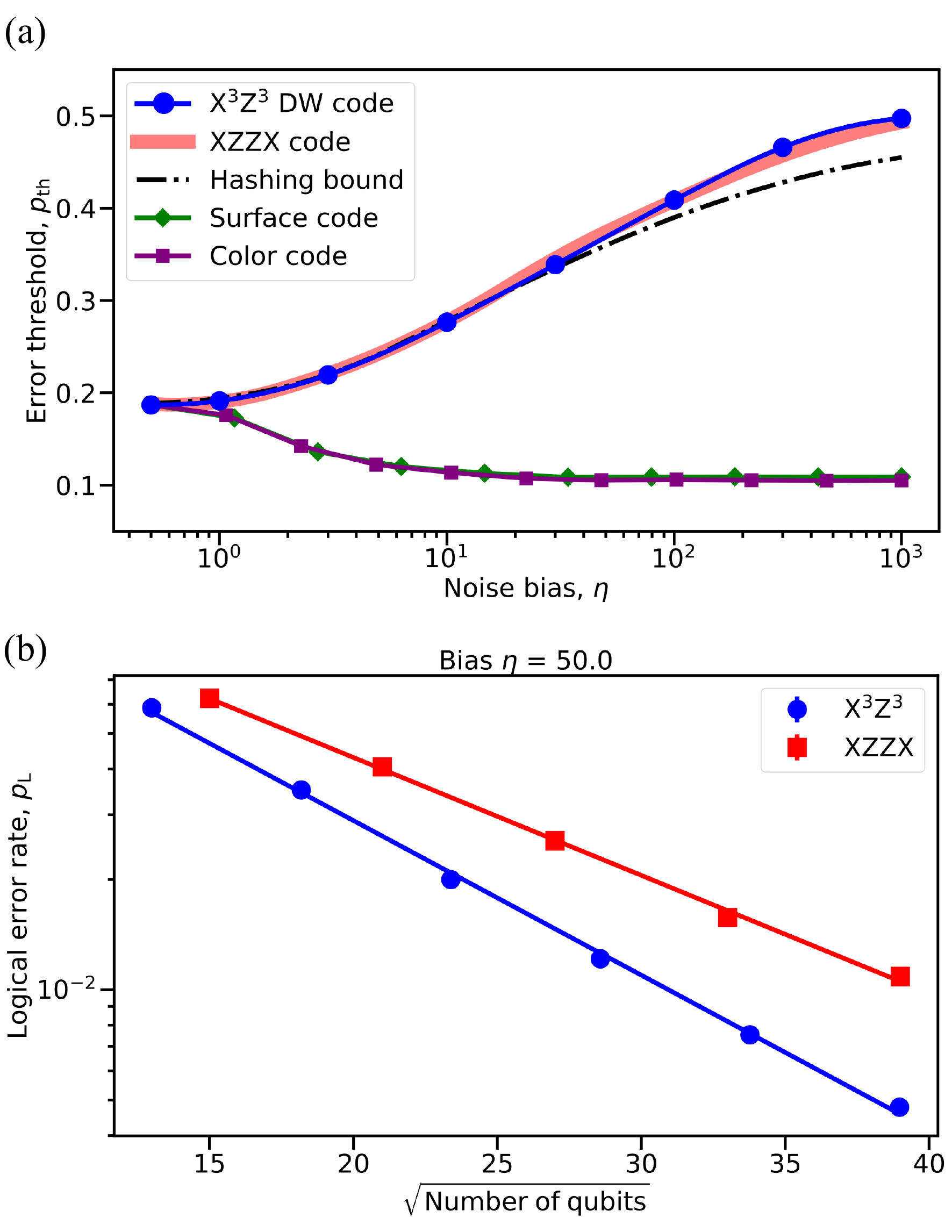}
\caption{\label{fig:2} Code-capacity thresholds of the \xz{} color~(blue solid line) and the XZZX surface~(thick salmon line) codes versus noise bias. For comparison, we also present the hashing bound~(black dash-dotted line) and thresholds of the standard CSS surface~(green diamonds) and color~(purple squares) codes derived in Ref.~\cite{PhysRevX.9.041031}. All curves are fitted using quadratic splines for a better visual appearance. For the \xz{} code, the calculated threshold values are shown explicitly with blue dots. (b)~Sub-threshold logical error rates of the \xz{} DW code~(blue circles) and the rotated squared-shape XZZX code~(red squares) calculated at bias $\eta=50$ and single-qubit error probability $p=25$\%. 
} 
\end{figure}

A common strategy for improving the error correction efficiency of a code is to modify its stabilizers by applying single-qubit Clifford rotations, collectively referred to as a \emph{Clifford deformation} of the code~\cite{https://doi.org/10.48550/arxiv.2201.07802,BiasedNonIID}. 
As an example, deforming the CSS surface code to the XZZX configuration doubles the amount of useful information associated with the dominant noise, which significantly enhances error-correcting capabilities of the code~\cite{XZZX}. 
A similar deformation of the color-code stabilizers maximizes the information content of the syndrome under biased noise, resulting in a substantial gain in performance.

Our \emph{DW color code} is locally equivalent to the conventional color code up to the Hadamard rotations applied to half of the qubits in the code. 
The DW code is therefore reminiscent of the XZZX code derived by applying the Hadamard rotation to every second qubit of the surface code~\cite{XZZX}. An example of the DW code is shown in Fig.~\ref{fig:1}~(b). Each 
stabilizer of weight six measures three Pauli-$X$
and three Pauli-$Z$ operators. 
Due to the structure of its stabilizers, we will refer to this instance of the DW code as the \emph{$X^3Z^3$ code}. As shown in the figure, stabilizers of the code naturally form domain walls, separating regions where one type of Pauli is measured on each stabilizer sub-lattice, hence the name DW code.
 
To understand the advantage of the \xz{} DW code over the standard color code, consider the propagation of code stabilizers flipped by Pauli errors. Flipped stabilizers are known to obey anyonic statistics~\cite{Kitaev-AnnPhys-2003,Kubica2018}, thus we will refer to them as 
anyons for brevity. In the CSS code of Fig.~\ref{fig:1}~(a), anyons due to Pauli-$Z$~($X$) errors can propagate in any direction within the two-dimensional lattice. Assume for instance a $Z$-type anyon, i.e., an anyon created
by Pauli-$Z$ errors in a primal stabilizer. Such an anyon can move freely in a 2D plane due to subsequent $Z$ errors. Hence, the number of possible shortest-path logical operators composed of Paulis of one type is highly degenerate. In the \xz{} color code, on the other hand, propagation of anyons under pure $Z$~(or, equivalently, pure $X$) noise is restricted to 1D domains bounded by the DWs. This is, because the transmission of anyons through a DW requires 
anyons to change type~\cite{Kesselring18,Condensation,BauerBoundaries}. 
Indeed, consider two stabilizers of the same color partially lying within the same domain, such 
as the two red stabilizers of  in Fig.~\ref{fig:1}~(c). Strings of single-qubit errors connecting two such anyons have to be composed of Paulis of one type. In contrast, strings connecting two anyons separated by a DW have to be composed of different Paulis, as in Fig.~\ref{fig:1}~(d). Hence, at infinite bias, anyons are not permitted to cross DWs since that would require a Pauli error of a different type. The only possible logical $\bar{Z}$ operator that does not cross a DW is shown with the red string in Fig.~\ref{fig:1}~(b), 
which is in contrast to a highly degenerate number of logical operators in the CSS code. 
Consequently, the probability of logical errors in the \xz{} code is greatly suppressed. 
Since at infinite bias, the propagation of anyons is restricted to 1D domains, the decoding problem reduces to decoding a series of repetition codes, resulting in a 50\% threshold. A detailed proof of this threshold is provided in Appendix~\ref{sec:dephasing_threshold}.

Clifford transformations similar to the one described above give rise to an entire family of DW codes. We denote different instances of the DW codes as \dw~($\kappa,\phi$), 
with the \emph{density of domain walls} $\kappa$ defined as the number of domain walls $n_{\textrm{DW}}$ per unit distance $d$,
\begin{equation}\label{eq:kappa}
    \kappa = 
    \lim_{d\to\infty}
    \frac{n_{\textrm{DW}}}{d}.
\end{equation}
The parameter $\phi$ denotes a clockwise rotation of DWs with respect to horizontal orientation, and the values it can take are determined by the symmetries of the code stabilizers. Using this nomenclature, the \xz{} code of Fig.~\ref{fig:1}~(b) corresponds to \dw($1, \pi/6$). {In DW codes characterized by $\kappa \leq 1$, the dynamics of anyons at infinite bias are restricted within regions bounded by DWs, similarly to the case of the \xz{} code. Such DW codes together with the XZZX code belong to a category of so-called lineon codes,
i.e., codes whose anyon dynamics is restricted to quasi-1D manifolds. We will refer to DW codes with $\kappa = 1$ as the \emph{dense codes} and $\kappa < 1$ as the \emph{underdense codes}, 
with $\kappa = 0$ being the CSS code. In contrast, in codes characterized by $\kappa > 1$, DWs are placed so close to each other that anyons can propagate both along and across domains even at infinite bias. Anyon dynamics are hence not restricted to one-dimensional manifolds. We will refer to such DW codes as \emph{overdense codes}, with a particularly notable example being the XYZ~\cite{miguel2022cellular} color code, see also Appendix~\ref{sec:hexaginal_codes}. Anyonic excitations in such codes propagate isotropically in two dimensions, giving rise to a type-II fracton syndrome. 
Here, we focus on the properties of the dense \xz{} code and present various alternative instances of DW codes in Appendix~\ref{sec:hexaginal_codes}.  

We numerically investigate the properties of DW codes under biased noise by performing comprehensive Monte Carlo simulations. For decoding, we adapt the approximate \emph{maximum-likelihood decoder}~\cite{PhysRevX.9.041031,PhysRevA.90.032326} 
and assume perfect stabilizer measurements}~\cite{note:alternative_decoders}. Figure~\ref{fig:2}~(a) 
shows the calculated code-capacity thresholds of the \xz{} code at different noise biases $\eta$. Interestingly, we find that for any bias, the threshold of the \xz{} color code perfectly matches that of the surface XZZX code and violates the zero-rate hashing bound of random codes in the strong-bias regime. This observation might imply the existence of a more general theoretical upper bound that holds for non-CSS codes. We note that the hexagonal $XYZ^2$ code of Ref.~\cite{Srivastava_2022} under biased noise exhibits a code-capacity threshold identical to those of the XZZX and \xz{} codes, which further motivates 
research on what is achievable with non-CSS codes.

Below the threshold, the logical failure rates $p_{\textrm{L}}$ of the \xz{} and the square-shaped XZZX codes demonstrate qualitatively similar scaling with the number of qubits, 
\begin{equation}
    \log{p_{\textrm{L}}} 
    \propto 
    -d
    \propto
    -\sqrt{N_{\textrm{q}}},
\end{equation} 
where $N_{\textrm{q}}$ is the number of qubits in the code. 
However, we observe the \xz{} code to be more resource-efficient than its surface-code counterpart, as shown in Fig.~\ref{fig:2}~(b). 
There, we also provide a version of the \xz{} code on a periodic hexagonal lattice with co-prime dimensions. We prove analytically the remarkable property of such a code at high bias: the logical failure rate scales as 
\begin{equation}
    \log{p_{\textrm{L}}} 
    \propto 
    -d^2
    \propto
    - N_{\textrm{q}},
\end{equation}
similarly to the case of the co-prime XZZX code~\cite{XZZX}.

Translating the high thresholds of Fig.~\ref{fig:2}~(a) into practice requires a scalable decoder, such as a commonly used one based on the matching algorithm~\cite{edmonds_1965}. To this end, we adopted a matching-based restriction decoder of Ref.~\cite{PhysRevA.89.012317,PhysRevA.89.012317} to the \xz{} code. As we show in Appendix~\ref{sec:matching}, error thresholds derived with the restriction decoder monotonically increase with the noise bias, however, are noticeably below the optimal thresholds, 
which we attribute to a non-optimal decoding algorithm. Further improvements may be possible using more advanced versions of a matching decoder, 
such as the M\"obius-strip decoder of Ref.~\cite{PRXQuantum.3.010310}. We leave the question of designing optimized scalable decoders as well as fault-tolerant threshold calculations outside of the scope of this work. 

Quantum error-correcting color codes are known to be more versatile than surface codes when it comes to fault-tolerant gates. Their thresholds, on the other hand, are conventionally believed to be lower than those of the surface codes for noise models other than depolarization. In this work, we have shown that the color code with a minor modification in fact demonstrates thresholds matching those achievable with tailored surface codes when qubits are subject to biased noise ubiquitous in many physical architectures. Practically, those high thresholds, accompanied by comparably resource-efficient scaling of the logical failure rate and transversality of the full Clifford gate set make the noise-tailored color codes an efficient quantum error-correcting scheme highly suitable for realistic noise. 

In this work, we have considered only the simplest color code with a hexagonal layout of physical qubits. Clifford deformation into a DW-type code can be directly implemented in alternative color-code configurations. As such, a few examples of 4.8.8 DW color codes are presented in Appendix~\ref{sec:4.8.8}. In the future, it will be important to determine the thresholds of such codes, as they typically require fewer qubits to achieve the desired failure rates and support a more diverse variety of physically distinct DW configurations due to higher-weight stabilizers. A more fundamental question is whether it is possible to impose a domain-wall structure on 3D color codes, i.e., to find a Clifford deformation that restricts propagation of anyons to lower-dimensional manifolds. The first adaptation of 3D topological codes to biased noise has very recently been reported in Ref.~\cite{PRXQuantum.4.030338}. {Such noise-tailored 3D codes, in conjunction with fault-tolerant code switching between the color codes of different dimensions, allows for transversal implementation of non-Clifford operations. Furthermore, the DW color code can be conjugated with the standard techniques for topological codes, such as magic state distillation~\cite{PRXQuantum.2.020341,PhysRevA.71.022316} and entangling gates via lattice surgery~\cite{Horsman_2012,landahl2014quantum,KesselringNew}, paving the way to a universal set of fault-tolerant gates. Combined with bias-preserving entangling gates~\cite{PRXQuantum.2.030345,doi:10.1126/sciadv.aay5901,PhysRevResearch.4.013082,PhysRevX.9.041053}, the logical qubit based on the DW color code becomes a promising candidate for the basic element of a universal quantum computer.}  On a higher level, this work reinforces the understanding that concepts from mathematical condensed matter physics are highly valuable in devising practical schemes for topological quantum error correction. 

\begin{acknowledgments}
This work has been supported by the BMBF (RealistiQ, MUNIQC-Atoms, PhoQuant, QSolid). For RealistiQ, this is collaborative joint-node work. It has also been funded by the Munich Quantum Valley (K-8), {the Quantum Flagship
(Millenion)}, and the Einstein Research Foundation {and the ERC (DebuQC)}. The authors thank the HPC Service of ZEDAT, Freie Universit{\"a}t Berlin, for offering computing time~\cite{Bennett2020}.
\end{acknowledgments}

\bibliographystyle{apsrev4-1}
\bibliography{reflist}

\begin{thebibliography}{72}%
\makeatletter
\providecommand \@ifxundefined [1]{%
 \@ifx{#1\undefined}
}%
\providecommand \@ifnum [1]{%
 \ifnum #1\expandafter \@firstoftwo
 \else \expandafter \@secondoftwo
 \fi
}%
\providecommand \@ifx [1]{%
 \ifx #1\expandafter \@firstoftwo
 \else \expandafter \@secondoftwo
 \fi
}%
\providecommand \natexlab [1]{#1}%
\providecommand \enquote  [1]{``#1''}%
\providecommand \bibnamefont  [1]{#1}%
\providecommand \bibfnamefont [1]{#1}%
\providecommand \citenamefont [1]{#1}%
\providecommand \href@noop [0]{\@secondoftwo}%
\providecommand \href [0]{\begingroup \@sanitize@url \@href}%
\providecommand \@href[1]{\@@startlink{#1}\@@href}%
\providecommand \@@href[1]{\endgroup#1\@@endlink}%
\providecommand \@sanitize@url [0]{\catcode `\\12\catcode `\$12\catcode
  `\&12\catcode `\#12\catcode `\^12\catcode `\_12\catcode `\%12\relax}%
\providecommand \@@startlink[1]{}%
\providecommand \@@endlink[0]{}%
\providecommand \url  [0]{\begingroup\@sanitize@url \@url }%
\providecommand \@url [1]{\endgroup\@href {#1}{\urlprefix }}%
\providecommand \urlprefix  [0]{URL }%
\providecommand \Eprint [0]{\href }%
\providecommand \doibase [0]{http://dx.doi.org/}%
\providecommand \selectlanguage [0]{\@gobble}%
\providecommand \bibinfo  [0]{\@secondoftwo}%
\providecommand \bibfield  [0]{\@secondoftwo}%
\providecommand \translation [1]{[#1]}%
\providecommand \BibitemOpen [0]{}%
\providecommand \bibitemStop [0]{}%
\providecommand \bibitemNoStop [0]{.\EOS\space}%
\providecommand \EOS [0]{\spacefactor3000\relax}%
\providecommand \BibitemShut  [1]{\csname bibitem#1\endcsname}%
\let\auto@bib@innerbib\@empty
\bibitem [{\citenamefont {Nielsen}\ and\ \citenamefont
  {Chuang}(2011)}]{9781107002173}%
  \BibitemOpen
  \bibfield  {author} {\bibinfo {author} {\bibfnamefont {M.~A.}\ \bibnamefont
  {Nielsen}}\ and\ \bibinfo {author} {\bibfnamefont {I.~L.}\ \bibnamefont
  {Chuang}},\ }\href@noop {} {\emph {\bibinfo {title} {Quantum Computation and
  Quantum Information: 10th Anniversary Edition}}}\ (\bibinfo  {publisher}
  {Cambridge University Press},\ \bibinfo {year} {2011})\BibitemShut {NoStop}%
\bibitem [{\citenamefont {Terhal}(2015)}]{RevModPhys.87.307}%
  \BibitemOpen
  \bibfield  {author} {\bibinfo {author} {\bibfnamefont {B.~M.}\ \bibnamefont
  {Terhal}},\ }\href {\doibase 10.1103/RevModPhys.87.307} {\bibfield  {journal}
  {\bibinfo  {journal} {Rev. Mod. Phys.}\ }\textbf {\bibinfo {volume} {87}},\
  \bibinfo {pages} {307} (\bibinfo {year} {2015})}\BibitemShut {NoStop}%
\bibitem [{\citenamefont {Gottesman}(2009)}]{QEC3}%
  \BibitemOpen
  \bibfield  {author} {\bibinfo {author} {\bibfnamefont {D.}~\bibnamefont
  {Gottesman}},\ }\href@noop {} {\enquote {\bibinfo {title} {An introduction to
  quantum error correction and fault-tolerant quantum computation},}\ }
  (\bibinfo {year} {2009}),\ \Eprint {http://arxiv.org/abs/0904.2557}
  {arXiv:0904.2557} \BibitemShut {NoStop}%
\bibitem [{\citenamefont {Devitt}\ \emph {et~al.}(2013)\citenamefont {Devitt},
  \citenamefont {Munro},\ and\ \citenamefont {Nemoto}}]{Devitt_2013}%
  \BibitemOpen
  \bibfield  {author} {\bibinfo {author} {\bibfnamefont {S.~J.}\ \bibnamefont
  {Devitt}}, \bibinfo {author} {\bibfnamefont {W.~J.}\ \bibnamefont {Munro}}, \
  and\ \bibinfo {author} {\bibfnamefont {K.}~\bibnamefont {Nemoto}},\ }\href
  {\doibase 10.1088/0034-4885/76/7/076001} {\bibfield  {journal} {\bibinfo
  {journal} {Rep. Prog. Phys.}\ }\textbf {\bibinfo {volume} {76}},\ \bibinfo
  {pages} {076001} (\bibinfo {year} {2013})}\BibitemShut {NoStop}%
\bibitem [{\citenamefont {Roffe}(2019)}]{doi:10.1080/00107514.2019.1667078}%
  \BibitemOpen
  \bibfield  {author} {\bibinfo {author} {\bibfnamefont {J.}~\bibnamefont
  {Roffe}},\ }\href {\doibase 10.1080/00107514.2019.1667078} {\bibfield
  {journal} {\bibinfo  {journal} {Contemp. Phys.}\ }\textbf {\bibinfo {volume}
  {60}},\ \bibinfo {pages} {226} (\bibinfo {year} {2019})}\BibitemShut
  {NoStop}%
\bibitem [{\citenamefont {Knill}\ \emph {et~al.}(1998)\citenamefont {Knill},
  \citenamefont {Laflamme},\ and\ \citenamefont
  {Zurek}}]{doi:10.1126/science.279.5349.342}%
  \BibitemOpen
  \bibfield  {author} {\bibinfo {author} {\bibfnamefont {E.}~\bibnamefont
  {Knill}}, \bibinfo {author} {\bibfnamefont {R.}~\bibnamefont {Laflamme}}, \
  and\ \bibinfo {author} {\bibfnamefont {W.~H.}\ \bibnamefont {Zurek}},\ }\href
  {\doibase 10.1126/science.279.5349.342} {\bibfield  {journal} {\bibinfo
  {journal} {Science}\ }\textbf {\bibinfo {volume} {279}},\ \bibinfo {pages}
  {342} (\bibinfo {year} {1998})}\BibitemShut {NoStop}%
\bibitem [{\citenamefont {Aliferis}\ \emph {et~al.}(2006)\citenamefont
  {Aliferis}, \citenamefont {Gottesman},\ and\ \citenamefont
  {Preskill}}]{10.5555/2011665.2011666}%
  \BibitemOpen
  \bibfield  {author} {\bibinfo {author} {\bibfnamefont {P.}~\bibnamefont
  {Aliferis}}, \bibinfo {author} {\bibfnamefont {D.}~\bibnamefont {Gottesman}},
  \ and\ \bibinfo {author} {\bibfnamefont {J.}~\bibnamefont {Preskill}},\
  }\href {\doibase 10.26421/QIC8.3-4-1} {\bibfield  {journal} {\bibinfo
  {journal} {Quantum Info. Comput.}\ }\textbf {\bibinfo {volume} {6}},\
  \bibinfo {pages} {97–165} (\bibinfo {year} {2006})}\BibitemShut {NoStop}%
\bibitem [{\citenamefont {Kitaev}(2003)}]{Kitaev-AnnPhys-2003}%
  \BibitemOpen
  \bibfield  {author} {\bibinfo {author} {\bibfnamefont {A.~Y.}\ \bibnamefont
  {Kitaev}},\ }\href {\doibase http://dx.doi.org/10.1016/S0003-4916(02)00018-0}
  {\bibfield  {journal} {\bibinfo  {journal} {Ann. Phys.}\ }\textbf {\bibinfo
  {volume} {303}},\ \bibinfo {pages} {2 } (\bibinfo {year} {2003})}\BibitemShut
  {NoStop}%
\bibitem [{\citenamefont {Dennis}\ \emph {et~al.}(2002)\citenamefont {Dennis},
  \citenamefont {Kitaev}, \citenamefont {Landahl},\ and\ \citenamefont
  {Preskill}}]{TopologicalQuantumMemory}%
  \BibitemOpen
  \bibfield  {author} {\bibinfo {author} {\bibfnamefont {E.}~\bibnamefont
  {Dennis}}, \bibinfo {author} {\bibfnamefont {A.}~\bibnamefont {Kitaev}},
  \bibinfo {author} {\bibfnamefont {A.}~\bibnamefont {Landahl}}, \ and\
  \bibinfo {author} {\bibfnamefont {J.}~\bibnamefont {Preskill}},\ }\href
  {\doibase 10.1063/1.1499754} {\bibfield  {journal} {\bibinfo  {journal} {J.
  Math. Phys.}\ }\textbf {\bibinfo {volume} {43}},\ \bibinfo {pages} {4452}
  (\bibinfo {year} {2002})}\BibitemShut {NoStop}%
\bibitem [{\citenamefont {Fowler}\ \emph
  {et~al.}(2012{\natexlab{a}})\citenamefont {Fowler}, \citenamefont
  {Whiteside},\ and\ \citenamefont {Hollenberg}}]{PhysRevLett.108.180501}%
  \BibitemOpen
  \bibfield  {author} {\bibinfo {author} {\bibfnamefont {A.~G.}\ \bibnamefont
  {Fowler}}, \bibinfo {author} {\bibfnamefont {A.~C.}\ \bibnamefont
  {Whiteside}}, \ and\ \bibinfo {author} {\bibfnamefont {L.~C.~L.}\
  \bibnamefont {Hollenberg}},\ }\href {\doibase 10.1103/PhysRevLett.108.180501}
  {\bibfield  {journal} {\bibinfo  {journal} {Phys. Rev. Lett.}\ }\textbf
  {\bibinfo {volume} {108}},\ \bibinfo {pages} {180501} (\bibinfo {year}
  {2012}{\natexlab{a}})}\BibitemShut {NoStop}%
\bibitem [{\citenamefont {Fowler}\ \emph
  {et~al.}(2012{\natexlab{b}})\citenamefont {Fowler}, \citenamefont
  {Mariantoni}, \citenamefont {Martinis},\ and\ \citenamefont
  {Cleland}}]{PhysRevA.86.032324}%
  \BibitemOpen
  \bibfield  {author} {\bibinfo {author} {\bibfnamefont {A.~G.}\ \bibnamefont
  {Fowler}}, \bibinfo {author} {\bibfnamefont {M.}~\bibnamefont {Mariantoni}},
  \bibinfo {author} {\bibfnamefont {J.~M.}\ \bibnamefont {Martinis}}, \ and\
  \bibinfo {author} {\bibfnamefont {A.~N.}\ \bibnamefont {Cleland}},\ }\href
  {\doibase 10.1103/PhysRevA.86.032324} {\bibfield  {journal} {\bibinfo
  {journal} {Phys. Rev. A}\ }\textbf {\bibinfo {volume} {86}},\ \bibinfo
  {pages} {032324} (\bibinfo {year} {2012}{\natexlab{b}})}\BibitemShut
  {NoStop}%
\bibitem [{\citenamefont {Bombin}\ and\ \citenamefont
  {Martin-Delgado}(2006)}]{PhysRevLett.97.180501}%
  \BibitemOpen
  \bibfield  {author} {\bibinfo {author} {\bibfnamefont {H.}~\bibnamefont
  {Bombin}}\ and\ \bibinfo {author} {\bibfnamefont {M.~A.}\ \bibnamefont
  {Martin-Delgado}},\ }\href {\doibase 10.1103/PhysRevLett.97.180501}
  {\bibfield  {journal} {\bibinfo  {journal} {Phys. Rev. Lett.}\ }\textbf
  {\bibinfo {volume} {97}},\ \bibinfo {pages} {180501} (\bibinfo {year}
  {2006})}\BibitemShut {NoStop}%
\bibitem [{\citenamefont {Landahl}\ \emph {et~al.}(2011)\citenamefont
  {Landahl}, \citenamefont {Anderson},\ and\ \citenamefont
  {Rice}}]{LandahlColor}%
  \BibitemOpen
  \bibfield  {author} {\bibinfo {author} {\bibfnamefont {A.~J.}\ \bibnamefont
  {Landahl}}, \bibinfo {author} {\bibfnamefont {J.~T.}\ \bibnamefont
  {Anderson}}, \ and\ \bibinfo {author} {\bibfnamefont {P.~R.}\ \bibnamefont
  {Rice}},\ }\href@noop {} {\enquote {\bibinfo {title} {Fault-tolerant quantum
  computing with color codes},}\ } (\bibinfo {year} {2011}),\ \Eprint
  {http://arxiv.org/abs/1108.5738} {arXiv:1108.5738} \BibitemShut {NoStop}%
\bibitem [{\citenamefont {Kubica}(2018)}]{Kubica2018}%
  \BibitemOpen
  \bibfield  {author} {\bibinfo {author} {\bibfnamefont {A.~M.}\ \bibnamefont
  {Kubica}},\ }\emph {\bibinfo {title} {The ABCs of the color code: A study of
  topological quantum codes as toy models for fault-tolerant quantum
  computation and quantum phases of matter}},\ \href {\doibase
  10.7907/059V-MG69} {Ph.D. thesis},\ \bibinfo  {school} {California Institute
  of Technology} (\bibinfo {year} {2018})\BibitemShut {NoStop}%
\bibitem [{\citenamefont {Bombín}(2015)}]{Bombin_2015}%
  \BibitemOpen
  \bibfield  {author} {\bibinfo {author} {\bibfnamefont {H.}~\bibnamefont
  {Bombín}},\ }\href {\doibase 10.1088/1367-2630/17/8/083002} {\bibfield
  {journal} {\bibinfo  {journal} {New J. Phys.}\ }\textbf {\bibinfo {volume}
  {17}},\ \bibinfo {pages} {083002} (\bibinfo {year} {2015})}\BibitemShut
  {NoStop}%
\bibitem [{\citenamefont {Litinski}\ \emph {et~al.}(2017)\citenamefont
  {Litinski}, \citenamefont {Kesselring}, \citenamefont {Eisert},\ and\
  \citenamefont {von Oppen}}]{PhysRevX.7.031048}%
  \BibitemOpen
  \bibfield  {author} {\bibinfo {author} {\bibfnamefont {D.}~\bibnamefont
  {Litinski}}, \bibinfo {author} {\bibfnamefont {M.~S.}\ \bibnamefont
  {Kesselring}}, \bibinfo {author} {\bibfnamefont {J.}~\bibnamefont {Eisert}},
  \ and\ \bibinfo {author} {\bibfnamefont {F.}~\bibnamefont {von Oppen}},\
  }\href {\doibase 10.1103/PhysRevX.7.031048} {\bibfield  {journal} {\bibinfo
  {journal} {Phys. Rev. X}\ }\textbf {\bibinfo {volume} {7}},\ \bibinfo {pages}
  {031048} (\bibinfo {year} {2017})}\BibitemShut {NoStop}%
\bibitem [{\citenamefont {Kubica}\ and\ \citenamefont
  {Beverland}(2015)}]{PhysRevA.91.032330}%
  \BibitemOpen
  \bibfield  {author} {\bibinfo {author} {\bibfnamefont {A.}~\bibnamefont
  {Kubica}}\ and\ \bibinfo {author} {\bibfnamefont {M.~E.}\ \bibnamefont
  {Beverland}},\ }\href {\doibase 10.1103/PhysRevA.91.032330} {\bibfield
  {journal} {\bibinfo  {journal} {Phys. Rev. A}\ }\textbf {\bibinfo {volume}
  {91}},\ \bibinfo {pages} {032330} (\bibinfo {year} {2015})}\BibitemShut
  {NoStop}%
\bibitem [{\citenamefont {Eastin}\ and\ \citenamefont
  {Knill}(2009)}]{PhysRevLett.102.110502}%
  \BibitemOpen
  \bibfield  {author} {\bibinfo {author} {\bibfnamefont {B.}~\bibnamefont
  {Eastin}}\ and\ \bibinfo {author} {\bibfnamefont {E.}~\bibnamefont {Knill}},\
  }\href {\doibase 10.1103/PhysRevLett.102.110502} {\bibfield  {journal}
  {\bibinfo  {journal} {Phys. Rev. Lett.}\ }\textbf {\bibinfo {volume} {102}},\
  \bibinfo {pages} {110502} (\bibinfo {year} {2009})}\BibitemShut {NoStop}%
\bibitem [{\citenamefont {Zeng}\ \emph {et~al.}(2011)\citenamefont {Zeng},
  \citenamefont {Cross},\ and\ \citenamefont {Chuang}}]{6006592}%
  \BibitemOpen
  \bibfield  {author} {\bibinfo {author} {\bibfnamefont {B.}~\bibnamefont
  {Zeng}}, \bibinfo {author} {\bibfnamefont {A.}~\bibnamefont {Cross}}, \ and\
  \bibinfo {author} {\bibfnamefont {I.~L.}\ \bibnamefont {Chuang}},\ }\href
  {\doibase 10.1109/TIT.2011.2161917} {\bibfield  {journal} {\bibinfo
  {journal} {IEEE Trans. Inf. Th.}\ }\textbf {\bibinfo {volume} {57}},\
  \bibinfo {pages} {6272} (\bibinfo {year} {2011})}\BibitemShut {NoStop}%
\bibitem [{\citenamefont {Bombin}\ and\ \citenamefont
  {Martin-Delgado}(2007{\natexlab{a}})}]{PhysRevLett.98.160502}%
  \BibitemOpen
  \bibfield  {author} {\bibinfo {author} {\bibfnamefont {H.}~\bibnamefont
  {Bombin}}\ and\ \bibinfo {author} {\bibfnamefont {M.~A.}\ \bibnamefont
  {Martin-Delgado}},\ }\href {\doibase 10.1103/PhysRevLett.98.160502}
  {\bibfield  {journal} {\bibinfo  {journal} {Phys. Rev. Lett.}\ }\textbf
  {\bibinfo {volume} {98}},\ \bibinfo {pages} {160502} (\bibinfo {year}
  {2007}{\natexlab{a}})}\BibitemShut {NoStop}%
\bibitem [{\citenamefont {Beverland}\ \emph {et~al.}(2021)\citenamefont
  {Beverland}, \citenamefont {Kubica},\ and\ \citenamefont
  {Svore}}]{PRXQuantum.2.020341}%
  \BibitemOpen
  \bibfield  {author} {\bibinfo {author} {\bibfnamefont {M.~E.}\ \bibnamefont
  {Beverland}}, \bibinfo {author} {\bibfnamefont {A.}~\bibnamefont {Kubica}}, \
  and\ \bibinfo {author} {\bibfnamefont {K.~M.}\ \bibnamefont {Svore}},\ }\href
  {\doibase 10.1103/PRXQuantum.2.020341} {\bibfield  {journal} {\bibinfo
  {journal} {PRX Quantum}\ }\textbf {\bibinfo {volume} {2}},\ \bibinfo {pages}
  {020341} (\bibinfo {year} {2021})}\BibitemShut {NoStop}%
\bibitem [{\citenamefont {Anderson}\ \emph {et~al.}(2014)\citenamefont
  {Anderson}, \citenamefont {Duclos-Cianci},\ and\ \citenamefont
  {Poulin}}]{PhysRevLett.113.080501}%
  \BibitemOpen
  \bibfield  {author} {\bibinfo {author} {\bibfnamefont {J.~T.}\ \bibnamefont
  {Anderson}}, \bibinfo {author} {\bibfnamefont {G.}~\bibnamefont
  {Duclos-Cianci}}, \ and\ \bibinfo {author} {\bibfnamefont {D.}~\bibnamefont
  {Poulin}},\ }\href {\doibase 10.1103/PhysRevLett.113.080501} {\bibfield
  {journal} {\bibinfo  {journal} {Phys. Rev. Lett.}\ }\textbf {\bibinfo
  {volume} {113}},\ \bibinfo {pages} {080501} (\bibinfo {year}
  {2014})}\BibitemShut {NoStop}%
\bibitem [{\citenamefont {Bombin}\ and\ \citenamefont
  {Martin-Delgado}(2007{\natexlab{b}})}]{PhysRevA.76.012305}%
  \BibitemOpen
  \bibfield  {author} {\bibinfo {author} {\bibfnamefont {H.}~\bibnamefont
  {Bombin}}\ and\ \bibinfo {author} {\bibfnamefont {M.~A.}\ \bibnamefont
  {Martin-Delgado}},\ }\href {\doibase 10.1103/PhysRevA.76.012305} {\bibfield
  {journal} {\bibinfo  {journal} {Phys. Rev. A}\ }\textbf {\bibinfo {volume}
  {76}},\ \bibinfo {pages} {012305} (\bibinfo {year}
  {2007}{\natexlab{b}})}\BibitemShut {NoStop}%
\bibitem [{\citenamefont {Thomsen}\ \emph {et~al.}(2022)\citenamefont
  {Thomsen}, \citenamefont {Kesselring}, \citenamefont {Bartlett},\ and\
  \citenamefont {Brown}}]{KesselringNew}%
  \BibitemOpen
  \bibfield  {author} {\bibinfo {author} {\bibfnamefont {F.}~\bibnamefont
  {Thomsen}}, \bibinfo {author} {\bibfnamefont {M.~S.}\ \bibnamefont
  {Kesselring}}, \bibinfo {author} {\bibfnamefont {S.~D.}\ \bibnamefont
  {Bartlett}}, \ and\ \bibinfo {author} {\bibfnamefont {B.~J.}\ \bibnamefont
  {Brown}},\ }\href@noop {} {\enquote {\bibinfo {title} {Low-overhead quantum
  computing with the color code},}\ } (\bibinfo {year} {2022}),\ \Eprint
  {http://arxiv.org/abs/2201.07806} {arXiv:2201.07806} \BibitemShut {NoStop}%
\bibitem [{\citenamefont {Kesselring}\ \emph {et~al.}(2018)\citenamefont
  {Kesselring}, \citenamefont {Pastawski}, \citenamefont {Eisert},\ and\
  \citenamefont {Brown}}]{Kesselring18}%
  \BibitemOpen
  \bibfield  {author} {\bibinfo {author} {\bibfnamefont {M.~S.}\ \bibnamefont
  {Kesselring}}, \bibinfo {author} {\bibfnamefont {F.}~\bibnamefont
  {Pastawski}}, \bibinfo {author} {\bibfnamefont {J.}~\bibnamefont {Eisert}}, \
  and\ \bibinfo {author} {\bibfnamefont {B.~J.}\ \bibnamefont {Brown}},\ }\href
  {\doibase 10.22331/q-2018-10-19-101} {\bibfield  {journal} {\bibinfo
  {journal} {{Quantum}}\ }\textbf {\bibinfo {volume} {2}},\ \bibinfo {pages}
  {101} (\bibinfo {year} {2018})}\BibitemShut {NoStop}%
\bibitem [{\citenamefont {Lavasani}\ and\ \citenamefont
  {Barkeshli}(2018)}]{PhysRevA.98.052319}%
  \BibitemOpen
  \bibfield  {author} {\bibinfo {author} {\bibfnamefont {A.}~\bibnamefont
  {Lavasani}}\ and\ \bibinfo {author} {\bibfnamefont {M.}~\bibnamefont
  {Barkeshli}},\ }\href {\doibase 10.1103/PhysRevA.98.052319} {\bibfield
  {journal} {\bibinfo  {journal} {Phys. Rev. A}\ }\textbf {\bibinfo {volume}
  {98}},\ \bibinfo {pages} {052319} (\bibinfo {year} {2018})}\BibitemShut
  {NoStop}%
\bibitem [{\citenamefont {Landahl}\ and\ \citenamefont
  {Ryan-Anderson}(2014{\natexlab{a}})}]{https://doi.org/10.48550/arxiv.1407.5103}%
  \BibitemOpen
  \bibfield  {author} {\bibinfo {author} {\bibfnamefont {A.~J.}\ \bibnamefont
  {Landahl}}\ and\ \bibinfo {author} {\bibfnamefont {C.}~\bibnamefont
  {Ryan-Anderson}},\ }\href@noop {} {\enquote {\bibinfo {title} {Quantum
  computing by color-code lattice surgery},}\ } (\bibinfo {year}
  {2014}{\natexlab{a}}),\ \Eprint {http://arxiv.org/abs/1407.5103}
  {arXiv:1407.5103} \BibitemShut {NoStop}%
\bibitem [{\citenamefont {Fowler}(2011)}]{PhysRevA.83.042310}%
  \BibitemOpen
  \bibfield  {author} {\bibinfo {author} {\bibfnamefont {A.~G.}\ \bibnamefont
  {Fowler}},\ }\href {\doibase 10.1103/PhysRevA.83.042310} {\bibfield
  {journal} {\bibinfo  {journal} {Phys. Rev. A}\ }\textbf {\bibinfo {volume}
  {83}},\ \bibinfo {pages} {042310} (\bibinfo {year} {2011})}\BibitemShut
  {NoStop}%
\bibitem [{\citenamefont {Aliferis}\ \emph {et~al.}(2009)\citenamefont
  {Aliferis}, \citenamefont {Brito}, \citenamefont {DiVincenzo}, \citenamefont
  {Preskill}, \citenamefont {Steffen},\ and\ \citenamefont
  {Terhal}}]{Aliferis_2009}%
  \BibitemOpen
  \bibfield  {author} {\bibinfo {author} {\bibfnamefont {P.}~\bibnamefont
  {Aliferis}}, \bibinfo {author} {\bibfnamefont {F.}~\bibnamefont {Brito}},
  \bibinfo {author} {\bibfnamefont {D.~P.}\ \bibnamefont {DiVincenzo}},
  \bibinfo {author} {\bibfnamefont {J.}~\bibnamefont {Preskill}}, \bibinfo
  {author} {\bibfnamefont {M.}~\bibnamefont {Steffen}}, \ and\ \bibinfo
  {author} {\bibfnamefont {B.~M.}\ \bibnamefont {Terhal}},\ }\href {\doibase
  10.1088/1367-2630/11/1/013061} {\bibfield  {journal} {\bibinfo  {journal}
  {New J. Phys.}\ }\textbf {\bibinfo {volume} {11}},\ \bibinfo {pages} {013061}
  (\bibinfo {year} {2009})}\BibitemShut {NoStop}%
\bibitem [{\citenamefont {Brito}\ \emph {et~al.}(2008)\citenamefont {Brito},
  \citenamefont {DiVincenzo}, \citenamefont {Koch},\ and\ \citenamefont
  {Steffen}}]{Brito_2008}%
  \BibitemOpen
  \bibfield  {author} {\bibinfo {author} {\bibfnamefont {F.}~\bibnamefont
  {Brito}}, \bibinfo {author} {\bibfnamefont {D.~P.}\ \bibnamefont
  {DiVincenzo}}, \bibinfo {author} {\bibfnamefont {R.~H.}\ \bibnamefont
  {Koch}}, \ and\ \bibinfo {author} {\bibfnamefont {M.}~\bibnamefont
  {Steffen}},\ }\href {\doibase 10.1088/1367-2630/10/3/033027} {\bibfield
  {journal} {\bibinfo  {journal} {New J. Phys.}\ }\textbf {\bibinfo {volume}
  {10}},\ \bibinfo {pages} {033027} (\bibinfo {year} {2008})}\BibitemShut
  {NoStop}%
\bibitem [{\citenamefont {Nigg}\ \emph {et~al.}(2014)\citenamefont {Nigg},
  \citenamefont {M{\"u}ller}, \citenamefont {Martinez}, \citenamefont
  {Schindler}, \citenamefont {Hennrich}, \citenamefont {Monz}, \citenamefont
  {Martin-Delgado},\ and\ \citenamefont {Blatt}}]{doi:10.1126/science.1253742}%
  \BibitemOpen
  \bibfield  {author} {\bibinfo {author} {\bibfnamefont {D.}~\bibnamefont
  {Nigg}}, \bibinfo {author} {\bibfnamefont {M.}~\bibnamefont {M{\"u}ller}},
  \bibinfo {author} {\bibfnamefont {E.~A.}\ \bibnamefont {Martinez}}, \bibinfo
  {author} {\bibfnamefont {P.}~\bibnamefont {Schindler}}, \bibinfo {author}
  {\bibfnamefont {M.}~\bibnamefont {Hennrich}}, \bibinfo {author}
  {\bibfnamefont {T.}~\bibnamefont {Monz}}, \bibinfo {author} {\bibfnamefont
  {M.~A.}\ \bibnamefont {Martin-Delgado}}, \ and\ \bibinfo {author}
  {\bibfnamefont {R.}~\bibnamefont {Blatt}},\ }\href {\doibase
  10.1126/science.1253742} {\bibfield  {journal} {\bibinfo  {journal}
  {Science}\ }\textbf {\bibinfo {volume} {345}},\ \bibinfo {pages} {302}
  (\bibinfo {year} {2014})}\BibitemShut {NoStop}%
\bibitem [{\citenamefont {Shulman}\ \emph {et~al.}(2012)\citenamefont
  {Shulman}, \citenamefont {Dial}, \citenamefont {Harvey}, \citenamefont
  {Bluhm}, \citenamefont {Umansky},\ and\ \citenamefont
  {Yacoby}}]{doi:10.1126/science.1217692}%
  \BibitemOpen
  \bibfield  {author} {\bibinfo {author} {\bibfnamefont {M.~D.}\ \bibnamefont
  {Shulman}}, \bibinfo {author} {\bibfnamefont {O.~E.}\ \bibnamefont {Dial}},
  \bibinfo {author} {\bibfnamefont {S.~P.}\ \bibnamefont {Harvey}}, \bibinfo
  {author} {\bibfnamefont {H.}~\bibnamefont {Bluhm}}, \bibinfo {author}
  {\bibfnamefont {V.}~\bibnamefont {Umansky}}, \ and\ \bibinfo {author}
  {\bibfnamefont {A.}~\bibnamefont {Yacoby}},\ }\href {\doibase
  10.1126/science.1217692} {\bibfield  {journal} {\bibinfo  {journal}
  {Science}\ }\textbf {\bibinfo {volume} {336}},\ \bibinfo {pages} {202}
  (\bibinfo {year} {2012})}\BibitemShut {NoStop}%
\bibitem [{\citenamefont {Guillaud}\ and\ \citenamefont
  {Mirrahimi}(2019)}]{PhysRevX.9.041053}%
  \BibitemOpen
  \bibfield  {author} {\bibinfo {author} {\bibfnamefont {J.}~\bibnamefont
  {Guillaud}}\ and\ \bibinfo {author} {\bibfnamefont {M.}~\bibnamefont
  {Mirrahimi}},\ }\href {\doibase 10.1103/PhysRevX.9.041053} {\bibfield
  {journal} {\bibinfo  {journal} {Phys. Rev. X}\ }\textbf {\bibinfo {volume}
  {9}},\ \bibinfo {pages} {041053} (\bibinfo {year} {2019})}\BibitemShut
  {NoStop}%
\bibitem [{\citenamefont {Darmawan}\ \emph {et~al.}(2021)\citenamefont
  {Darmawan}, \citenamefont {Brown}, \citenamefont {Grimsmo}, \citenamefont
  {Tuckett},\ and\ \citenamefont {Puri}}]{PRXQuantum.2.030345}%
  \BibitemOpen
  \bibfield  {author} {\bibinfo {author} {\bibfnamefont {A.~S.}\ \bibnamefont
  {Darmawan}}, \bibinfo {author} {\bibfnamefont {B.~J.}\ \bibnamefont {Brown}},
  \bibinfo {author} {\bibfnamefont {A.~L.}\ \bibnamefont {Grimsmo}}, \bibinfo
  {author} {\bibfnamefont {D.~K.}\ \bibnamefont {Tuckett}}, \ and\ \bibinfo
  {author} {\bibfnamefont {S.}~\bibnamefont {Puri}},\ }\href {\doibase
  10.1103/PRXQuantum.2.030345} {\bibfield  {journal} {\bibinfo  {journal} {PRX
  Quantum}\ }\textbf {\bibinfo {volume} {2}},\ \bibinfo {pages} {030345}
  (\bibinfo {year} {2021})}\BibitemShut {NoStop}%
\bibitem [{\citenamefont {Ataides}\ \emph {et~al.}(2021)\citenamefont
  {Ataides}, \citenamefont {Tuckett}, \citenamefont {Bartlett}, \citenamefont
  {Flammia},\ and\ \citenamefont {Brown}}]{XZZX}%
  \BibitemOpen
  \bibfield  {author} {\bibinfo {author} {\bibfnamefont {J.~P.~B.}\
  \bibnamefont {Ataides}}, \bibinfo {author} {\bibfnamefont {D.~K.}\
  \bibnamefont {Tuckett}}, \bibinfo {author} {\bibfnamefont {S.~D.}\
  \bibnamefont {Bartlett}}, \bibinfo {author} {\bibfnamefont {S.~T.}\
  \bibnamefont {Flammia}}, \ and\ \bibinfo {author} {\bibfnamefont {B.~J.}\
  \bibnamefont {Brown}},\ }\href {\doibase 10.1038/s41467-021-22274-1}
  {\bibfield  {journal} {\bibinfo  {journal} {Nature Comm.}\ }\textbf {\bibinfo
  {volume} {12}},\ \bibinfo {pages} {2172} (\bibinfo {year}
  {2021})}\BibitemShut {NoStop}%
\bibitem [{\citenamefont {Tuckett}\ \emph {et~al.}(2019)\citenamefont
  {Tuckett}, \citenamefont {Darmawan}, \citenamefont {Chubb}, \citenamefont
  {Bravyi}, \citenamefont {Bartlett},\ and\ \citenamefont
  {Flammia}}]{PhysRevX.9.041031}%
  \BibitemOpen
  \bibfield  {author} {\bibinfo {author} {\bibfnamefont {D.~K.}\ \bibnamefont
  {Tuckett}}, \bibinfo {author} {\bibfnamefont {A.~S.}\ \bibnamefont
  {Darmawan}}, \bibinfo {author} {\bibfnamefont {C.~T.}\ \bibnamefont {Chubb}},
  \bibinfo {author} {\bibfnamefont {S.}~\bibnamefont {Bravyi}}, \bibinfo
  {author} {\bibfnamefont {S.~D.}\ \bibnamefont {Bartlett}}, \ and\ \bibinfo
  {author} {\bibfnamefont {S.~T.}\ \bibnamefont {Flammia}},\ }\href {\doibase
  10.1103/PhysRevX.9.041031} {\bibfield  {journal} {\bibinfo  {journal} {Phys.
  Rev. X}\ }\textbf {\bibinfo {volume} {9}},\ \bibinfo {pages} {041031}
  (\bibinfo {year} {2019})}\BibitemShut {NoStop}%
\bibitem [{\citenamefont {Tuckett}\ \emph {et~al.}(2018)\citenamefont
  {Tuckett}, \citenamefont {Bartlett},\ and\ \citenamefont
  {Flammia}}]{PhysRevLett.120.050505}%
  \BibitemOpen
  \bibfield  {author} {\bibinfo {author} {\bibfnamefont {D.~K.}\ \bibnamefont
  {Tuckett}}, \bibinfo {author} {\bibfnamefont {S.~D.}\ \bibnamefont
  {Bartlett}}, \ and\ \bibinfo {author} {\bibfnamefont {S.~T.}\ \bibnamefont
  {Flammia}},\ }\href {\doibase 10.1103/PhysRevLett.120.050505} {\bibfield
  {journal} {\bibinfo  {journal} {Phys. Rev. Lett.}\ }\textbf {\bibinfo
  {volume} {120}},\ \bibinfo {pages} {050505} (\bibinfo {year}
  {2018})}\BibitemShut {NoStop}%
\bibitem [{\citenamefont {Tuckett}\ \emph {et~al.}(2020)\citenamefont
  {Tuckett}, \citenamefont {Bartlett}, \citenamefont {Flammia},\ and\
  \citenamefont {Brown}}]{PhysRevLett.124.130501}%
  \BibitemOpen
  \bibfield  {author} {\bibinfo {author} {\bibfnamefont {D.~K.}\ \bibnamefont
  {Tuckett}}, \bibinfo {author} {\bibfnamefont {S.~D.}\ \bibnamefont
  {Bartlett}}, \bibinfo {author} {\bibfnamefont {S.~T.}\ \bibnamefont
  {Flammia}}, \ and\ \bibinfo {author} {\bibfnamefont {B.~J.}\ \bibnamefont
  {Brown}},\ }\href {\doibase 10.1103/PhysRevLett.124.130501} {\bibfield
  {journal} {\bibinfo  {journal} {Phys. Rev. Lett.}\ }\textbf {\bibinfo
  {volume} {124}},\ \bibinfo {pages} {130501} (\bibinfo {year}
  {2020})}\BibitemShut {NoStop}%
\bibitem [{\citenamefont {Tiurev}\ \emph {et~al.}(2023)\citenamefont {Tiurev},
  \citenamefont {Derks}, \citenamefont {Roffe}, \citenamefont {Eisert},\ and\
  \citenamefont {Reiner}}]{BiasedNonIID}%
  \BibitemOpen
  \bibfield  {author} {\bibinfo {author} {\bibfnamefont {K.}~\bibnamefont
  {Tiurev}}, \bibinfo {author} {\bibfnamefont {P.-J. H.~S.}\ \bibnamefont
  {Derks}}, \bibinfo {author} {\bibfnamefont {J.}~\bibnamefont {Roffe}},
  \bibinfo {author} {\bibfnamefont {J.}~\bibnamefont {Eisert}}, \ and\ \bibinfo
  {author} {\bibfnamefont {J.-M.}\ \bibnamefont {Reiner}},\ }\href {\doibase
  10.22331/q-2023-09-26-1123} {\bibfield  {journal} {\bibinfo  {journal}
  {Quantum}\ }\textbf {\bibinfo {volume} {7}},\ \bibinfo {pages} {1123}
  (\bibinfo {year} {2023})}\BibitemShut {NoStop}%
\bibitem [{\citenamefont {Miguel}\ \emph {et~al.}(2023)\citenamefont {Miguel},
  \citenamefont {Williamson},\ and\ \citenamefont
  {Brown}}]{miguel2022cellular}%
  \BibitemOpen
  \bibfield  {author} {\bibinfo {author} {\bibfnamefont {J.~F.~S.}\
  \bibnamefont {Miguel}}, \bibinfo {author} {\bibfnamefont {D.~J.}\
  \bibnamefont {Williamson}}, \ and\ \bibinfo {author} {\bibfnamefont {B.~J.}\
  \bibnamefont {Brown}},\ }\href {\doibase 10.22331/q-2023-03-09-940}
  {\bibfield  {journal} {\bibinfo  {journal} {Quantum}\ }\textbf {\bibinfo
  {volume} {7}},\ \bibinfo {pages} {940} (\bibinfo {year} {2023})}\BibitemShut
  {NoStop}%
\bibitem [{\citenamefont {Bennett}\ \emph {et~al.}(1996)\citenamefont
  {Bennett}, \citenamefont {DiVincenzo}, \citenamefont {Smolin},\ and\
  \citenamefont {Wootters}}]{PhysRevA.54.3824}%
  \BibitemOpen
  \bibfield  {author} {\bibinfo {author} {\bibfnamefont {C.~H.}\ \bibnamefont
  {Bennett}}, \bibinfo {author} {\bibfnamefont {D.~P.}\ \bibnamefont
  {DiVincenzo}}, \bibinfo {author} {\bibfnamefont {J.~A.}\ \bibnamefont
  {Smolin}}, \ and\ \bibinfo {author} {\bibfnamefont {W.~K.}\ \bibnamefont
  {Wootters}},\ }\href {\doibase 10.1103/PhysRevA.54.3824} {\bibfield
  {journal} {\bibinfo  {journal} {Phys. Rev. A}\ }\textbf {\bibinfo {volume}
  {54}},\ \bibinfo {pages} {3824} (\bibinfo {year} {1996})}\BibitemShut
  {NoStop}%
\bibitem [{\citenamefont {Fern}\ and\ \citenamefont
  {Whaley}(2008)}]{PhysRevA.78.062335}%
  \BibitemOpen
  \bibfield  {author} {\bibinfo {author} {\bibfnamefont {J.}~\bibnamefont
  {Fern}}\ and\ \bibinfo {author} {\bibfnamefont {K.~B.}\ \bibnamefont
  {Whaley}},\ }\href {\doibase 10.1103/PhysRevA.78.062335} {\bibfield
  {journal} {\bibinfo  {journal} {Phys. Rev. A}\ }\textbf {\bibinfo {volume}
  {78}},\ \bibinfo {pages} {062335} (\bibinfo {year} {2008})}\BibitemShut
  {NoStop}%
\bibitem [{\citenamefont {Bausch}\ and\ \citenamefont
  {Leditzky}(2021)}]{doi:10.1137/20M1337375}%
  \BibitemOpen
  \bibfield  {author} {\bibinfo {author} {\bibfnamefont {J.}~\bibnamefont
  {Bausch}}\ and\ \bibinfo {author} {\bibfnamefont {F.}~\bibnamefont
  {Leditzky}},\ }\href {\doibase 10.1137/20M1337375} {\bibfield  {journal}
  {\bibinfo  {journal} {SIAM J. Comp.}\ }\textbf {\bibinfo {volume} {50}},\
  \bibinfo {pages} {1410} (\bibinfo {year} {2021})}\BibitemShut {NoStop}%
\bibitem [{\citenamefont {Smith}\ and\ \citenamefont
  {Yard}(2008)}]{doi:10.1126/science.1162242}%
  \BibitemOpen
  \bibfield  {author} {\bibinfo {author} {\bibfnamefont {G.}~\bibnamefont
  {Smith}}\ and\ \bibinfo {author} {\bibfnamefont {J.}~\bibnamefont {Yard}},\
  }\href {\doibase 10.1126/science.1162242} {\bibfield  {journal} {\bibinfo
  {journal} {Science}\ }\textbf {\bibinfo {volume} {321}},\ \bibinfo {pages}
  {1812} (\bibinfo {year} {2008})}\BibitemShut {NoStop}%
\bibitem [{\citenamefont {Dua}\ \emph {et~al.}(2022)\citenamefont {Dua},
  \citenamefont {Kubica}, \citenamefont {Jiang}, \citenamefont {Flammia},\ and\
  \citenamefont {Gullans}}]{https://doi.org/10.48550/arxiv.2201.07802}%
  \BibitemOpen
  \bibfield  {author} {\bibinfo {author} {\bibfnamefont {A.}~\bibnamefont
  {Dua}}, \bibinfo {author} {\bibfnamefont {A.}~\bibnamefont {Kubica}},
  \bibinfo {author} {\bibfnamefont {L.}~\bibnamefont {Jiang}}, \bibinfo
  {author} {\bibfnamefont {S.~T.}\ \bibnamefont {Flammia}}, \ and\ \bibinfo
  {author} {\bibfnamefont {M.~J.}\ \bibnamefont {Gullans}},\ }\href@noop {}
  {\enquote {\bibinfo {title} {Clifford-deformed surface codes},}\ } (\bibinfo
  {year} {2022}),\ \Eprint {http://arxiv.org/abs/2201.07802} {arXiv:2201.07802}
  \BibitemShut {NoStop}%
\bibitem [{\citenamefont {Kesselring}\ \emph {et~al.}(2022)\citenamefont
  {Kesselring}, \citenamefont {de~la Fuente}, \citenamefont {Thomsen},
  \citenamefont {Eisert}, \citenamefont {Bartlett},\ and\ \citenamefont
  {Brown}}]{Condensation}%
  \BibitemOpen
  \bibfield  {author} {\bibinfo {author} {\bibfnamefont {M.~S.}\ \bibnamefont
  {Kesselring}}, \bibinfo {author} {\bibfnamefont {J.~C.~M.}\ \bibnamefont
  {de~la Fuente}}, \bibinfo {author} {\bibfnamefont {F.}~\bibnamefont
  {Thomsen}}, \bibinfo {author} {\bibfnamefont {J.}~\bibnamefont {Eisert}},
  \bibinfo {author} {\bibfnamefont {S.~D.}\ \bibnamefont {Bartlett}}, \ and\
  \bibinfo {author} {\bibfnamefont {B.~J.}\ \bibnamefont {Brown}},\ }\href@noop
  {} {\enquote {\bibinfo {title} {Anyon condensation and the color code},}\ }
  (\bibinfo {year} {2022}),\ \Eprint {http://arxiv.org/abs/2212.00042}
  {arXiv:2212.00042} \BibitemShut {NoStop}%
\bibitem [{\citenamefont {de~la Fuente}\ \emph {et~al.}(2023)\citenamefont
  {de~la Fuente}, \citenamefont {Eisert},\ and\ \citenamefont
  {Bauer}}]{BauerBoundaries}%
  \BibitemOpen
  \bibfield  {author} {\bibinfo {author} {\bibfnamefont {J.~C.~M.}\
  \bibnamefont {de~la Fuente}}, \bibinfo {author} {\bibfnamefont
  {J.}~\bibnamefont {Eisert}}, \ and\ \bibinfo {author} {\bibfnamefont
  {A.}~\bibnamefont {Bauer}},\ }\href {\doibase 10.48550/arXiv.2302.01835}
  {\bibfield  {journal} {\bibinfo  {journal} {J. Math. Phys.}\ }\textbf
  {\bibinfo {volume} {64}},\ \bibinfo {pages} {111904} (\bibinfo {year}
  {2023})}\BibitemShut {NoStop}%
\bibitem [{\citenamefont {Bravyi}\ \emph {et~al.}(2014)\citenamefont {Bravyi},
  \citenamefont {Suchara},\ and\ \citenamefont {Vargo}}]{PhysRevA.90.032326}%
  \BibitemOpen
  \bibfield  {author} {\bibinfo {author} {\bibfnamefont {S.}~\bibnamefont
  {Bravyi}}, \bibinfo {author} {\bibfnamefont {M.}~\bibnamefont {Suchara}}, \
  and\ \bibinfo {author} {\bibfnamefont {A.}~\bibnamefont {Vargo}},\ }\href
  {\doibase 10.1103/PhysRevA.90.032326} {\bibfield  {journal} {\bibinfo
  {journal} {Phys. Rev. A}\ }\textbf {\bibinfo {volume} {90}},\ \bibinfo
  {pages} {032326} (\bibinfo {year} {2014})}\BibitemShut {NoStop}%
\bibitem [{not()}]{note:alternative_decoders}%
  \BibitemOpen
  \bibinfo {note} {For alternative decoders for color codes, see
  Refs.~\cite{PhysRevLett.103.090501,LightsOut}}\BibitemShut {NoStop}%
\bibitem [{\citenamefont {Srivastava}\ \emph {et~al.}(2022)\citenamefont
  {Srivastava}, \citenamefont {Kockum},\ and\ \citenamefont
  {Granath}}]{Srivastava_2022}%
  \BibitemOpen
  \bibfield  {author} {\bibinfo {author} {\bibfnamefont {B.}~\bibnamefont
  {Srivastava}}, \bibinfo {author} {\bibfnamefont {A.~F.}\ \bibnamefont
  {Kockum}}, \ and\ \bibinfo {author} {\bibfnamefont {M.}~\bibnamefont
  {Granath}},\ }\href {\doibase 10.22331/q-2022-04-27-698} {\bibfield
  {journal} {\bibinfo  {journal} {Quantum}\ }\textbf {\bibinfo {volume} {6}},\
  \bibinfo {pages} {698} (\bibinfo {year} {2022})}\BibitemShut {NoStop}%
\bibitem [{\citenamefont {Edmonds}(1965)}]{edmonds_1965}%
  \BibitemOpen
  \bibfield  {author} {\bibinfo {author} {\bibfnamefont {J.}~\bibnamefont
  {Edmonds}},\ }\href {\doibase 10.4153/CJM-1965-045-4} {\bibfield  {journal}
  {\bibinfo  {journal} {Canad. J. Math.}\ }\textbf {\bibinfo {volume} {17}},\
  \bibinfo {pages} {449–467} (\bibinfo {year} {1965})}\BibitemShut {NoStop}%
\bibitem [{\citenamefont {Delfosse}(2014{\natexlab{a}})}]{PhysRevA.89.012317}%
  \BibitemOpen
  \bibfield  {author} {\bibinfo {author} {\bibfnamefont {N.}~\bibnamefont
  {Delfosse}},\ }\href {\doibase 10.1103/PhysRevA.89.012317} {\bibfield
  {journal} {\bibinfo  {journal} {Phys. Rev. A}\ }\textbf {\bibinfo {volume}
  {89}},\ \bibinfo {pages} {012317} (\bibinfo {year}
  {2014}{\natexlab{a}})}\BibitemShut {NoStop}%
\bibitem [{\citenamefont {Sahay}\ and\ \citenamefont
  {Brown}(2022)}]{PRXQuantum.3.010310}%
  \BibitemOpen
  \bibfield  {author} {\bibinfo {author} {\bibfnamefont {K.}~\bibnamefont
  {Sahay}}\ and\ \bibinfo {author} {\bibfnamefont {B.~J.}\ \bibnamefont
  {Brown}},\ }\href {\doibase 10.1103/PRXQuantum.3.010310} {\bibfield
  {journal} {\bibinfo  {journal} {PRX Quantum}\ }\textbf {\bibinfo {volume}
  {3}},\ \bibinfo {pages} {010310} (\bibinfo {year} {2022})}\BibitemShut
  {NoStop}%
\bibitem [{\citenamefont {Huang}\ \emph {et~al.}(2023)\citenamefont {Huang},
  \citenamefont {Pesah}, \citenamefont {Chubb}, \citenamefont {Vasmer},\ and\
  \citenamefont {Dua}}]{PRXQuantum.4.030338}%
  \BibitemOpen
  \bibfield  {author} {\bibinfo {author} {\bibfnamefont {E.}~\bibnamefont
  {Huang}}, \bibinfo {author} {\bibfnamefont {A.}~\bibnamefont {Pesah}},
  \bibinfo {author} {\bibfnamefont {C.~T.}\ \bibnamefont {Chubb}}, \bibinfo
  {author} {\bibfnamefont {M.}~\bibnamefont {Vasmer}}, \ and\ \bibinfo {author}
  {\bibfnamefont {A.}~\bibnamefont {Dua}},\ }\href {\doibase
  10.1103/PRXQuantum.4.030338} {\bibfield  {journal} {\bibinfo  {journal} {PRX
  Quantum}\ }\textbf {\bibinfo {volume} {4}},\ \bibinfo {pages} {030338}
  (\bibinfo {year} {2023})}\BibitemShut {NoStop}%
\bibitem [{\citenamefont {Bravyi}\ and\ \citenamefont
  {Kitaev}(2005)}]{PhysRevA.71.022316}%
  \BibitemOpen
  \bibfield  {author} {\bibinfo {author} {\bibfnamefont {S.}~\bibnamefont
  {Bravyi}}\ and\ \bibinfo {author} {\bibfnamefont {A.}~\bibnamefont
  {Kitaev}},\ }\href {\doibase 10.1103/PhysRevA.71.022316} {\bibfield
  {journal} {\bibinfo  {journal} {Phys. Rev. A}\ }\textbf {\bibinfo {volume}
  {71}},\ \bibinfo {pages} {022316} (\bibinfo {year} {2005})}\BibitemShut
  {NoStop}%
\bibitem [{\citenamefont {Horsman}\ \emph {et~al.}(2012)\citenamefont
  {Horsman}, \citenamefont {Fowler}, \citenamefont {Devitt},\ and\
  \citenamefont {Meter}}]{Horsman_2012}%
  \BibitemOpen
  \bibfield  {author} {\bibinfo {author} {\bibfnamefont {D.}~\bibnamefont
  {Horsman}}, \bibinfo {author} {\bibfnamefont {A.~G.}\ \bibnamefont {Fowler}},
  \bibinfo {author} {\bibfnamefont {S.}~\bibnamefont {Devitt}}, \ and\ \bibinfo
  {author} {\bibfnamefont {R.~V.}\ \bibnamefont {Meter}},\ }\href {\doibase
  10.1088/1367-2630/14/12/123011} {\bibfield  {journal} {\bibinfo  {journal}
  {New J. Phys.}\ }\textbf {\bibinfo {volume} {14}},\ \bibinfo {pages} {123011}
  (\bibinfo {year} {2012})}\BibitemShut {NoStop}%
\bibitem [{\citenamefont {Landahl}\ and\ \citenamefont
  {Ryan-Anderson}(2014{\natexlab{b}})}]{landahl2014quantum}%
  \BibitemOpen
  \bibfield  {author} {\bibinfo {author} {\bibfnamefont {A.~J.}\ \bibnamefont
  {Landahl}}\ and\ \bibinfo {author} {\bibfnamefont {C.}~\bibnamefont
  {Ryan-Anderson}},\ }\href@noop {} {\enquote {\bibinfo {title} {Quantum
  computing by color-code lattice surgery},}\ } (\bibinfo {year}
  {2014}{\natexlab{b}}),\ \Eprint {http://arxiv.org/abs/1407.5103}
  {arXiv:1407.5103} \BibitemShut {NoStop}%
\bibitem [{\citenamefont {Puri}\ \emph {et~al.}(2020)\citenamefont {Puri},
  \citenamefont {St-Jean}, \citenamefont {Gross}, \citenamefont {Grimm},
  \citenamefont {Frattini}, \citenamefont {Iyer}, \citenamefont {Krishna},
  \citenamefont {Touzard}, \citenamefont {Jiang}, \citenamefont {Blais},
  \citenamefont {Flammia},\ and\ \citenamefont
  {Girvin}}]{doi:10.1126/sciadv.aay5901}%
  \BibitemOpen
  \bibfield  {author} {\bibinfo {author} {\bibfnamefont {S.}~\bibnamefont
  {Puri}}, \bibinfo {author} {\bibfnamefont {L.}~\bibnamefont {St-Jean}},
  \bibinfo {author} {\bibfnamefont {J.~A.}\ \bibnamefont {Gross}}, \bibinfo
  {author} {\bibfnamefont {A.}~\bibnamefont {Grimm}}, \bibinfo {author}
  {\bibfnamefont {N.~E.}\ \bibnamefont {Frattini}}, \bibinfo {author}
  {\bibfnamefont {P.~S.}\ \bibnamefont {Iyer}}, \bibinfo {author}
  {\bibfnamefont {A.}~\bibnamefont {Krishna}}, \bibinfo {author} {\bibfnamefont
  {S.}~\bibnamefont {Touzard}}, \bibinfo {author} {\bibfnamefont
  {L.}~\bibnamefont {Jiang}}, \bibinfo {author} {\bibfnamefont
  {A.}~\bibnamefont {Blais}}, \bibinfo {author} {\bibfnamefont {S.~T.}\
  \bibnamefont {Flammia}}, \ and\ \bibinfo {author} {\bibfnamefont {S.~M.}\
  \bibnamefont {Girvin}},\ }\href {\doibase 10.1126/sciadv.aay5901} {\bibfield
  {journal} {\bibinfo  {journal} {Science Adv.}\ }\textbf {\bibinfo {volume}
  {6}},\ \bibinfo {pages} {eaay5901} (\bibinfo {year} {2020})}\BibitemShut
  {NoStop}%
\bibitem [{\citenamefont {Xu}\ \emph {et~al.}(2022)\citenamefont {Xu},
  \citenamefont {Iverson}, \citenamefont {Brand\~ao},\ and\ \citenamefont
  {Jiang}}]{PhysRevResearch.4.013082}%
  \BibitemOpen
  \bibfield  {author} {\bibinfo {author} {\bibfnamefont {Q.}~\bibnamefont
  {Xu}}, \bibinfo {author} {\bibfnamefont {J.~K.}\ \bibnamefont {Iverson}},
  \bibinfo {author} {\bibfnamefont {F.~G. S.~L.}\ \bibnamefont {Brand\~ao}}, \
  and\ \bibinfo {author} {\bibfnamefont {L.}~\bibnamefont {Jiang}},\ }\href
  {\doibase 10.1103/PhysRevResearch.4.013082} {\bibfield  {journal} {\bibinfo
  {journal} {Phys. Rev. Res.}\ }\textbf {\bibinfo {volume} {4}},\ \bibinfo
  {pages} {013082} (\bibinfo {year} {2022})}\BibitemShut {NoStop}%
\bibitem [{\citenamefont {Bennett}\ \emph {et~al.}(2020)\citenamefont
  {Bennett}, \citenamefont {Melchers},\ and\ \citenamefont
  {Proppe}}]{Bennett2020}%
  \BibitemOpen
  \bibfield  {author} {\bibinfo {author} {\bibfnamefont {L.}~\bibnamefont
  {Bennett}}, \bibinfo {author} {\bibfnamefont {B.}~\bibnamefont {Melchers}}, \
  and\ \bibinfo {author} {\bibfnamefont {B.}~\bibnamefont {Proppe}},\ }\href
  {http://dx.doi.org/10.17169/refubium-26754} {\enquote {\bibinfo {title}
  {Curta: A general-purpose high-performance computer at {ZEDAT}, {F}reie
  {U}niversität {B}erlin},}\ } (\bibinfo {year} {2020})\BibitemShut {NoStop}%
\bibitem [{\citenamefont {Katzgraber}\ \emph {et~al.}(2009)\citenamefont
  {Katzgraber}, \citenamefont {Bombin},\ and\ \citenamefont
  {Martin-Delgado}}]{PhysRevLett.103.090501}%
  \BibitemOpen
  \bibfield  {author} {\bibinfo {author} {\bibfnamefont {H.~G.}\ \bibnamefont
  {Katzgraber}}, \bibinfo {author} {\bibfnamefont {H.}~\bibnamefont {Bombin}},
  \ and\ \bibinfo {author} {\bibfnamefont {M.~A.}\ \bibnamefont
  {Martin-Delgado}},\ }\href {\doibase 10.1103/PhysRevLett.103.090501}
  {\bibfield  {journal} {\bibinfo  {journal} {Phys. Rev. Lett.}\ }\textbf
  {\bibinfo {volume} {103}},\ \bibinfo {pages} {090501} (\bibinfo {year}
  {2009})}\BibitemShut {NoStop}%
\bibitem [{\citenamefont {Berent}\ \emph {et~al.}(2023)\citenamefont {Berent},
  \citenamefont {Burgholzer}, \citenamefont {Derks}, \citenamefont {Eisert},\
  and\ \citenamefont {Wille}}]{LightsOut}%
  \BibitemOpen
  \bibfield  {author} {\bibinfo {author} {\bibfnamefont {L.}~\bibnamefont
  {Berent}}, \bibinfo {author} {\bibfnamefont {L.}~\bibnamefont {Burgholzer}},
  \bibinfo {author} {\bibfnamefont {P.-J. H.~S.}\ \bibnamefont {Derks}},
  \bibinfo {author} {\bibfnamefont {J.}~\bibnamefont {Eisert}}, \ and\ \bibinfo
  {author} {\bibfnamefont {R.}~\bibnamefont {Wille}},\ }\href@noop {} {\enquote
  {\bibinfo {title} {{Decoding quantum color codes with MaxSAT}},}\ } (\bibinfo
  {year} {2023}),\ \Eprint {http://arxiv.org/abs/2303.14237} {arXiv:2303.14237}
  \BibitemShut {NoStop}%
\bibitem [{\citenamefont {Brown}\ and\ \citenamefont
  {Williamson}(2020)}]{brown2020parallelized}%
  \BibitemOpen
  \bibfield  {author} {\bibinfo {author} {\bibfnamefont {B.~J.}\ \bibnamefont
  {Brown}}\ and\ \bibinfo {author} {\bibfnamefont {D.~J.}\ \bibnamefont
  {Williamson}},\ }\href {\doibase 10.1103/physrevresearch.2.013303} {\bibfield
   {journal} {\bibinfo  {journal} {Phys. Rev. Res.}\ }\textbf {\bibinfo
  {volume} {2}},\ \bibinfo {pages} {013303} (\bibinfo {year}
  {2020})}\BibitemShut {NoStop}%
\bibitem [{\citenamefont {Brown}(2022)}]{brown2022conservation}%
  \BibitemOpen
  \bibfield  {author} {\bibinfo {author} {\bibfnamefont {B.~J.}\ \bibnamefont
  {Brown}},\ }\href@noop {} {\enquote {\bibinfo {title} {Conservation laws and
  quantum error correction: towards a generalised matching decoder},}\ }
  (\bibinfo {year} {2022}),\ \Eprint {http://arxiv.org/abs/2207.06428}
  {arXiv:2207.06428} \BibitemShut {NoStop}%
\bibitem [{\citenamefont {Huang}\ \emph {et~al.}(2022)\citenamefont {Huang},
  \citenamefont {Pesah}, \citenamefont {Chubb}, \citenamefont {Vasmer},\ and\
  \citenamefont {Dua}}]{huang2022tailoring}%
  \BibitemOpen
  \bibfield  {author} {\bibinfo {author} {\bibfnamefont {E.}~\bibnamefont
  {Huang}}, \bibinfo {author} {\bibfnamefont {A.}~\bibnamefont {Pesah}},
  \bibinfo {author} {\bibfnamefont {C.~T.}\ \bibnamefont {Chubb}}, \bibinfo
  {author} {\bibfnamefont {M.}~\bibnamefont {Vasmer}}, \ and\ \bibinfo {author}
  {\bibfnamefont {A.}~\bibnamefont {Dua}},\ }\href@noop {} {\enquote {\bibinfo
  {title} {Tailoring three-dimensional topological codes for biased noise},}\ }
  (\bibinfo {year} {2022}),\ \Eprint {http://arxiv.org/abs/2211.02116}
  {arXiv:2211.02116} \BibitemShut {NoStop}%
\bibitem [{\citenamefont {Miguel}\ \emph {et~al.}(2022)\citenamefont {Miguel},
  \citenamefont {Williamson},\ and\ \citenamefont
  {Brown}}]{https://doi.org/10.48550/arxiv.2203.16534}%
  \BibitemOpen
  \bibfield  {author} {\bibinfo {author} {\bibfnamefont {J.~F.~S.}\
  \bibnamefont {Miguel}}, \bibinfo {author} {\bibfnamefont {D.~J.}\
  \bibnamefont {Williamson}}, \ and\ \bibinfo {author} {\bibfnamefont {B.~J.}\
  \bibnamefont {Brown}},\ }\href@noop {} {\enquote {\bibinfo {title} {A
  cellular automaton decoder for a noise-bias tailored color code},}\ }
  (\bibinfo {year} {2022}),\ \Eprint {http://arxiv.org/abs/2203.16534}
  {arXiv:2203.16534} \BibitemShut {NoStop}%
\bibitem [{\citenamefont
  {Delfosse}(2014{\natexlab{b}})}]{delfosse-restriction}%
  \BibitemOpen
  \bibfield  {author} {\bibinfo {author} {\bibfnamefont {N.}~\bibnamefont
  {Delfosse}},\ }\href {\doibase 10.1103/PhysRevA.89.012317} {\bibfield
  {journal} {\bibinfo  {journal} {Phys. Rev. A}\ }\textbf {\bibinfo {volume}
  {89}},\ \bibinfo {pages} {012317} (\bibinfo {year}
  {2014}{\natexlab{b}})}\BibitemShut {NoStop}%
\bibitem [{\citenamefont {Karp}(1972)}]{karp1972reducibility}%
  \BibitemOpen
  \bibfield  {author} {\bibinfo {author} {\bibfnamefont {R.~M.}\ \bibnamefont
  {Karp}},\ }in\ \href@noop {} {\emph {\bibinfo {booktitle} {Complexity of
  computer computations}}}\ (\bibinfo  {publisher} {Springer},\ \bibinfo {year}
  {1972})\ pp.\ \bibinfo {pages} {85--103}\BibitemShut {NoStop}%
\bibitem [{\citenamefont {Chamberland}\ \emph {et~al.}(2020)\citenamefont
  {Chamberland}, \citenamefont {Kubica}, \citenamefont {Yoder},\ and\
  \citenamefont {Zhu}}]{chamberland2020triangular}%
  \BibitemOpen
  \bibfield  {author} {\bibinfo {author} {\bibfnamefont {C.}~\bibnamefont
  {Chamberland}}, \bibinfo {author} {\bibfnamefont {A.}~\bibnamefont {Kubica}},
  \bibinfo {author} {\bibfnamefont {T.~J.}\ \bibnamefont {Yoder}}, \ and\
  \bibinfo {author} {\bibfnamefont {G.}~\bibnamefont {Zhu}},\ }\href {\doibase
  10.1088/1367-2630/ab68fd} {\bibfield  {journal} {\bibinfo  {journal} {New J.
  Phys.}\ }\textbf {\bibinfo {volume} {22}},\ \bibinfo {pages} {023019}
  (\bibinfo {year} {2020})}\BibitemShut {NoStop}%
\bibitem [{\citenamefont {Higgott}(2022)}]{higgott2022pymatching}%
  \BibitemOpen
  \bibfield  {author} {\bibinfo {author} {\bibfnamefont {O.}~\bibnamefont
  {Higgott}},\ }\href {\doibase 10.1145/3505637} {\bibfield  {journal}
  {\bibinfo  {journal} {ACM Trans. Quant. Comp.}\ }\textbf {\bibinfo {volume}
  {3}} (\bibinfo {year} {2022}),\ 10.1145/3505637}\BibitemShut {NoStop}%
\bibitem [{dwc()}]{dwcc_repo}%
  \BibitemOpen
  \bibinfo {note} {The data obtained from simulating the DW codes using the
  restriction decoder is available at
  \url{https://github.com/peter-janderks/restriction_decoder_domain_wall_colour_code}}\BibitemShut
  {NoStop}%
\bibitem [{\citenamefont {Harrington}(2004)}]{harrington2004analysis}%
  \BibitemOpen
  \bibfield  {author} {\bibinfo {author} {\bibfnamefont {J.~W.}\ \bibnamefont
  {Harrington}},\ }\emph {\bibinfo {title} {Analysis of quantum
  error-correcting codes: symplectic lattice codes and toric codes}},\
  \href@noop {} {Ph.D. thesis},\ \bibinfo  {school} {California Institute of
  Technology} (\bibinfo {year} {2004})\BibitemShut {NoStop}%
\end{thebibliography}%

\clearpage
\onecolumngrid
\appendix

\section{The \xz{} color code has a 50\% threshold} \label{sec:dephasing_threshold}

In Ref.~\cite[Lemma 1]{miguel2022cellular}, it is shown that if the number of pure $Z$ logical configurations is bounded by a constant, the threshold error rate at pure dephasing is $50\%$. The \xz{} DW code hence has a $ 50\%$ threshold at infinite bias. 
While it is folklore knowledge that the boundary conditions should not impact the threshold of a code, we rigorously show this to be true in the case of the \xz{} color code at infinite bias, by considering the code with periodic boundary conditions. We believe that the proof technique presented here can be of independent interest and reused for the study of different bias-tailored codes.

\begin{figure}[h!]
    \centering
    \subfloat[]{
        \includegraphics[width=0.24\columnwidth]{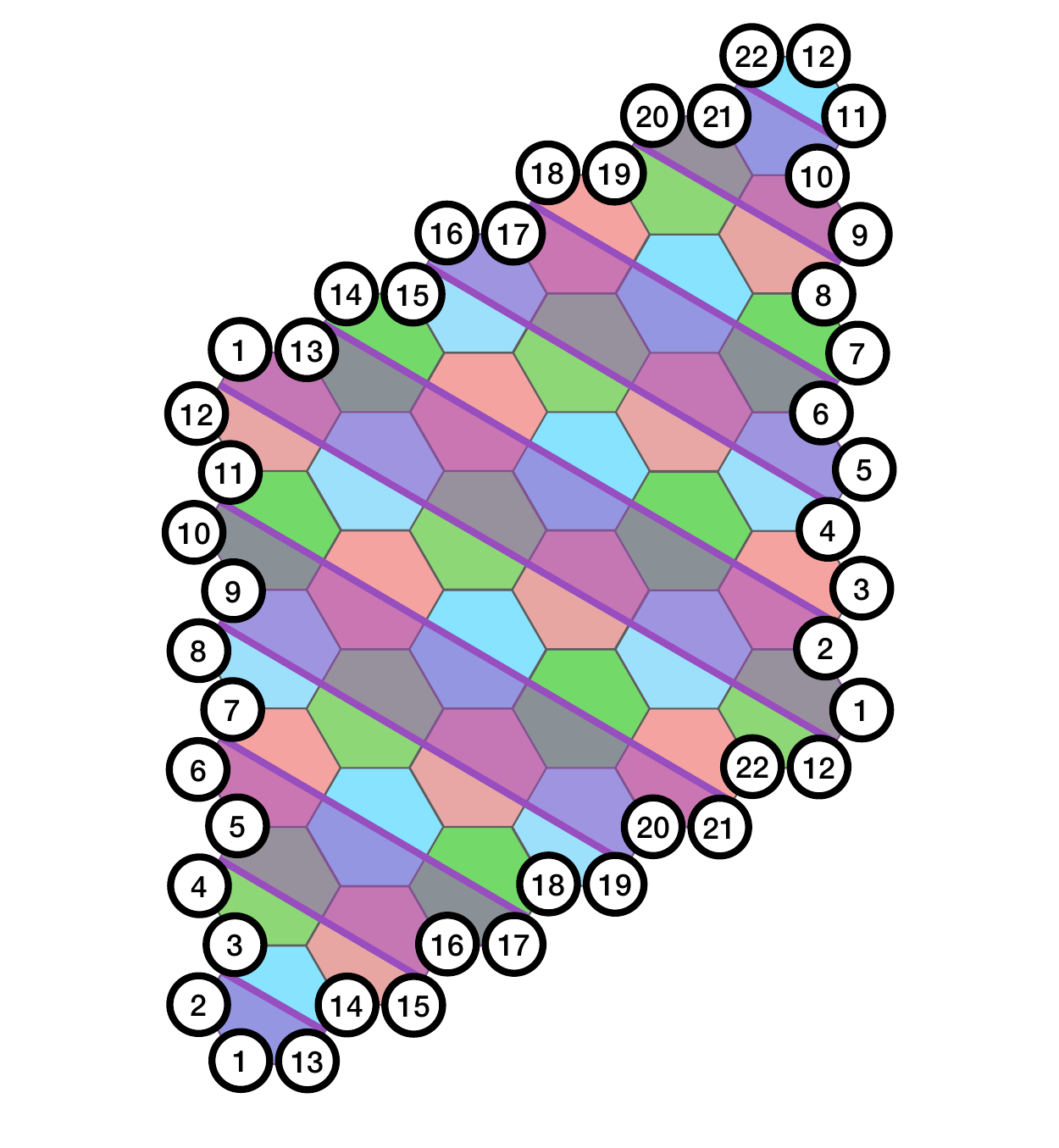}
        \label{fig:linear-symmetries-0}
    }
    \subfloat[]{
        \includegraphics[width=0.24\columnwidth]{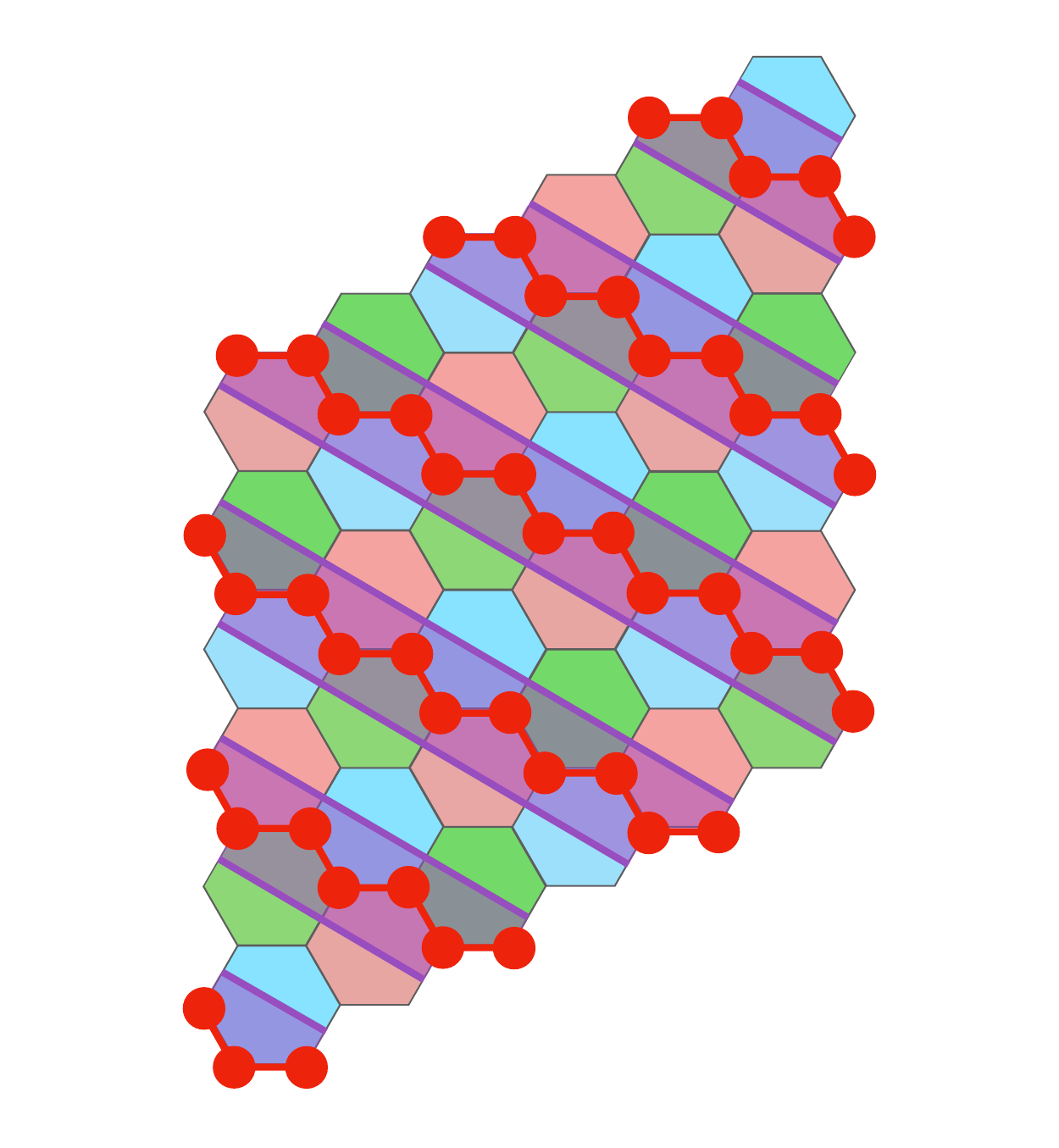}
        \label{fig:linear-symmetries-a}
    }
    \subfloat[]{
        \includegraphics[width=0.24\columnwidth]{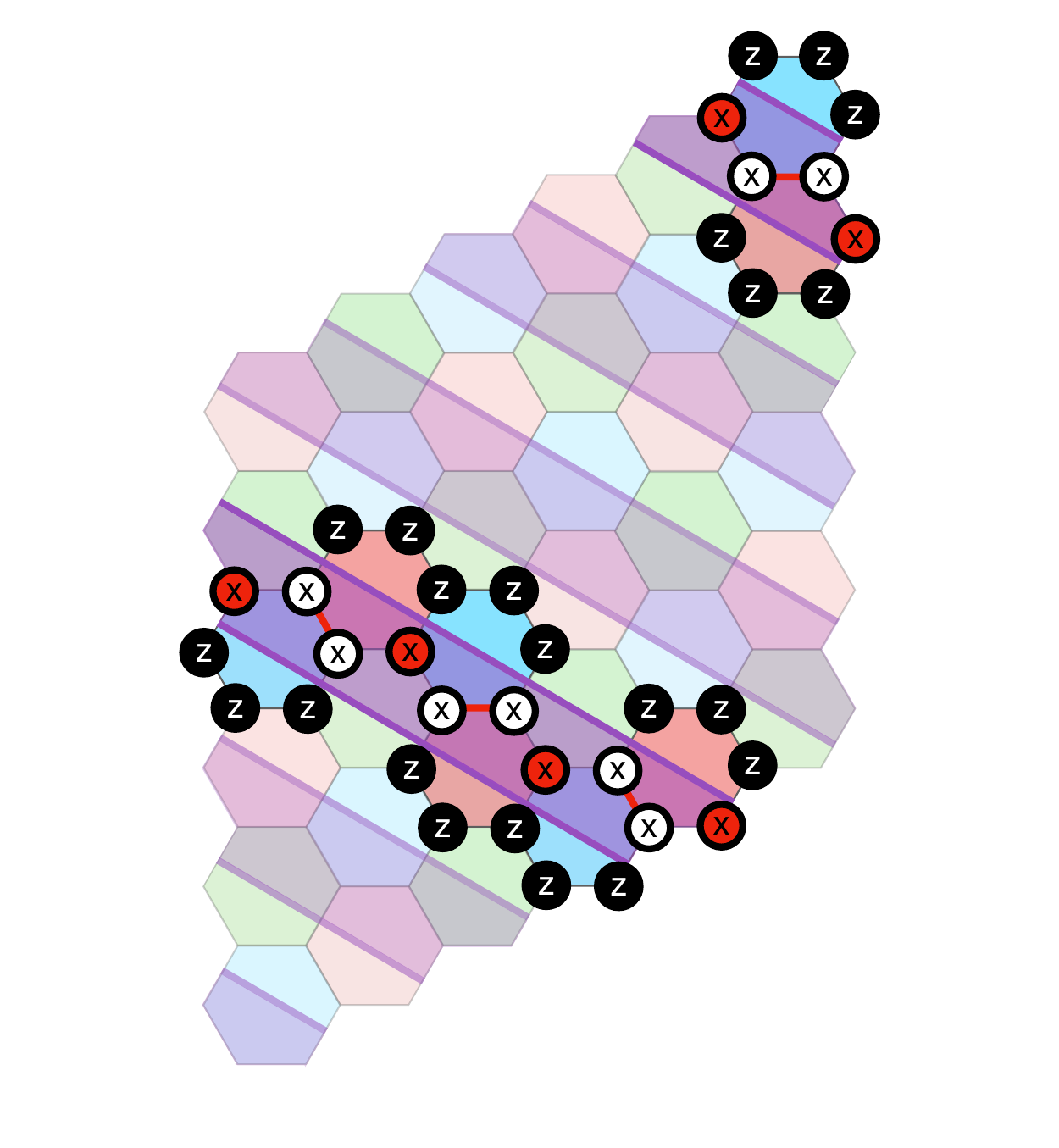}
        \label{fig:linear-symmetries-b}
    }
    \subfloat[]{
        \includegraphics[width=0.24\columnwidth]{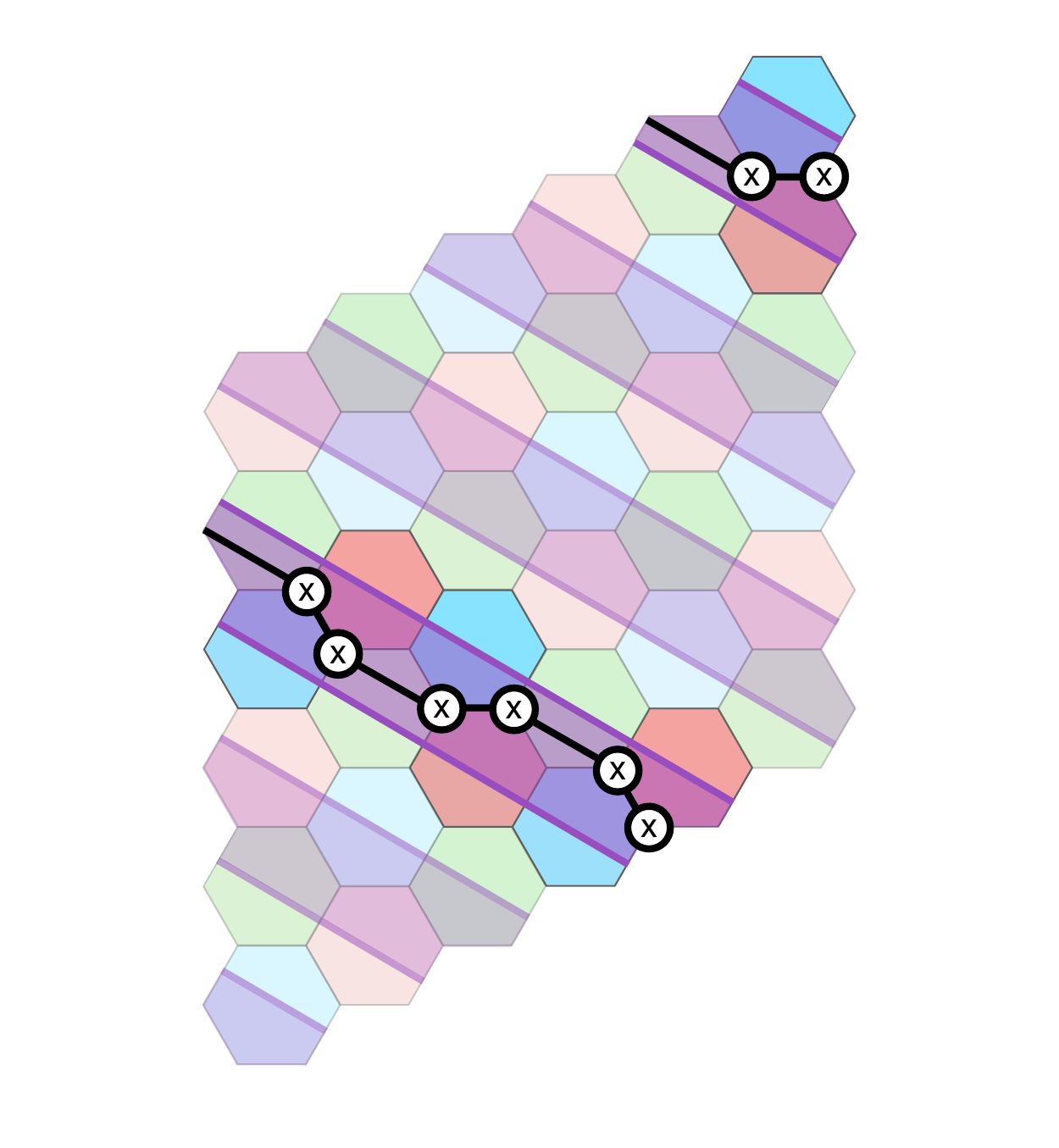}
        \label{fig:linear-symmetries-c}
    }
    \caption{
    Decoding of a domain wall color code at infinite $Z$ bias.
    \textbf{(a)} Color-6.6.6 code on a periodic lattice, with boundary qubits identified as shown in the figure. 
    \textbf{(b)} Clifford-deformation of a $6\times 6$ color code with periodic boundary conditions. Red vertices correspond to the application of a Hadamard operator, which turns the CSS color code to the \xz{} code.
    \textbf{(c)} Linear symmetries of the first decoding step. Multiplying cells of two different colors (red and blue here) along a curved line gives the identity operator in the infinite bias regime, when considering the $X$ part of the stabilizers only. Performing matching along this symmetry line enables the decoding of one third of the qubits (red vertices), and to obtain new edge checks connecting pairs of the remaining qubits (by weight-reduction).
    \textbf{(d)} Linear symmetries of the second decoding step. The green stabilizers are now effectively supported on two qubits (errors on the third one having been corrected). Combining the edge checks obtained in the previous step with green stabilizers gives us a new linear symmetry, that we can use to correct errors on all the qubits.
    }
    \label{fig:linear-symmetries} 
\end{figure}

To define the \xz{} color code on a periodic lattice, we consider a hexagonal lattice with $L \times L$ hexagons and periodic boundary conditions, as shown in Fig.~\ref{fig:linear-symmetries-0}. For the lattice to be three-colorable, we need $L$ to be a multiple of $3$, and for the Clifford deformation to be well-defined, we need $L$ to be a multiple of $2$. To satisfy those constraints, we impose $L$ to be a multiple of $6$. The Clifford deformation of this code is shown in Fig.~\ref{fig:linear-symmetries-a}. Similarly to the open-boundary version of the code, excitations created by pure $X$ or pure $Z$ errors are confined within domains.

We prove here that the periodic \xz{} code has a $50\%$ threshold at infinite bias based on materialized symmetry considerations: we show that at infinite bias, decoding the code is equivalent to decoding repetition codes along certain symmetries that we exhibit. We begin by introducing some terminology. We bipartition the physical qubits into type-A and type-B qubits, according to~\ref{fig:linear-symmetries-a}, where type-A qubits are coloured red. We observe that the Clifford-deformed $Z$ stabilizers act with $X$ operators on type-A qubits and $Z$ operators on type-B qubits, as can be seen in Fig.~\ref{fig:linear-symmetries-a}. In the infinite $Z$ bias regime, the $Z$ part of the stabilizers can be ignored, and those stabilizers effectively become classical parity check operators supported on three type-A qubits. We, therefore, call them type-A checks, while type-B checks are the classical checks induced by the Clifford-deformed $X$ stabilizers. We now construct a decoder that corrects errors supported on type-A qubits, using type-A checks. 
By symmetry, type-B qubits can be decoded with the same strategy, using type-B checks.

To construct our decoder, we start by finding \textit{materialized linear symmetries} of the code~\cite{brown2020parallelized, brown2022conservation, huang2022tailoring}. 
A materialized symmetry is a set of stabilizers $\{ S_i \}$ whose product is the identity operator,
\begin{equation}\label{eq:symmetry_op}
    \prod_{i} S_i = \id.
\end{equation}
Any materialized symmetry leads to a conservation law for the syndrome, as the product of all stabilizer measurements $\{s_i\}$ 
\begin{equation}\label{eq:symmetry_meas}
    \prod_{i} s_i = 1
\end{equation}
along the symmetry must be equal to 1.
Therefore, there is an even number of $i$s such that $s_i=-1$, or in other words, there is an even number of excitations along the symmetry. 
Therefore, when such a symmetry is present in a code, a \emph{minimum-weight perfect matching}~(MWPM) decoder can be used to pair up and annihilate the excitations. 
For instance, in the 2D toric code, all the vertex 
(plaquette) stabilizers multiply to the identity, meaning that there is always an even number of vertex (plaquette) excitations, and MWPM can be used to decode them.
We say that a materialized symmetry is \emph{linear} if it is supported on a one-dimensional sub-manifold of the lattice. Decoding excitations supported on a linear symmetry is equivalent to decoding a repetition code.

The strategy for decoding the \xz{} color code will be to find a first set of materialized linear symmetries and use a matching decoder along those symmetries. The result will be a new decoding problem with checks of reduced weights. Using the materialized linear symmetries of this new problem will then give us a correction operator for all the qubits of the code. A similar strategy, called the \textit{weight-reduction technique}, has  been introduced in Ref.~\cite{huang2022tailoring} to show thresholds of $50\%$ for several topological codes .

In the \xz{} color code, multiplying blue and red type-A checks on a diagonal line, as shown in Fig.~\ref{fig:linear-symmetries-b}, gives the identity operator. This means that there is always an even number of excitations supported on such line, and a MWPM decoder can therefore be used to match those excitations. However, there is a degeneracy in the matching solution: when two checks intersect on two qubits~(red edges in Fig.~\ref{fig:linear-symmetries-b}), either one of the two qubits can be used in the correction. Therefore, the only data that the matching decoder allows us to infer at those intersections is the parity of the errors on the two adjacent qubits. On the other hand, when two checks intersect on a single qubit~(red vertices in Fig.~\ref{fig:linear-symmetries-b}), the presence of an error on this qubit can be inferred directly.

Therefore, we are left with a second decoding problem, represented in Fig.~\ref{fig:linear-symmetries-c}, where the red edges of Fig.~\ref{fig:linear-symmetries-b} now support weight-two parity-check operators. Green checks have also been reduced to a two-body term, since the third qubit of each green plaquette~(the red vertices of Fig.~\ref{fig:linear-symmetries-b}) has been decoded in the previous step. This leaves us with a materialized linear symmetry represented in black in Fig.~\ref{fig:linear-symmetries-c}. Using a matching decoder along this line allows us to infer a correction operator on all the remaining type-A qubit.

Let us now show that this decoder has a threshold error rate of 50\%. The probability of success of our strategy is lower bounded by the probability that each repetition code decoder succeed. Since each step of our decoder consists of decoding $L/2$ repetition codes, we are decoding $L$ repetition codes in total. As the success probability of a given repetition code is of the form $1-Ae^{-\alpha L}$, $0 < A < 1$ for any physical error rate below $50\%$, we have the lower bound 
\begin{equation}
    p_{\text{success}} \geq \left(1-e^{-\alpha L}\right)^L > 1-Le^{-\alpha L} \xrightarrow[L \rightarrow \infty]{} 1
\end{equation}
on the total success probability. 
Therefore, for any physical error rate below $50\%$, the error probability goes to zero as we increase the lattice size. This achieves the proof that the \xz{} color code has a threshold of $50\%$. 



\section{Noise-tailored codes on the hexagonal lattice}
\label{sec:hexaginal_codes}

\subsection{Underdense and overdense DW codes}

\begin{figure}[h!]
    \centering
    \subfloat[]{
        \includegraphics[width=0.35\columnwidth]
        {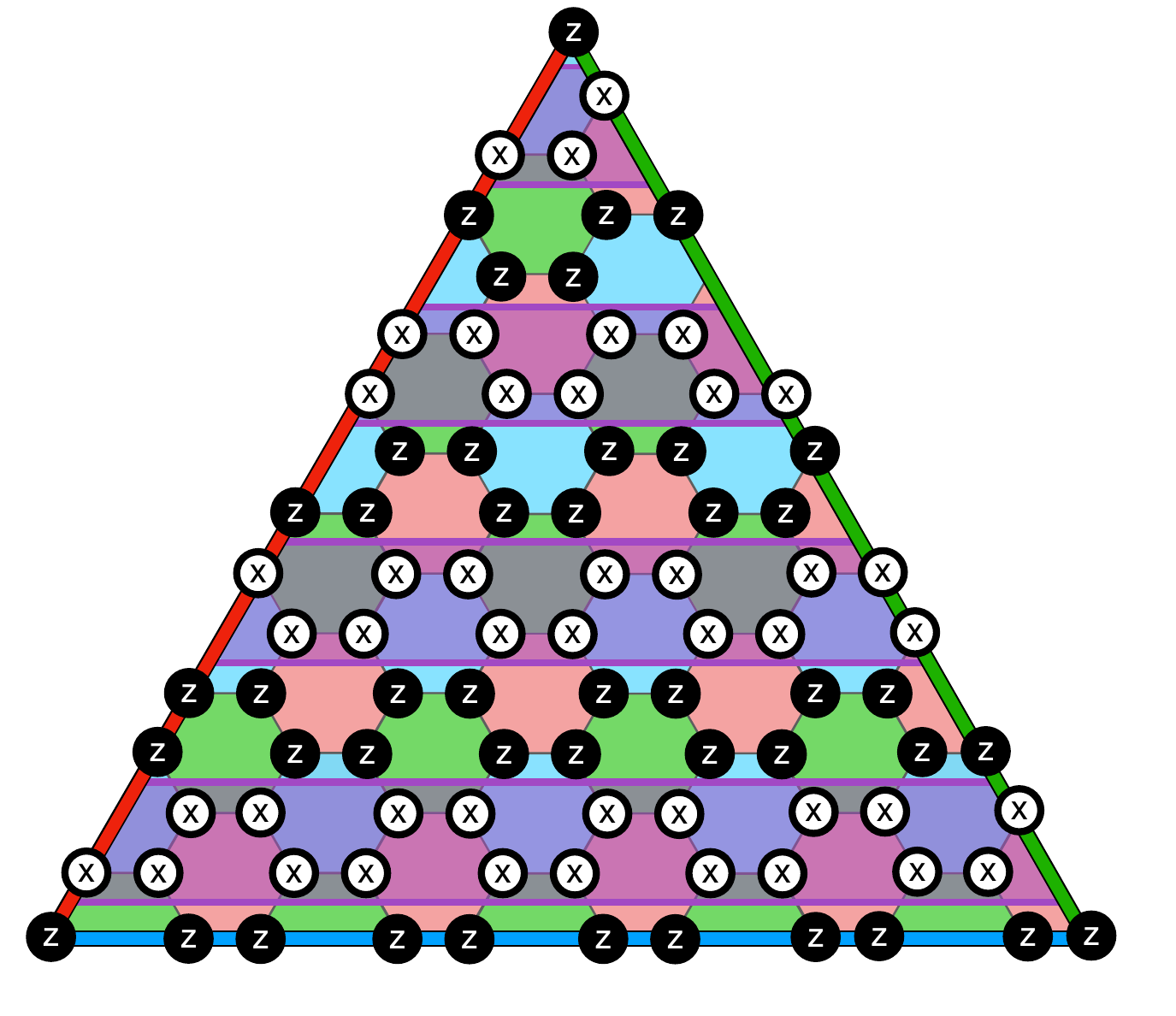}
        \label{fig:underdense-code-a}
    }
    \subfloat[]{
        \includegraphics[width=0.35\columnwidth]
        {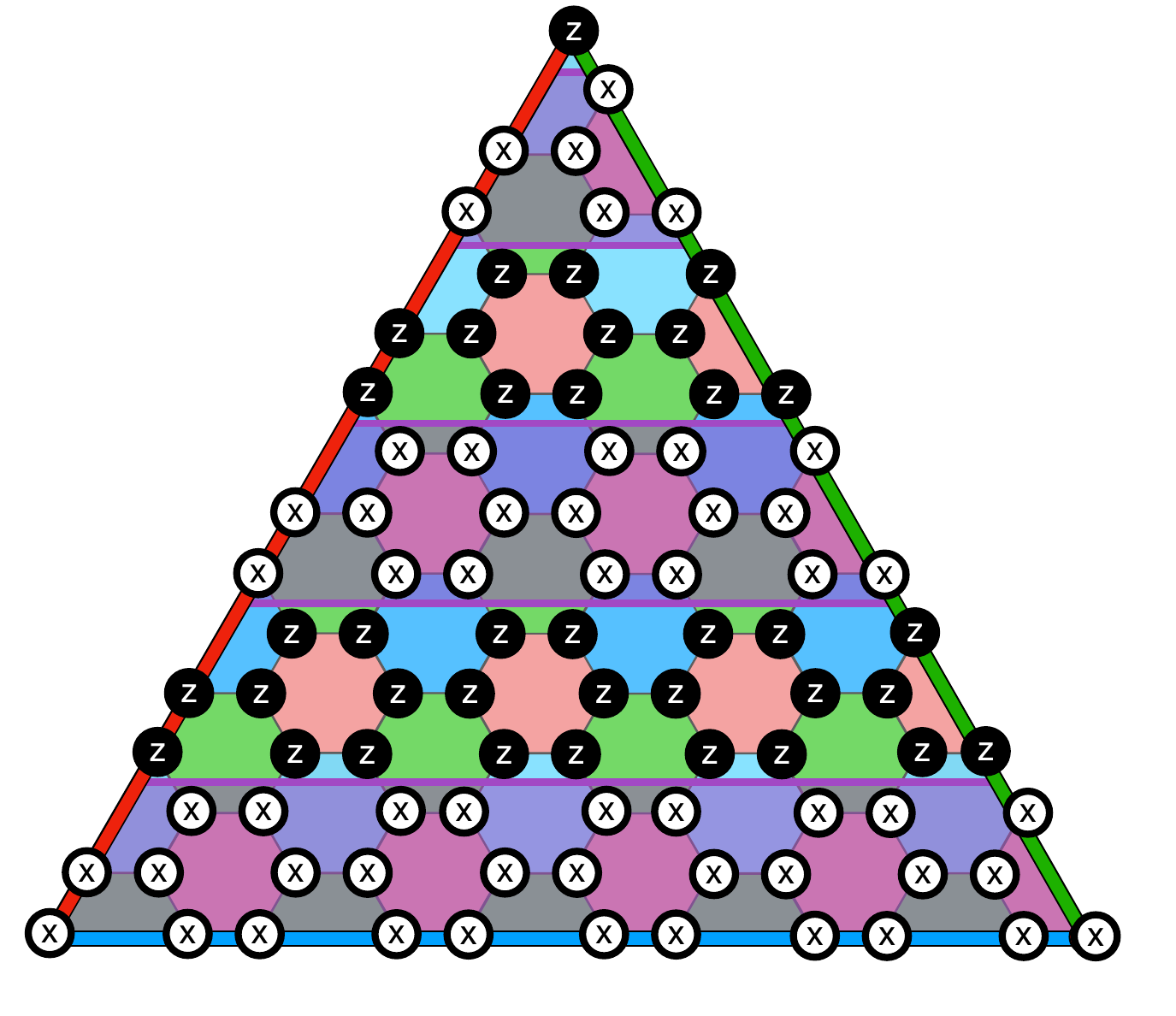}
        \label{fig:underdense-code-b}
    }
    \hfill
    \subfloat[]{
        \includegraphics[width=0.35\columnwidth]
        {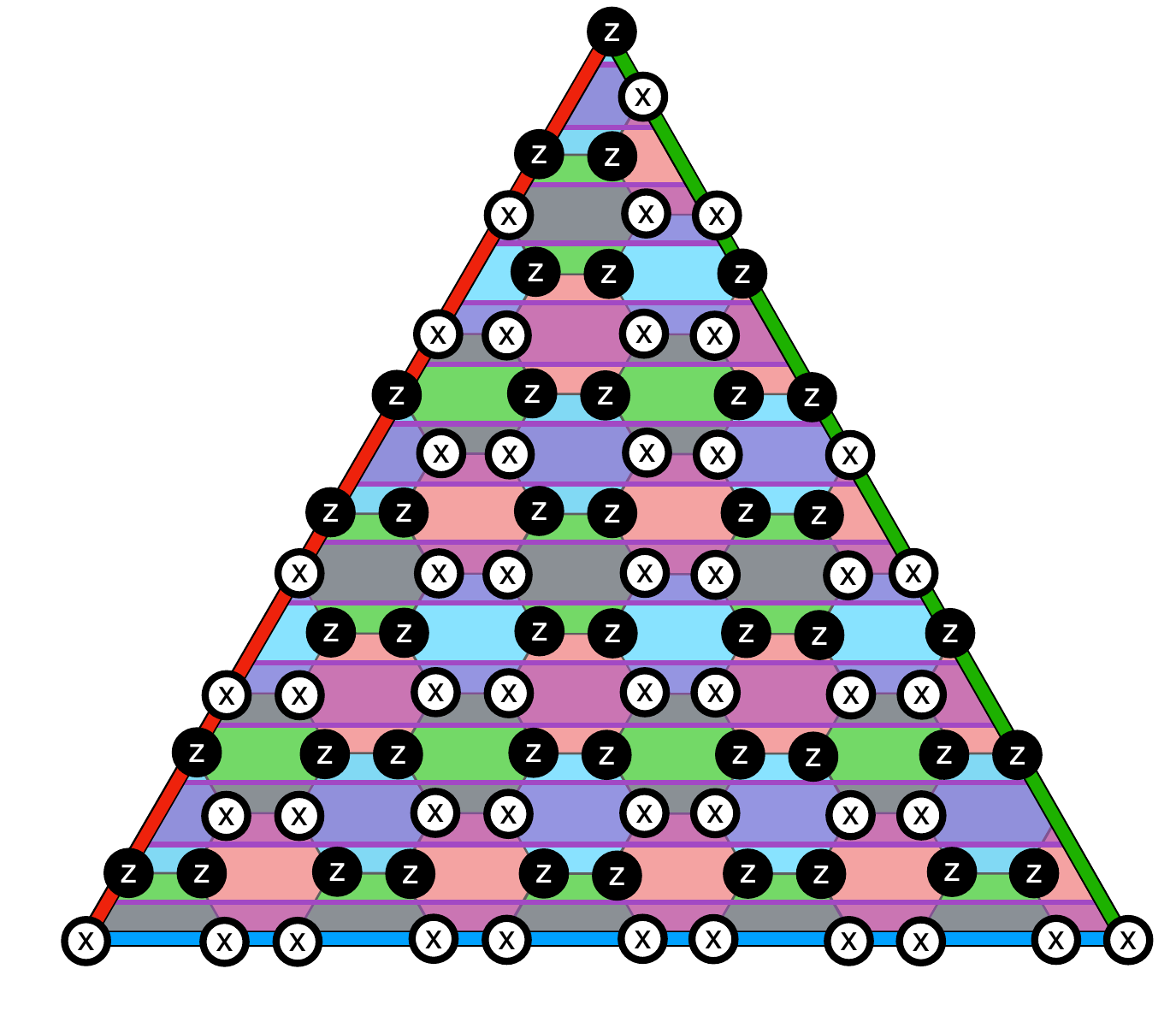}
        \label{fig:overdense-code-a}
    }
    \subfloat[]{
        \includegraphics[width=0.35\columnwidth]
        {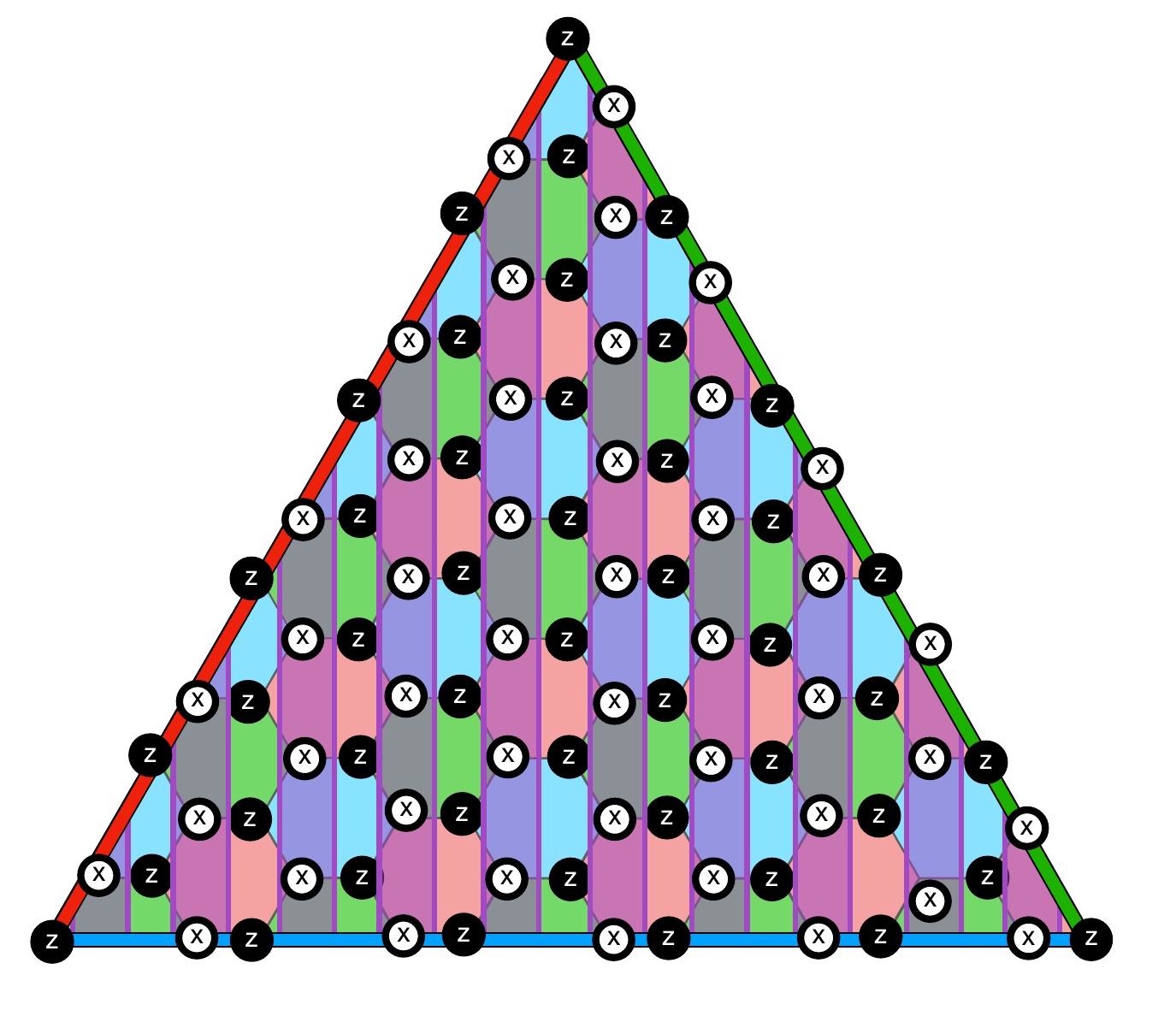}
        \label{fig:overdense-code-b}
    }
    \caption{
    DW codes on a hexagonal lattice. At infinite bias, the anyon dynamics are restricted withing their respective domains of (a)~$\mathcal{DW}$(2/3, 0$)$ and (b)~$\mathcal{DW}$(1/2, 0$)$ underdense DW codes. In contrast, in (c)~$\mathcal{DW}$(3/2, 0$)$ and (d)~$\mathcal{DW}$(2, $\pi/2$) overdense DW codes anyons are allowed to move between domains.
    }
    \label{fig:alternative-codes} 
\end{figure}

In the main text, we primarily focus on the \xz{} color code, a representative of dense DW codes on the hexagonal lattice. Figure~\ref{fig:alternative-codes} shows a few instances of underdense and overdense codes. The $\mathcal{DW}$(2, $\pi/2$) code shown in Fig.~\ref{fig:overdense-code-b} is of particular interest since for $Z$- and $X$-biased noise it is equivalent to the so-called \emph{XYZ} code of Ref.~\cite{https://doi.org/10.48550/arxiv.2203.16534}. In contrast to the \xz{} code, this code has no string-like logical operators at infinite bias. Instead, it reproduces the behavior of a type-II fracton code, whose logical operators have a fractal-like support. As shown in Ref.~\cite{https://doi.org/10.48550/arxiv.2203.16534}, this code reproduces the behavior of a partially self-correcting memory.
 

\begin{figure}[h!]
    \centering
    \includegraphics[width=0.5\textwidth]{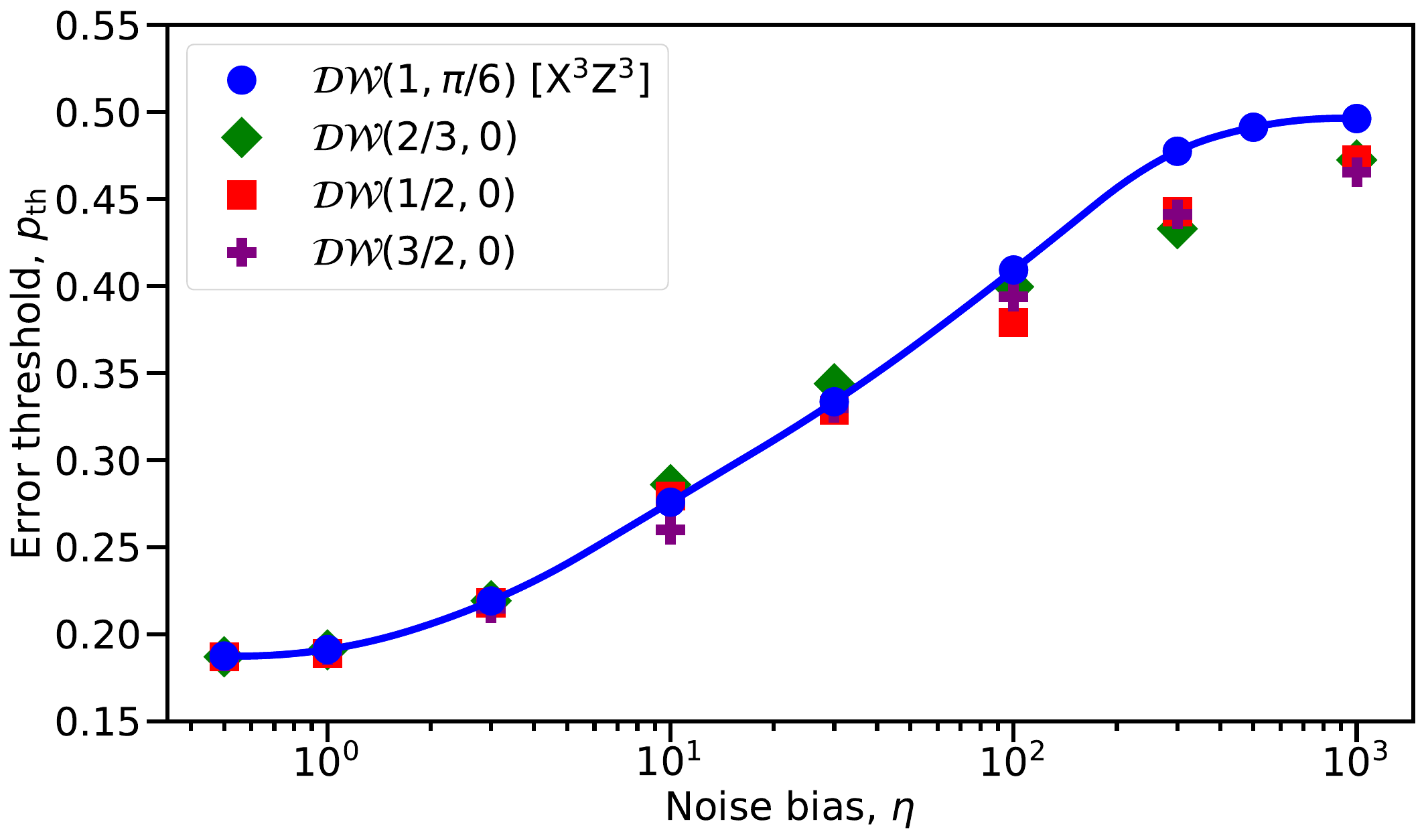}
    \caption{Code-capacity error thresholds of the \xz{} code and various DW codes of Fig.~\ref{fig:alternative-codes}. The curve corresponding to the \xz{} code is interpolated using quadratic splines.}
    \label{fig:codes_thresholds}
\end{figure}

\subsection{Effect of boundaries}

\begin{figure}[h!]
    \centering
    \subfloat[]{
        \includegraphics[width=0.3\columnwidth]{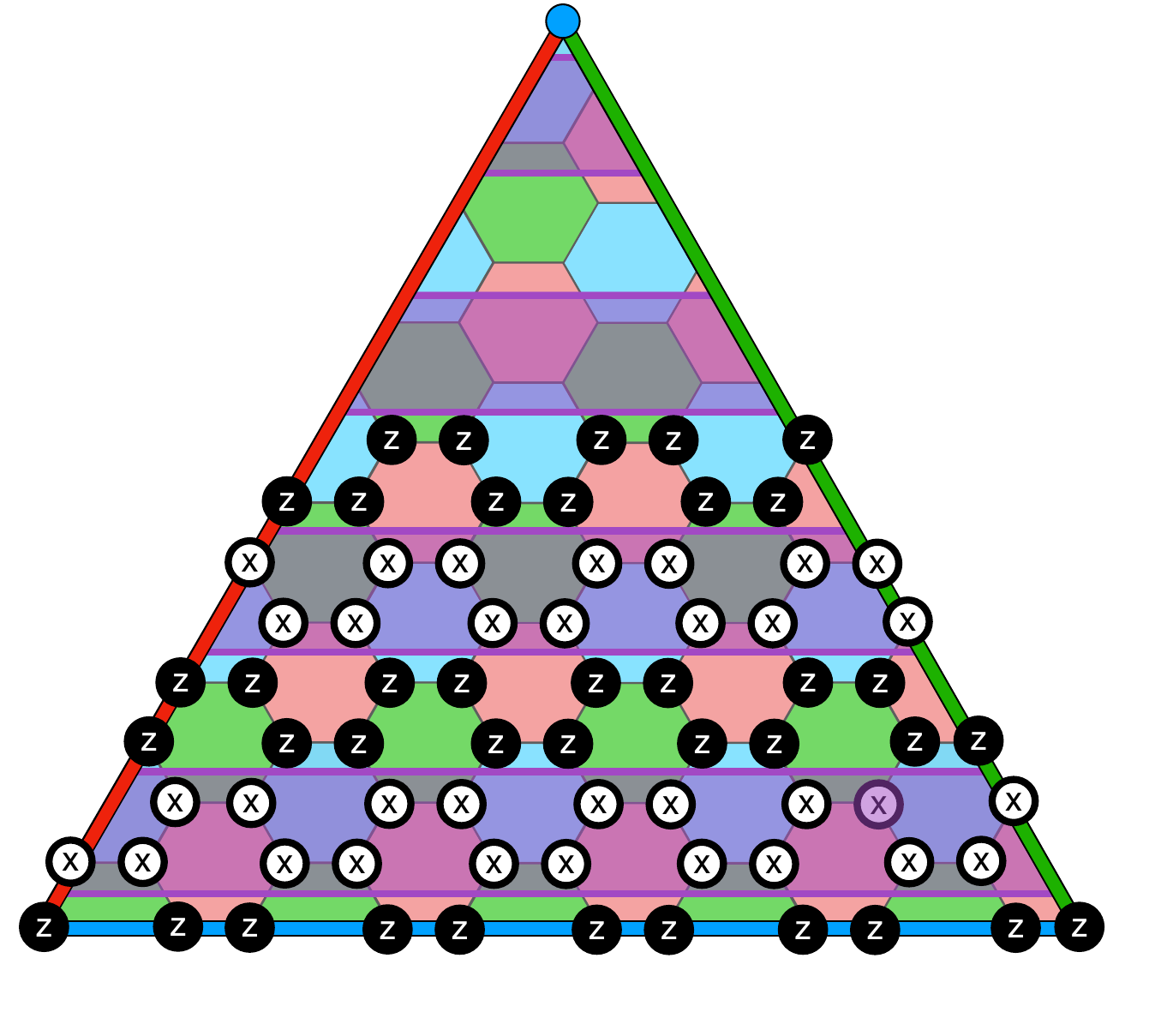}
        \label{fig:effect-of-boundaries-a}
    }
    \subfloat[]{
        \includegraphics[width=0.3\columnwidth]{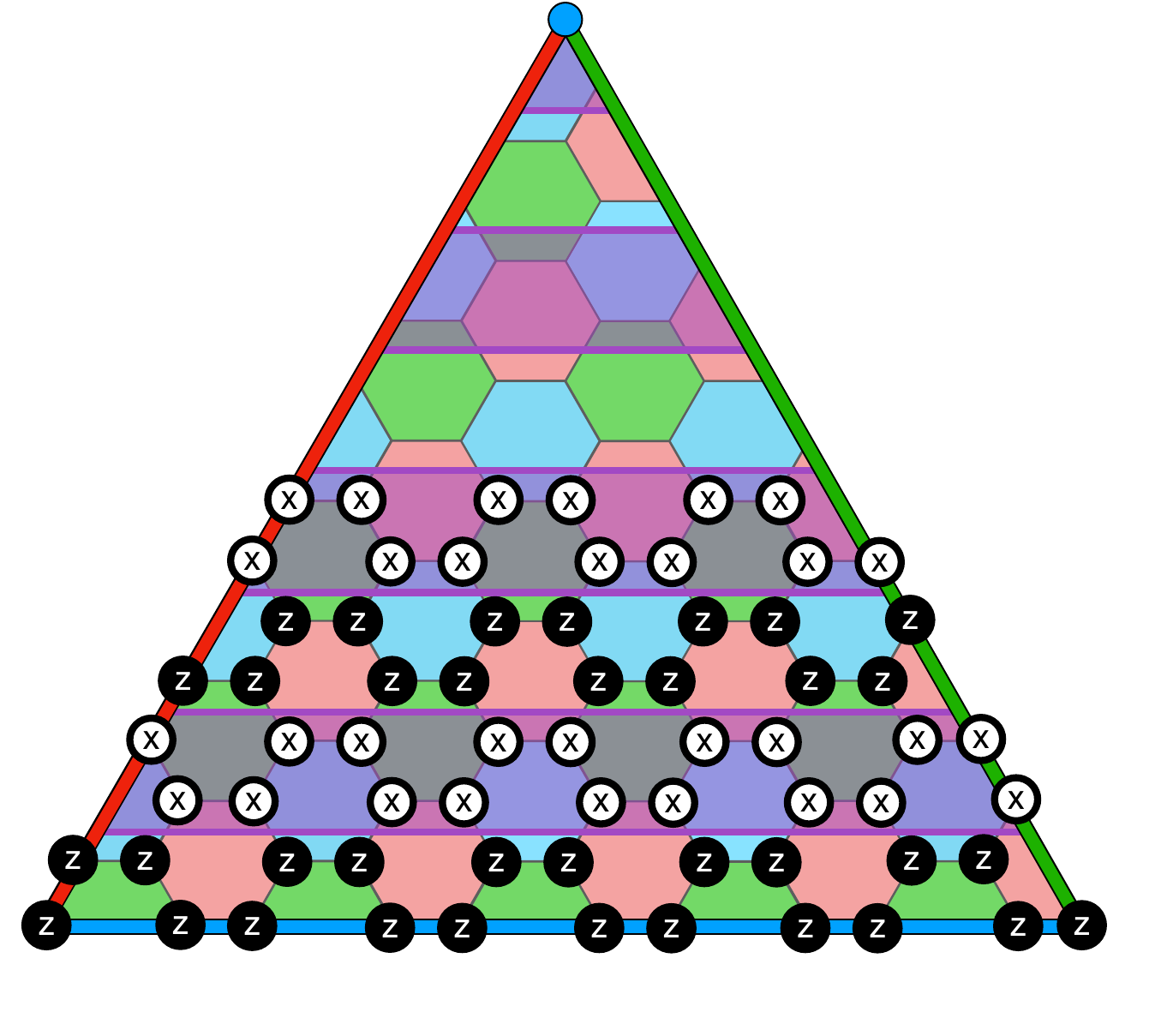}
        \label{fig:effect-of-boundaries-b}
    }
    \subfloat[]{
        \includegraphics[width=0.3\columnwidth]{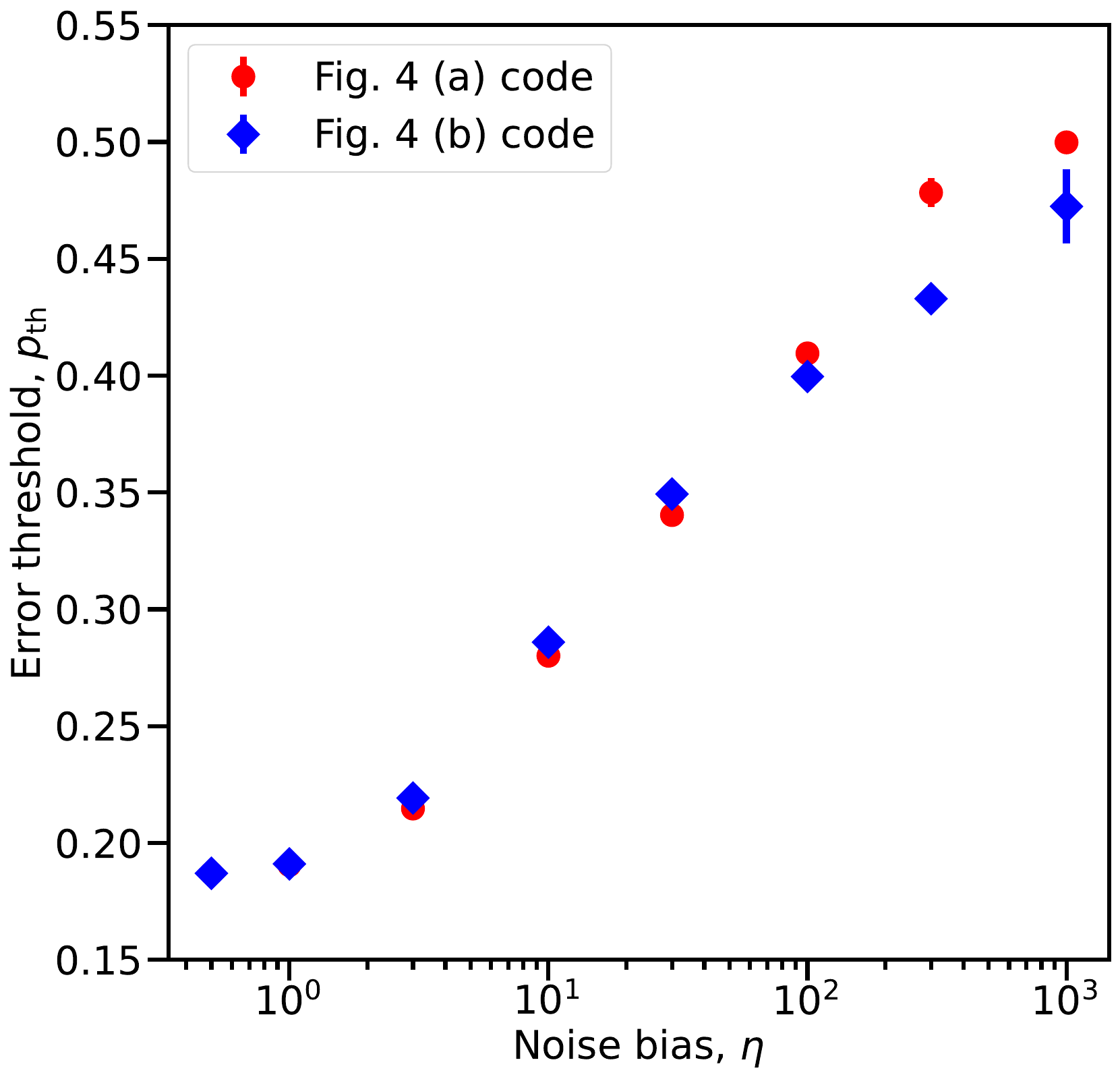}
        \label{fig:effect-of-boundaries-c}
    }
    \caption{
    Distance-11 logical qubits built upon \dw~$(2/3,0)$ codes. (a,b)~Depending on the boundary conditions, a number of pure-$Z$ logical operators can be (a)~1 or (b)~16. (c)~Calculated error thresholds the of the codes shown in (a) and (b).
    }
\label{fig:effect-of-boundaries} 
\end{figure}

We note that the set of parameters $(\kappa, \phi)$ only describes the \emph{bulk structure} of the DW code on a hexagonal lattice. Upon imposing different boundary conditions, logical qubits characterized by the same parameters ($\kappa$, $\phi$) can demonstrate distinct physical properties. As an example, compare the two DW codes of Fig.~\ref{fig:effect-of-boundaries}, both characterized by \dw~$(2/3,0)$. From materialized symmetry argument explained in Appendix \ref{sec:dephasing_threshold}, both codes turn into repetition codes, since multiplying plaquettes of two colors within each domain results in an identity operator. However, the number of infinite-bias logical operators depends on the choice of boundary conditions, with codes of panels~\ref{fig:effect-of-boundaries-a} and \ref{fig:effect-of-boundaries-b} supporting, respectively, 1 and 16 length-$d$ logical operators at infinite bias. 
As a result, the two codes yield noticeably different thresholds at a strong-bias regime~($\eta \geq 10^2$) as demonstrated in Fig.~\ref{fig:effect-of-boundaries-c}. We expect the choice of boundaries to become more important as DWs become less dense because the number of shortest-path logicals
grow fast with the width of domains. When such ambiguity is present, we choose the code boundaries such that the number of logical operators is minimized, hence optimizing the code performance. 

\subsection{Radial DW codes}

\begin{figure}[h!]
    \centering
    \subfloat[]{
        \includegraphics[width=0.3 \columnwidth]{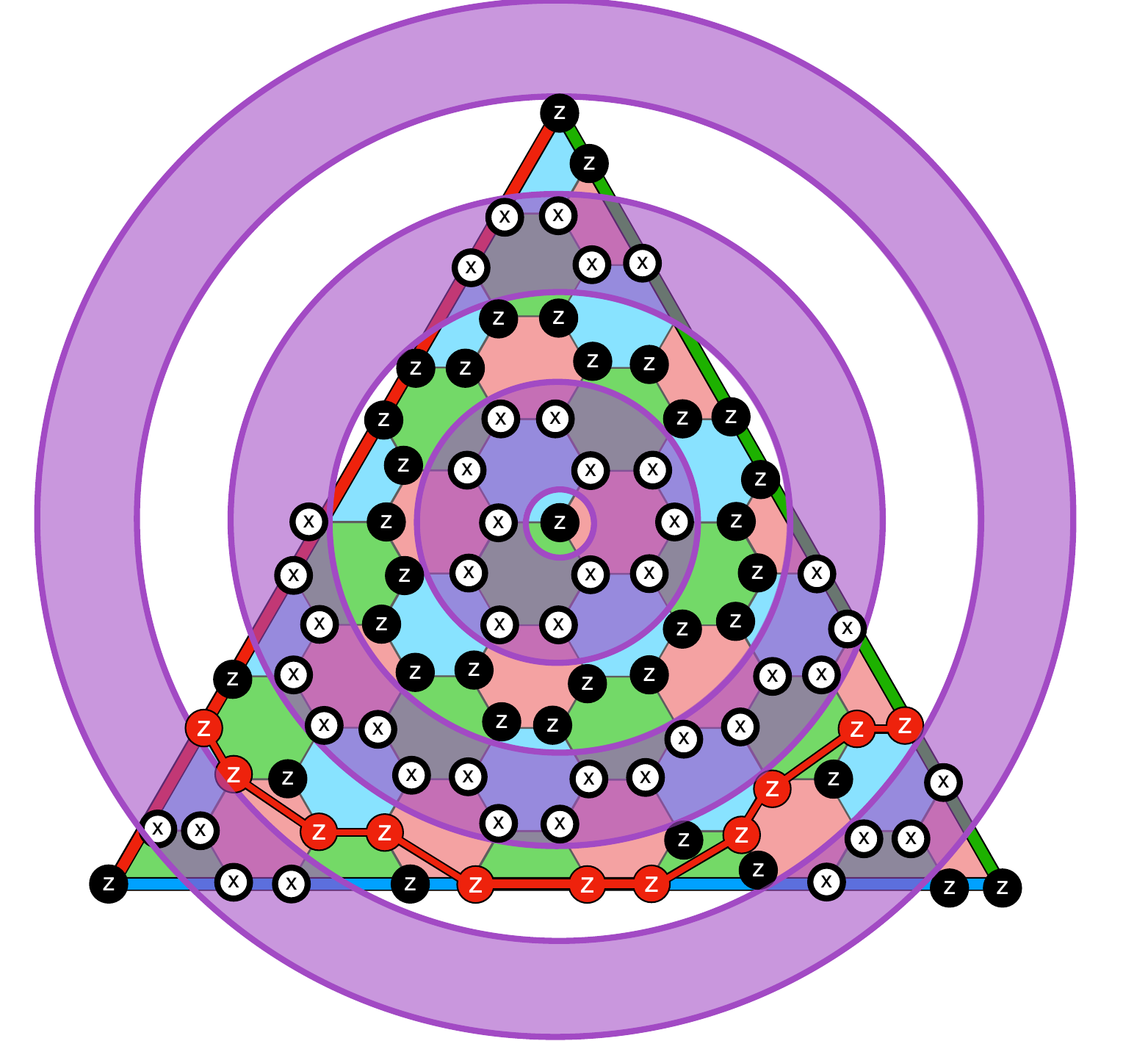}
        \label{fig:radial-a}
    }
    \subfloat[]{
        \includegraphics[width=0.3\columnwidth]{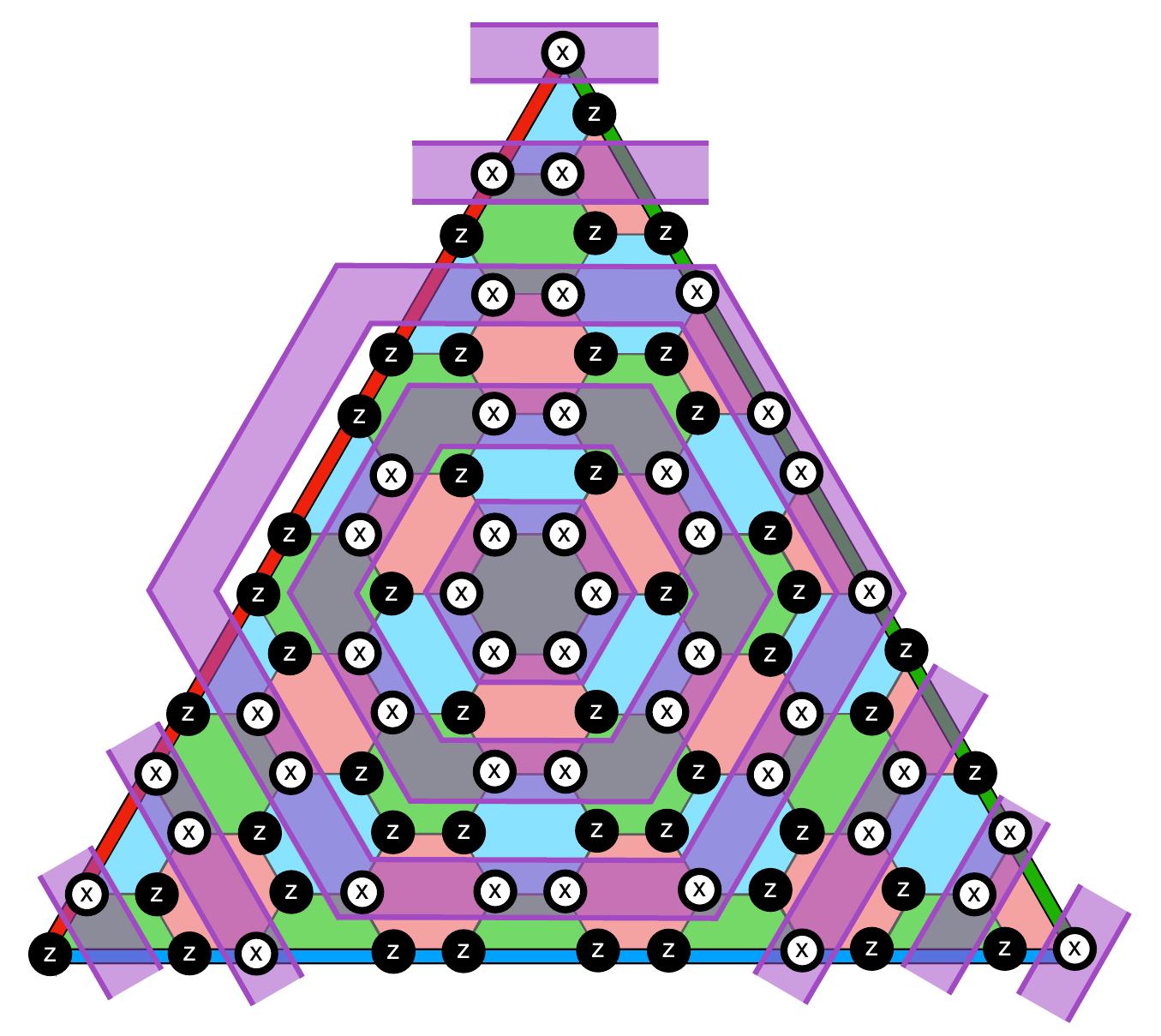}
        \label{fig:radial-b}
    }
    \caption{Radial DW codes with (a)~dense and (b)~overdense packing of domains.}
\label{fig:radial} 
\end{figure}

So far we have constructed various DW codes by introducing linear domains to the code stabilizers. Stabilizer patterns constructed in such a way obey translational symmetry along the direction orthogonal to the domain walls. One can also construct more exotic DW configurations, for example those obeying rotational symmetry. A few instances of such codes are presented in Fig.~\ref{fig:radial}. 
In particular, Figs.~\ref{fig:radial-a} and \ref{fig:radial-b} show examples of dense and overdense radial DW codes.

\section{The XZZX surface code as a DW code}
\label{sec:xzzx}

\begin{figure}[h!]
    \centering
    \subfloat[]{
        \includegraphics[width=0.3 \columnwidth]{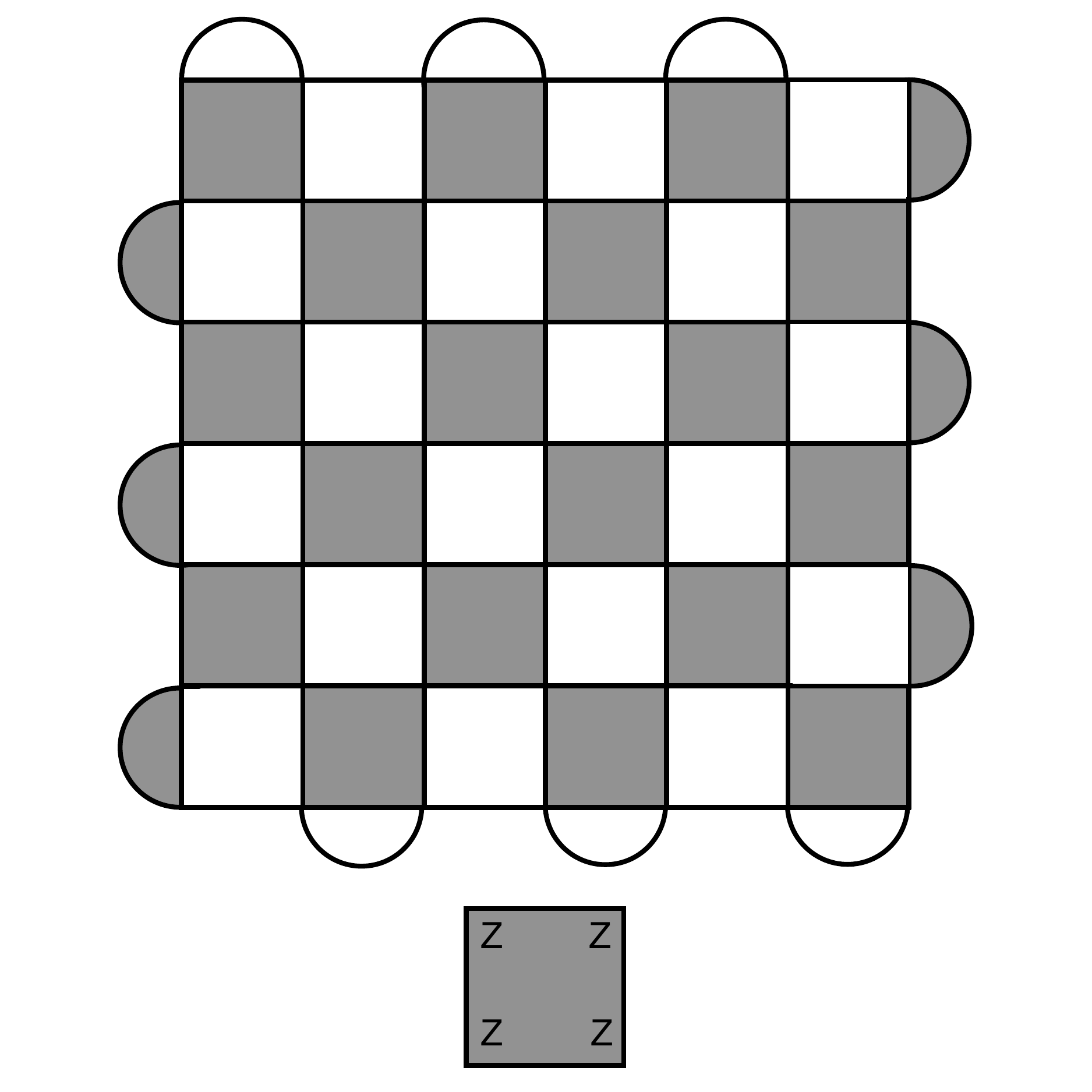}
        \label{fig:xzzx-a}
    }
    \subfloat[]{
        \includegraphics[width=0.3\columnwidth]{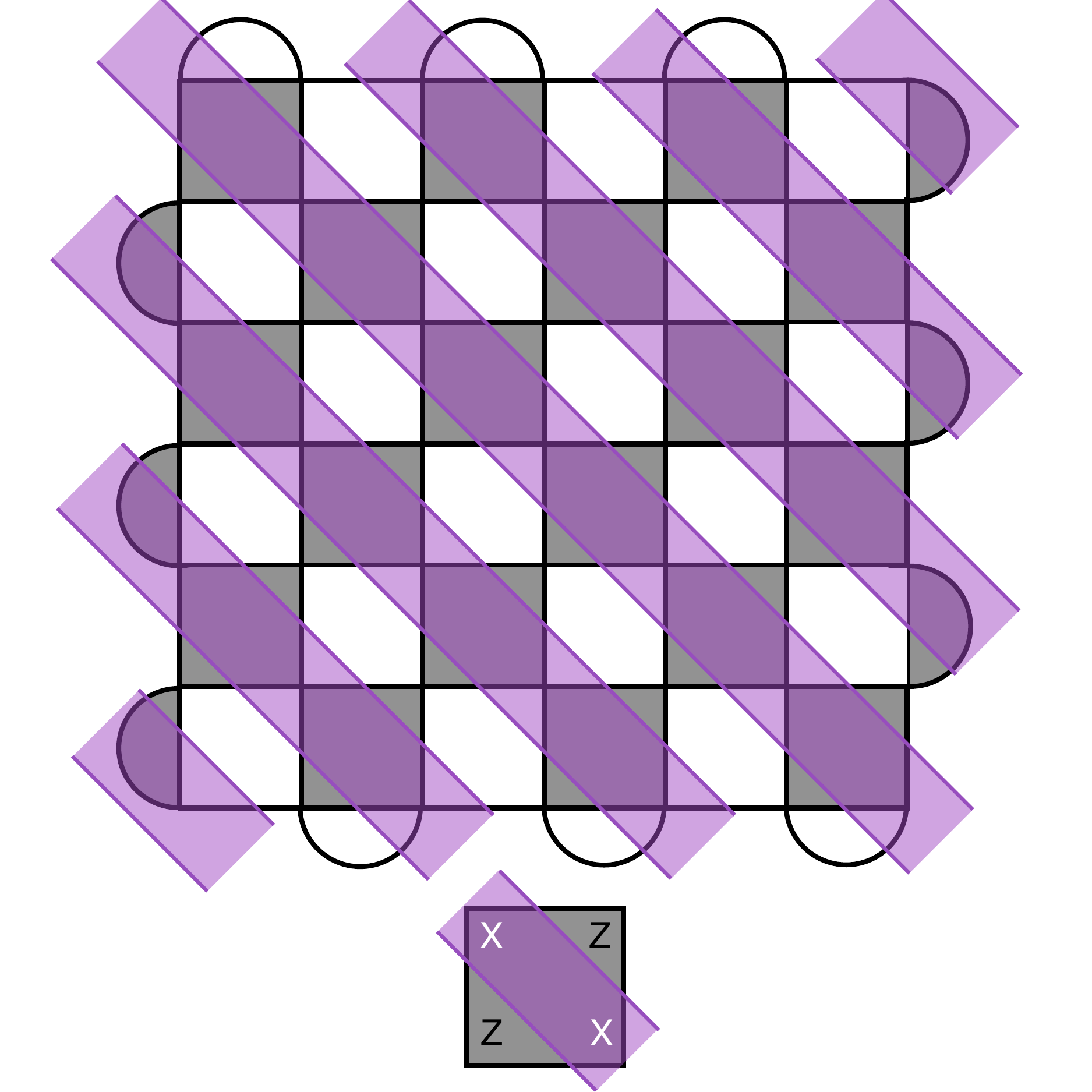}
        \label{fig:xzzx-b}
    }
    \caption{The surface code in its (a)~CSS and (b)~XZZX configurations. The two codes differ by application of Hadamard gates along the purple domains.}
\label{fig:xzzx} 
\end{figure}

The surface code in its XZZX configuration~\cite{XZZX} can also be thought as a representative of a DW code on the squared lattice. Indeed, the standard CSS code turns into the XZZX code when Hadamard gates are applied to qubits within domains located along even diagonals of the lattice, shown with purple in Fig.~\ref{fig:xzzx}. At pure dephasing, anyons can only propagate within their respective domains. We introduce the density of DWs $\kappa$ in the surface code analogously to the case of the color code,
\begin{equation}\label{eq:kappa_sc}
    \kappa = 
    \frac{n_{\textrm{DW}}}{d},
\end{equation}
where $n_{\textrm{DW}}$ denotes the number of DWs crossing any boundary of the code lattice. Hence the XZZX and the \xz{} codes are both characterized by $\kappa=1$ and represent the dense packing of domain walls on the squared and hexagonal lattices, respectively. Intuitively, this can partially explain the identical thresholds of the two codes for all noise channels, observed in Fig.~\ref{fig:2} of the main text. 




\clearpage

\section{Sub-threshold scaling and resource overhead}
\label{sec:subthreshold}

In this Section, we investigate sub-threshold performance of various noise-tailored QECCs. As expected, logical failure rates of the \xz{} code with triangular boundaries scale exponentially with the code distance $d$. We also introduce a version of the \xz{} code with modified periodic boundary conditions. We prove analytically that at pure dephasing, logical failure rates of such modified codes scale exponentially with $d^2$, i.e., with the total number of qubits in the code. 

\begin{figure}[h!]
    \centering
    \subfloat[]{
        \includegraphics[width=0.4\columnwidth]{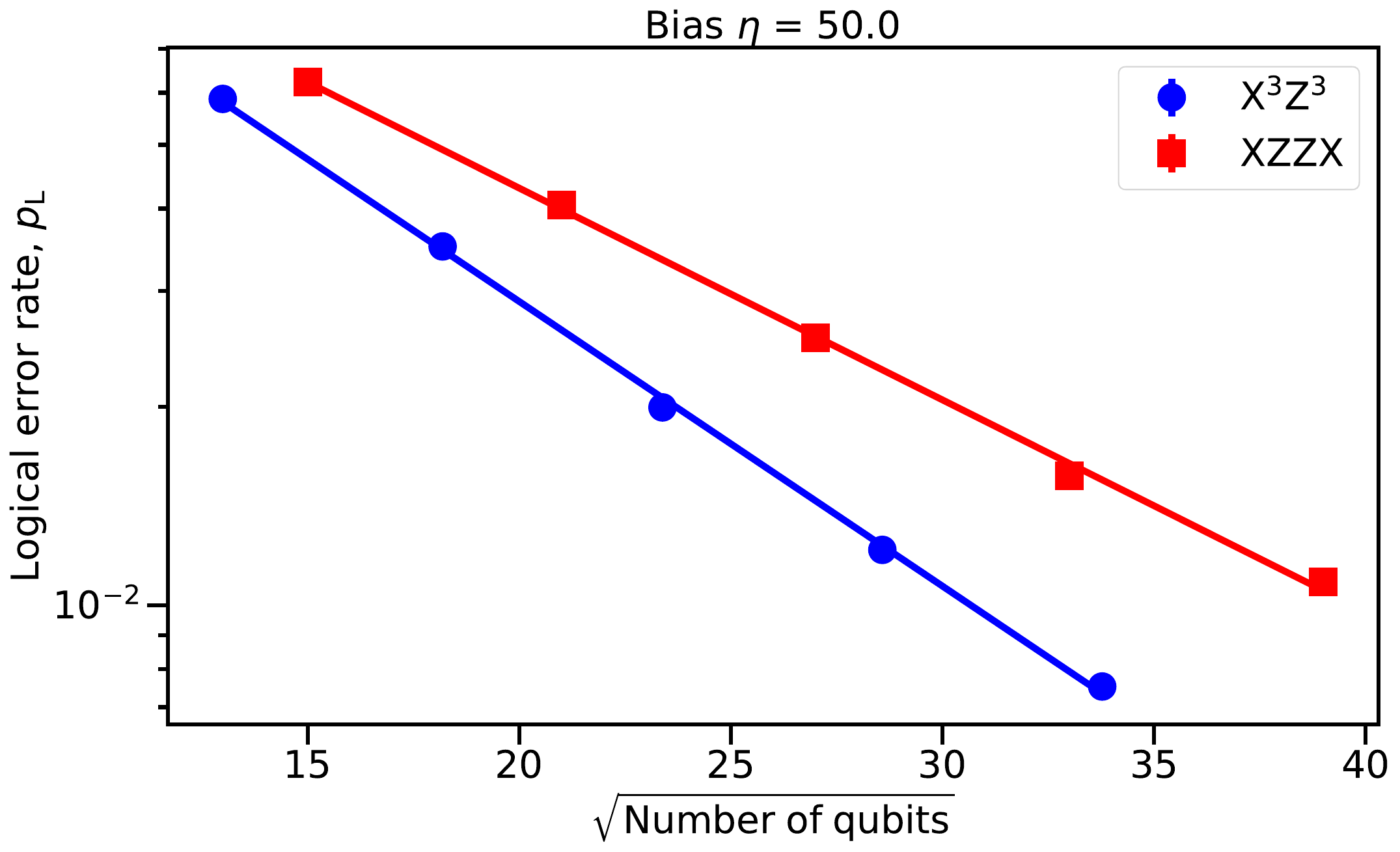}
        \label{fig:subthr-b}
    }
    \subfloat[]{
        \includegraphics[width=0.4\columnwidth]{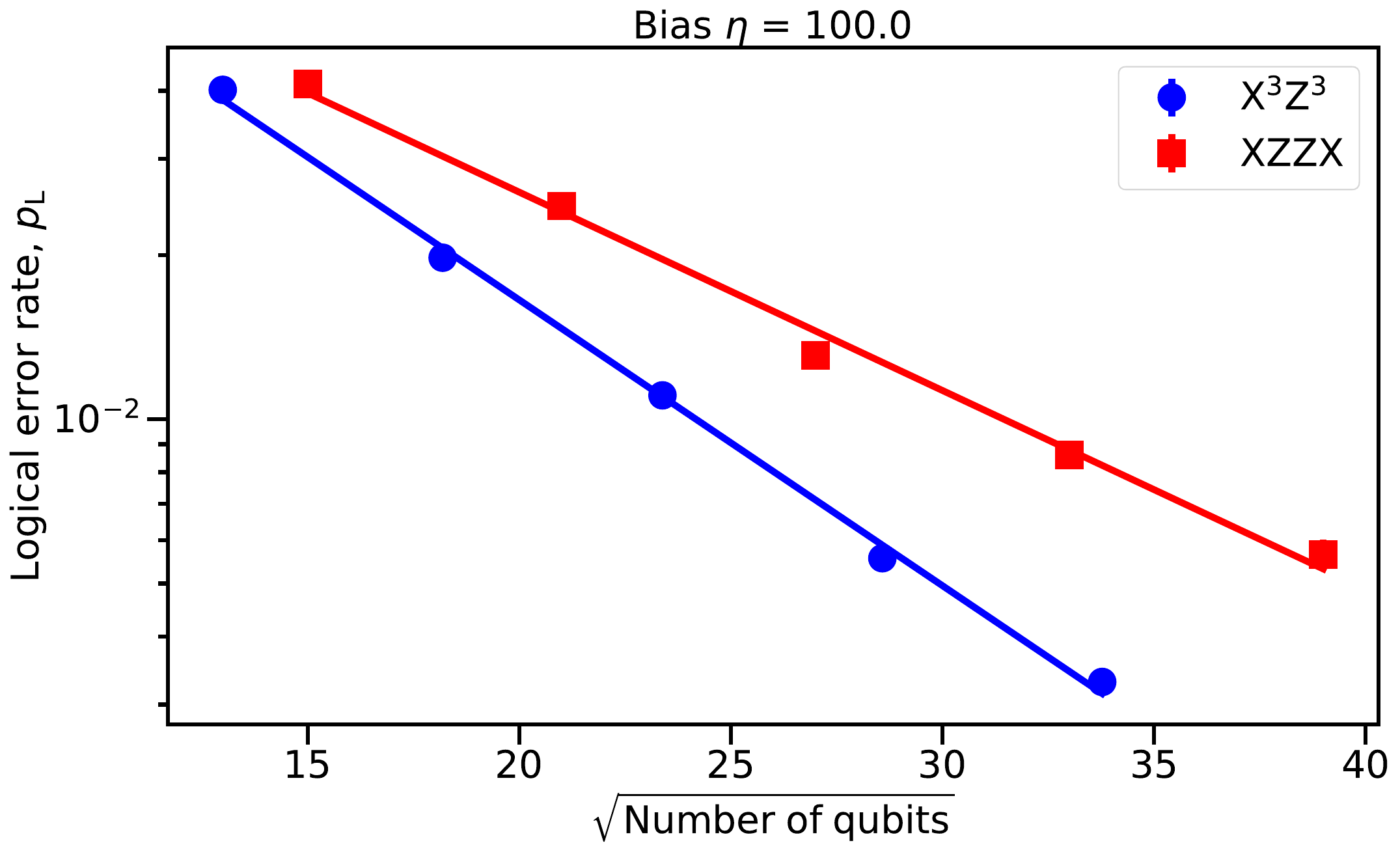}
        \label{fig:subthr-a}
    }
    \caption{
    {Sub-threshold scaling of the triangular \xz{} and squared-shaped XZZX codes with the code size for bias (a)~$\eta=50, p=25$\% and (b)~$\eta=100, p=25$\%.}
    }
\label{fig:subthr} 
\end{figure}

\emph{$\mathit{X^3Z^3}$ code with open boundaries}. In any noise regime, the shortest-path logical operators of the \xz{} code with open boundaries are length-$d$ strings of single-qubit Pauli errors. Therefore, sub-threshold logical failure rate scales exponentially with the code size, similarly to the case of a squared-shaped rotated XZZX code. Quantitatively, however, the \xz{} code is observed to be more resource-efficient, as shown in Fig.~\ref{fig:subthr}. The desired logical failure rate of the \xz{} code can hence be achieved with smaller codes compared to its surface-code counterpart. 

\begin{figure}[h!]
    \centering
    \subfloat[]{
        \includegraphics[width=0.35\columnwidth]{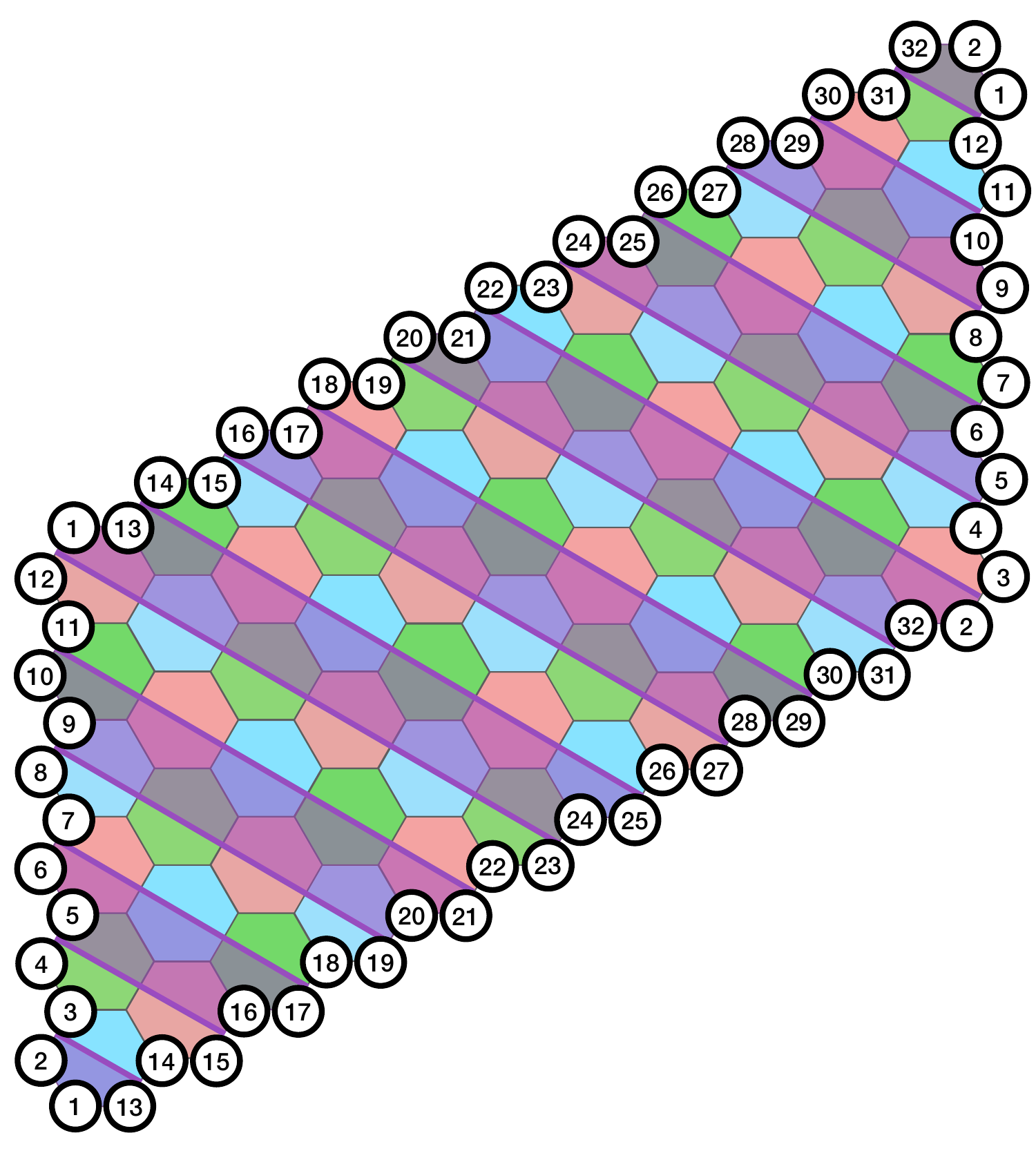}
        \label{fig:subthr_improved-a}
    }
    \subfloat[]{
        \includegraphics[width=0.35\columnwidth]{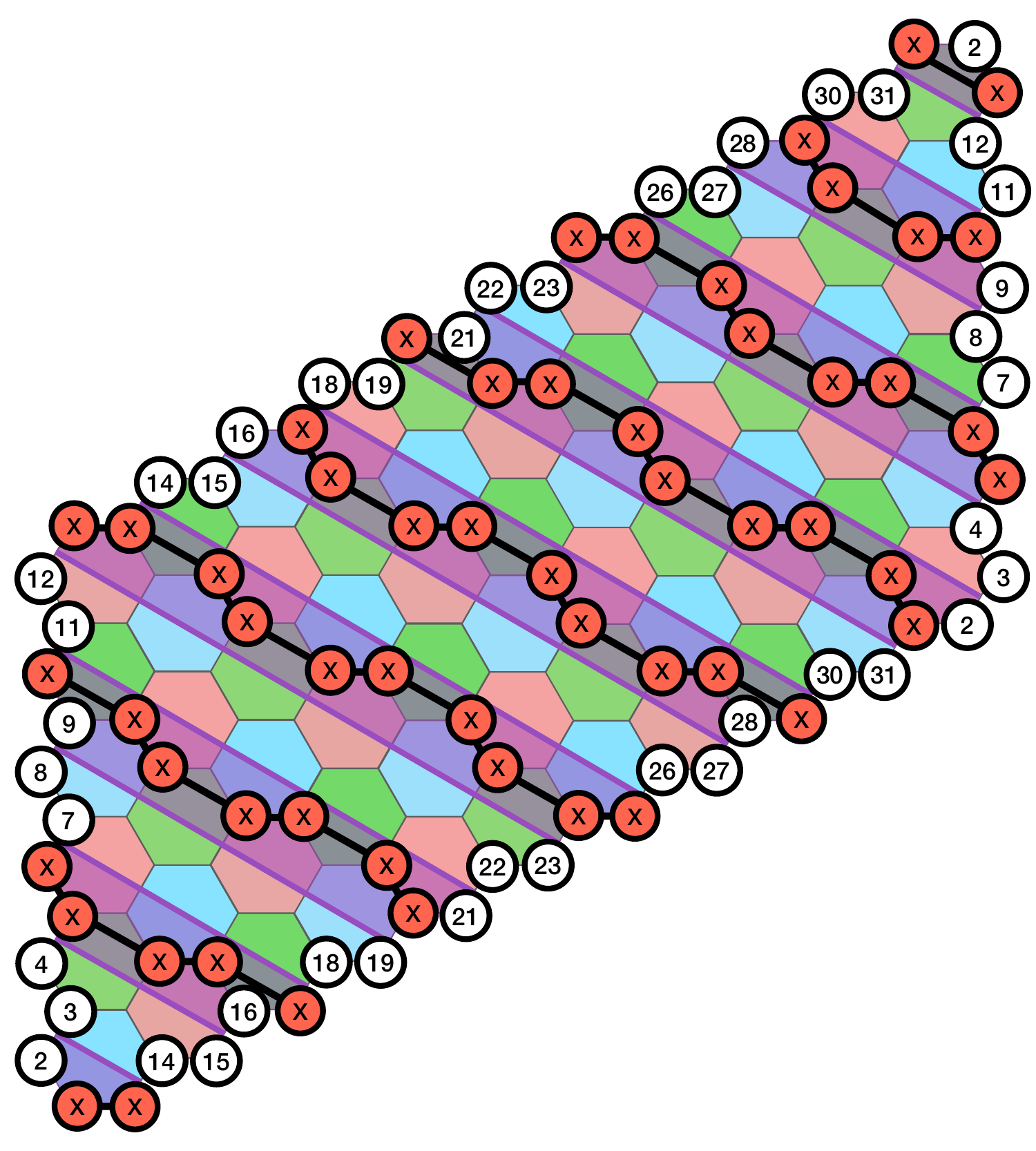}
        \label{fig:subthr_improved-b}
    }
    \caption{
    (a)~An \xz{} code with periodic boundary conditions on a periodic lattice of $11\times 6$ hexagons. As in Fig.~\ref{fig:linear-symmetries-0}, the enumerated qubits represent the boundary conditions. (b)~The shortest logical-$X$ operator in the limit of pure bit-flip noise has to wrap around the lattice, with a support on $1/3$ of all qubits of the code.}
\label{fig:subthr_improved} 
\end{figure}

\emph{$\mathit{X^3Z^3}$ code with periodic boundaries}. At pure dephasing, the \xz{} code with modified boundaries requires quadratically fewer qubits to achieve a desired logical failure rate compared to the triangular-shape code. Specifically, consider a version of the \xz{} code on a periodic lattice of dimensions $(6k,12k-1)$, $k \geq 1$, which we will refer to as a co-prime lattice in analogy with a co-prime squared-shape XZZX code introduced in Ref.~\cite{XZZX}. Note that in Sec.~\ref{sec:dephasing_threshold}, it has been argued that the code size in each direction has to be a multiple of 6 in order to meet both three-colorability and the Clifford-deformation conditions. Here, the co-prime dimensions are permitted due to an additional twist that we apply along the code boundary. An instance of a co-prime hexagonal code with $k=1$ is shown in Fig.~\ref{fig:subthr_improved}. As we show in the figure and discuss in the caption therein, the shortest-path logical operators consists of $N_{\textrm{q}}/3$ single-qubit errors in the limit of pure dephasing, where $N_{\textrm{q}}$ is the total number of data qubits in the code. Consequently, the logical failure probability scales as
\begin{equation}
    \log{p_{\textrm{L}}}
    \propto
    -N_{\textrm{q}}
    \propto
    -d^2, 
\end{equation}
resulting in significantly reduced qubit overhead, analogously to the case of the co-prime XZZX code~\cite{XZZX}. 


\section{DW codes on lattices with higher-weight stabilizers}
\label{sec:4.8.8}

\begin{figure}[h!]
    \centering
    \subfloat[]{
        \includegraphics[width=0.4\columnwidth]
        {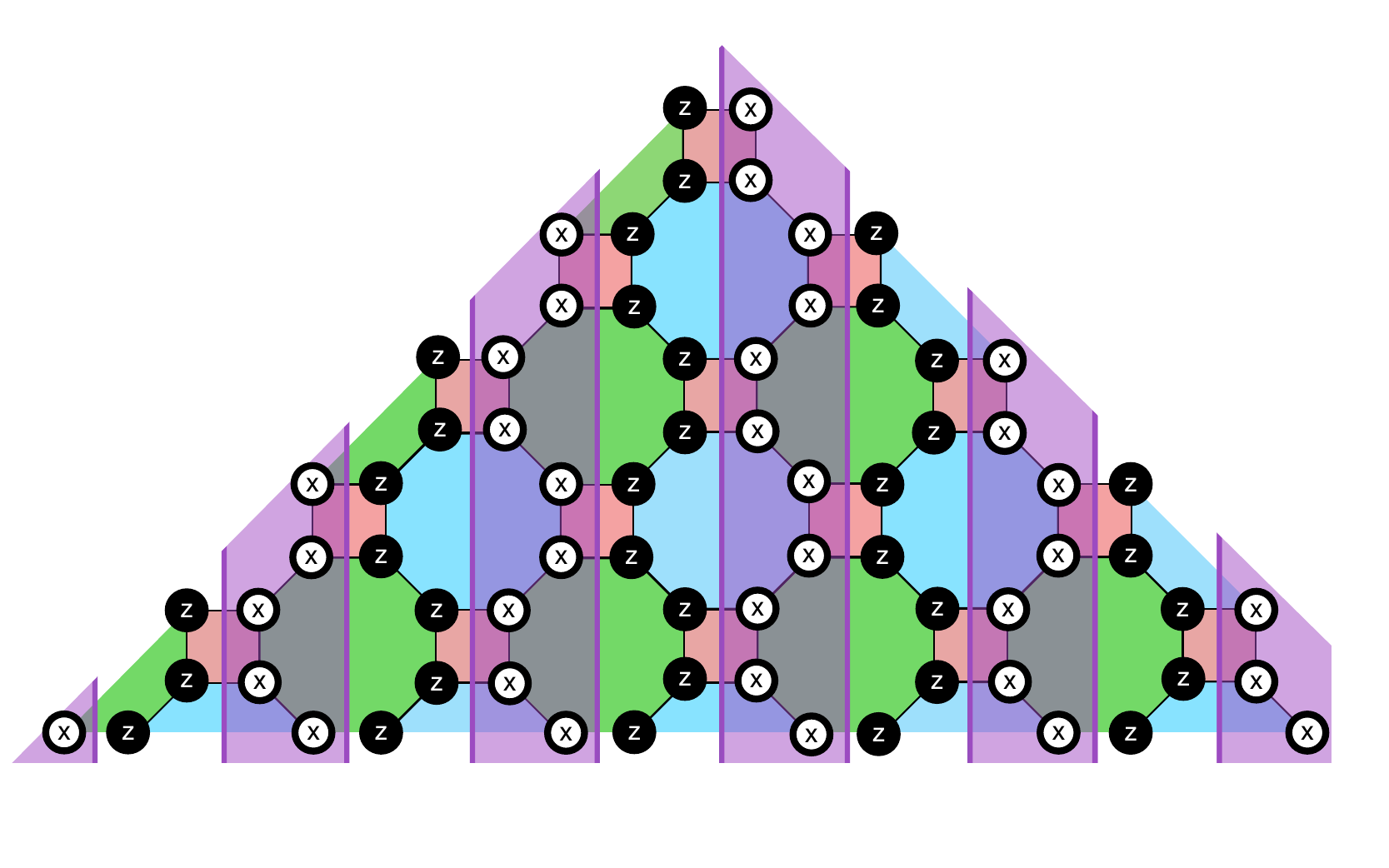}
        \label{fig:4.8.8-codes-a}
    }
    \subfloat[]{
        \includegraphics[width=0.4\columnwidth]
        {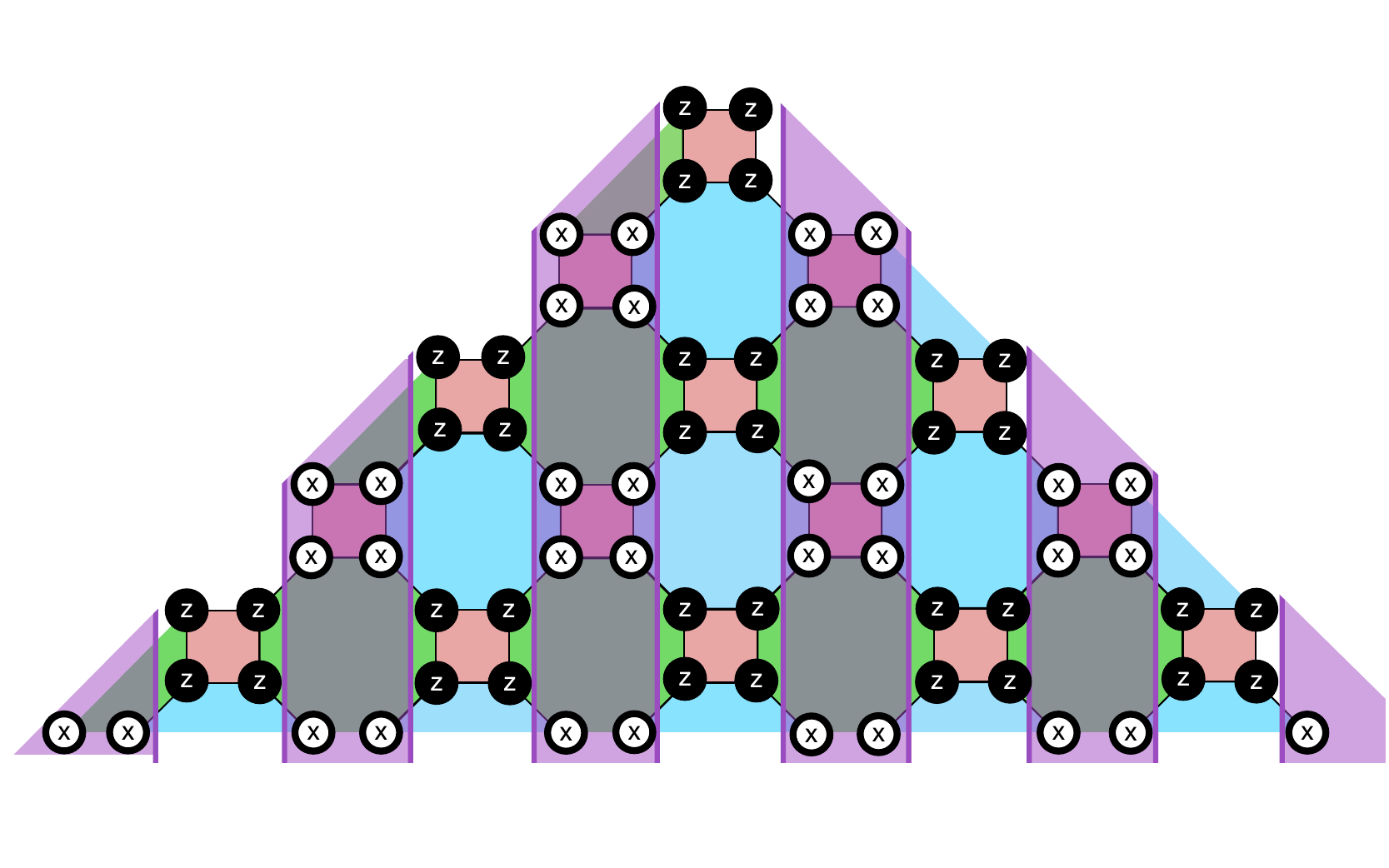}
        \label{fig:4.8.8-codes-b}
    }
    \hfill
    \subfloat[]{
        \includegraphics[width=0.4\columnwidth]
        {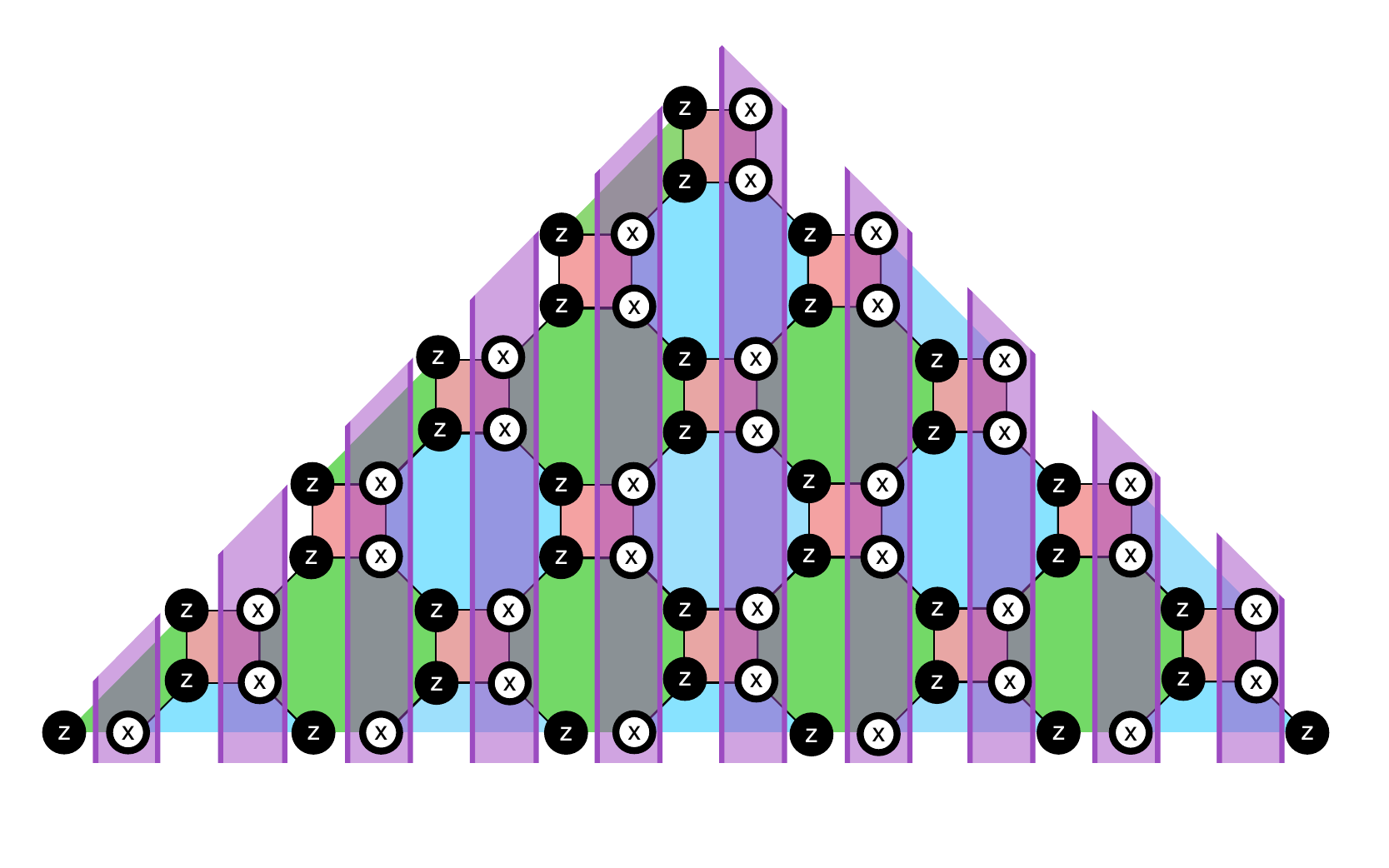}
        \label{fig:4.8.8-boundaries-a}
    }
    \subfloat[]{
        \includegraphics[width=0.4\columnwidth]
        {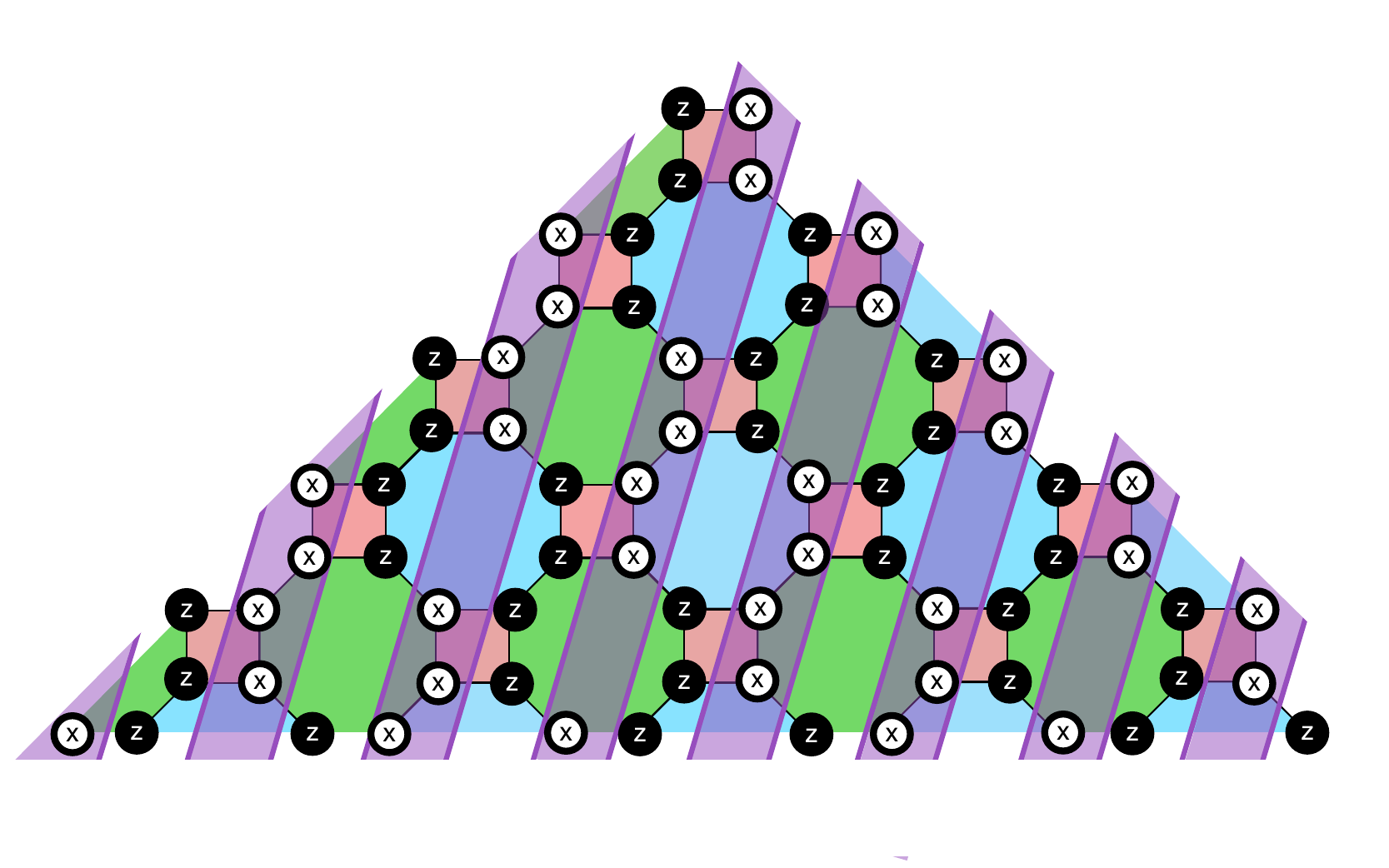}
        \label{fig:4.8.8-boundaries-b}
    }
    \caption{Examples of (a,b)~dense and (c,d)~overdense DW color codes on the 4.8.8 lattice.}
\label{fig:4.8.8-codes} 
\end{figure}

A DW code structure can be imposed on qubit lattices with higher-weight stabilizers. Figure~\ref{fig:4.8.8-codes} shows four examples of DW codes with various density and orientation on a 4.8.8 lattice. We note that due to a higher weight of stabilizers, specifying parameters $\kappa$ and $\phi$, as in the 6.6.6 case, is not sufficient to fully characterize the code structure, even in bulk. For instance, the codes of Fig.~\ref{fig:4.8.8-codes-a} and \ref{fig:4.8.8-codes-b} are characterised by the same set of parameters $\kappa$ and $\phi$, yet, give rise to different dynamics of anyons. 

\clearpage
\section{Matching-based decoder}
\label{sec:matching}

In Sec.~\ref{sec:dephasing_threshold}, we have proven that a matching decoder yields a threshold of 50\% at infinite bias. Here, we implement a matching-based restriction decoder for finite-bias noise and show that it yields thresholds that monotonically increase with noise bias. 

The restriction decoder efficiently decodes the color code by pair-wise matching syndromes on restricted lattices~\cite{delfosse-restriction}. The decoder constructs a syndrome graph, with graph vertices corresponding to the flipped stabilizers and edges corresponding to chains of errors connecting pairs of flipped stabilizers. The edges are weighted according to the probabilities of errors chains for a given noise model, such that the more probable error chains are characterized by smaller weights. The algorithm then searches for a perfect matching
such that the sum of the weights is minimal, corresponding to the most probable error configuration. Information about the noise bias can be exploited to determine the edge weights and fit the decoder to the noise model. 

For the color code, Pauli-$X$ and $Z$ errors anti-commute with three stabilizers, and therefore a naive syndrome graph contains hyperedges. Using this syndrome graph directly to decode is not feasible as this would require 3-dimensional matching which is NP-complete~\cite{karp1972reducibility}. This can be resolved by splitting the $X$ and $Z$ syndrome graph of the color code onto restricted graphs. Each of these graphs contains the syndromes on two of the three colors of plaquettes and is equivalent to the syndrome graph of a surface code. Consequently, single-qubit errors on the restricted graphs anti-commute with two stabilizers, creating anyonic excitations in pairs. Matching can thus be performed to pair the anyons  via the most probable error chains, e.g., using the MWPM algorithm~\cite{edmonds_1965}. Using the output of the matchings on the restricted graphs, a correction can be efficiently calculated as explained in Ref.~\cite{chamberland2020triangular}. 

\begin{figure}[h!]
    \centering    
    \includegraphics[width=0.5\textwidth]{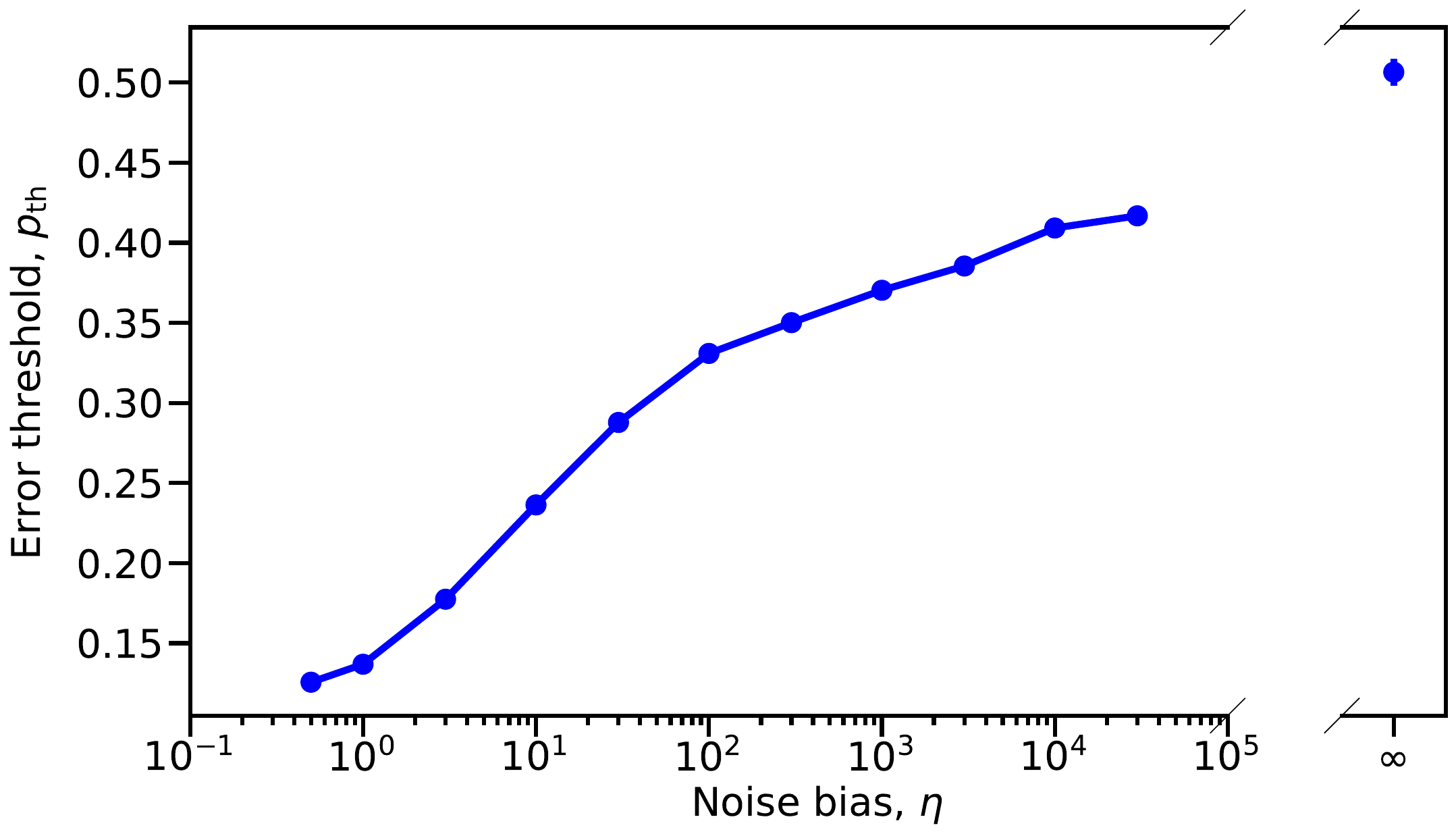}
    \caption{Numerically calculated noise thresholds of the \xz{} code with the restriction decoder at different noise biases. Each threshold is obtained using at least $10^6$ samples or 5000 logical errors. For $\eta < 10^4$ code distances $d=\{9, 11, 13, 17, 21 \}$ are used and for $\eta \geq 10^4$ code distances $d=\{17, 21, 25, 29\}$ are used.}

    \label{fig:restriction_decoder}
\end{figure}

We have implemented the restriction decoder using the PyMatching package~\cite{higgott2022pymatching}. Our implementation, generated data and threshold plots are available at~\cite{dwcc_repo}. The obtained thresholds for different biases are presented in Fig.~\ref{fig:restriction_decoder}. The threshold at depolarizing noise is found to be 12.6\%, exactly agreeing with the threshold found in Ref.~\cite{chamberland2020triangular}. The threshold monotonically increases up to 41.7\% at $\eta=30000$. As expected, the threshold at infinite bias (pure dephasing) is 50\%.

\clearpage
\section{Method for threshold estimation}

\begin{figure}[h!]
    \centering
    \subfloat[]{
        \includegraphics[width=0.4\columnwidth]{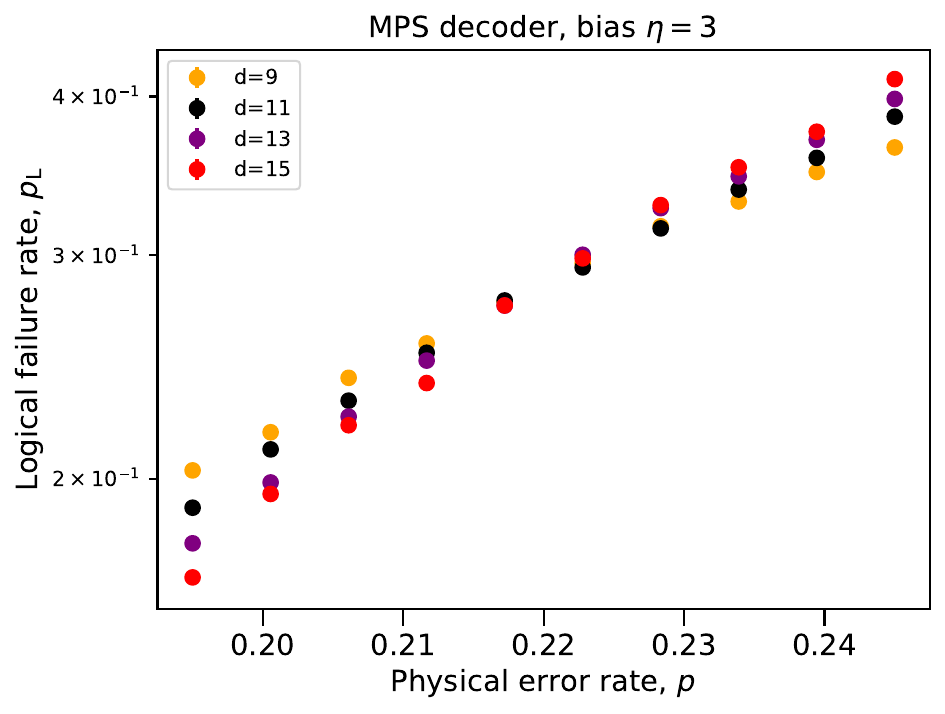}
        \label{fig:thresholds_extraction-a}
    }
    \subfloat[]{
        \includegraphics[width=0.4\columnwidth]{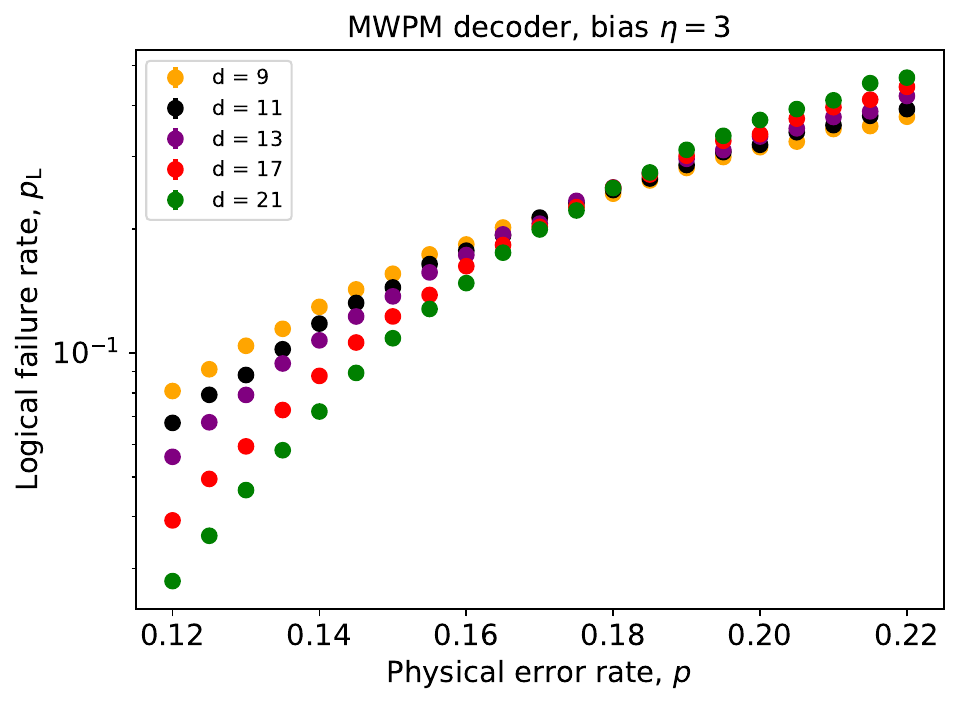}
        \label{fig:thresholds_extraction-b}
    }
    \caption{
    Data used for extracting code-capacity thresholds using (a)~the tensor-network and (b)~the matching-based restriction decoder at noise bias $\eta = 3$.
    }
\label{fig:thresholds_extraction} 
\end{figure}

With both the tensor-network decoder and the restriction decoder, the threshold estimates are calculated using the \emph{critical exponent method} described in Ref.~\cite{harrington2004analysis}. Specifically, for physical error rates near the threshold we fit the logical error rate to the 
function $P_{\textrm{fit}}: \mathbb{R}^+\rightarrow \mathbb{R}$ with
\begin{equation}
    P_{\textrm{fit}} (x) = B_0 + B_1 x + B_2 x^2,
\end{equation}
where
\begin{equation}
x := (p - p_{\textrm{th}})^{\alpha d ^{\beta}}. 
\end{equation}
Here, $d$ is the code distance, $p$ is the single-qubit error rate, and $B_0,B_1,B_2,\alpha,\beta$ are fitting parameters. Examples of data used for threshold extraction for both the restriction decoder and the tensor-network decoder are shown in Fig.~\ref{fig:thresholds_extraction}.

\clearpage
\section{Details of numerical simulations}

To decode the error syndrome, we adapt the tensor-network decoder of Ref.~\cite{PhysRevX.9.041031}.
To simulate the performance of Clifford-deformed DW codes, we use one of two techniques. The first method is to explicitly implement a Clifford transformation to the code stabilizers and logical operators. The applied Clifford transformation then has to be accounted for when constructing the decoder tensor network. Alternatively, one can simulate the dynamics of the DW codes with the CSS code by introducing an effective error model. In particular, to simulate the \xz{} code with the CSS code, we apply Hadamard transformation to the noise channel affecting qubits along even diagonals of the code, which effectively permutes the probabilities of $X$ and $Z$ errors. Hence, to simulate the \xz{} code with the CSS code we simply permute $p_X$ with $p_Z$ in the noise model affecting half of the codes qubits. A similar permutation is applied to the error probabilities fed to the tensor-network decoder. Our simulations confirm that the two methods return the same results. A similar effective error model has been used in the simulations of the XZZX code in Ref.~\cite{XZZX}. 

\begin{figure}[h!]
    \centering    \includegraphics[width=0.45\textwidth]{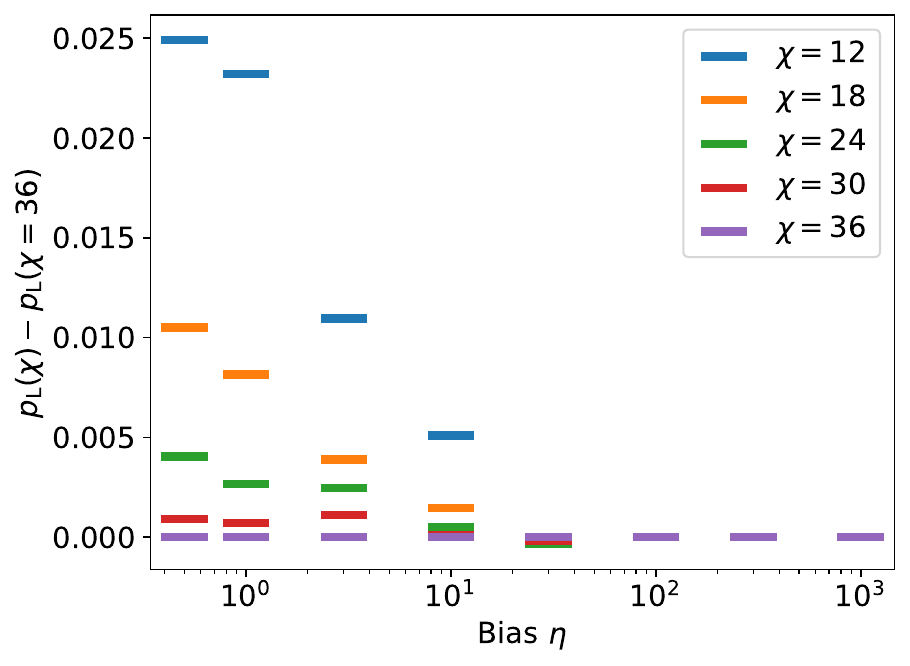}
    \caption{Convergence of the tensor-network decoder close to the threshold for different values of noise bias $\eta$. Different color bars indicate the error at a given $\eta$ relative to the logical error rate at $\eta = 36$.}
    \label{fig:chi_convergence}
\end{figure}

The performance of the tensor-network decoder depends on its truncated bond
dimension $\chi$, which determines the size of a local tensor; the accuracy of the algorithm improves with the bond dimension, but the run-time of the algorithm is exponential in $\chi$. Figure~\ref{fig:chi_convergence} shows the convergence of the solution as a function of $\chi$ for different values of noise bias. As seen from the figure, for small bias larger bond dimensions are required to derive a solution with high accuracy, while at strong bias small $\chi$ suffices.

Finally, in order to extract the noise threshold, we have to simulate codes of increasingly larger distances as we increase bias. In particular, our simulations show that at a given value of bias, the empirical threshold monotonically decreases as we increase the distances of simulated codes we extract it from. For large enough codes, the threshold saturates at a value which we identify as the true threshold of the code. As such, the threshold plots of Figs.~\ref{fig:codes_thresholds}, \ref{fig:effect-of-boundaries-c}, and Fig.~\ref{fig:2}(a) of the main text are extracted by simulating codes of sizes $d = \{ 9, 11, 13, 15 \}$~(for $\eta < 30$), $d = \{ 25, 27, 29, 31, 33, 35 \}$~(for $30 \leq \eta \leq 100$), and $d = \{ 39, 43, 47, 51 \}$~(for $\eta > 100$). We note that a similar effect has been pointed out by the authors of Ref.~\cite{XZZX} for the XZZX code.

We explain the ill-defined threshold at small distances by finite-size effects. At infinite bias, propagation of anyons in the \xz{} code is restricted within 1D domains bounded by domain walls, as described in the main text. Similarly, anyons in the XZZX code can only propagate along diagonals, see Ref.~\cite{XZZX} for a detailed discussion. Hence, both the \xz{} and the XZZX codes constitute a series of 1D repetition code and yield a threshold of 50\%. At finite bias, anyons are allowed to propagate across 1D manifolds, violating the repetition-code structure and leading to the reduced error thresholds. As our simulations show, the threshold is not well-defined for relatively small-distance codes at high~(but finite) bias, so we focus on this regime here. At high bias, it is reasonable to do a perturbation analysis up to first order in the probability of low-rate errors. For simplicity, we assume that high-rate Pauli-$Z$ errors take place with probability $p_Z$ and low-rate Pauli-$X$ errors take place with probability $p_X \ll p_Z$, while $p_Y = 0$. 

\begin{figure}[h!]
    \centering
    \subfloat[]{
        \includegraphics[width=0.33\columnwidth]{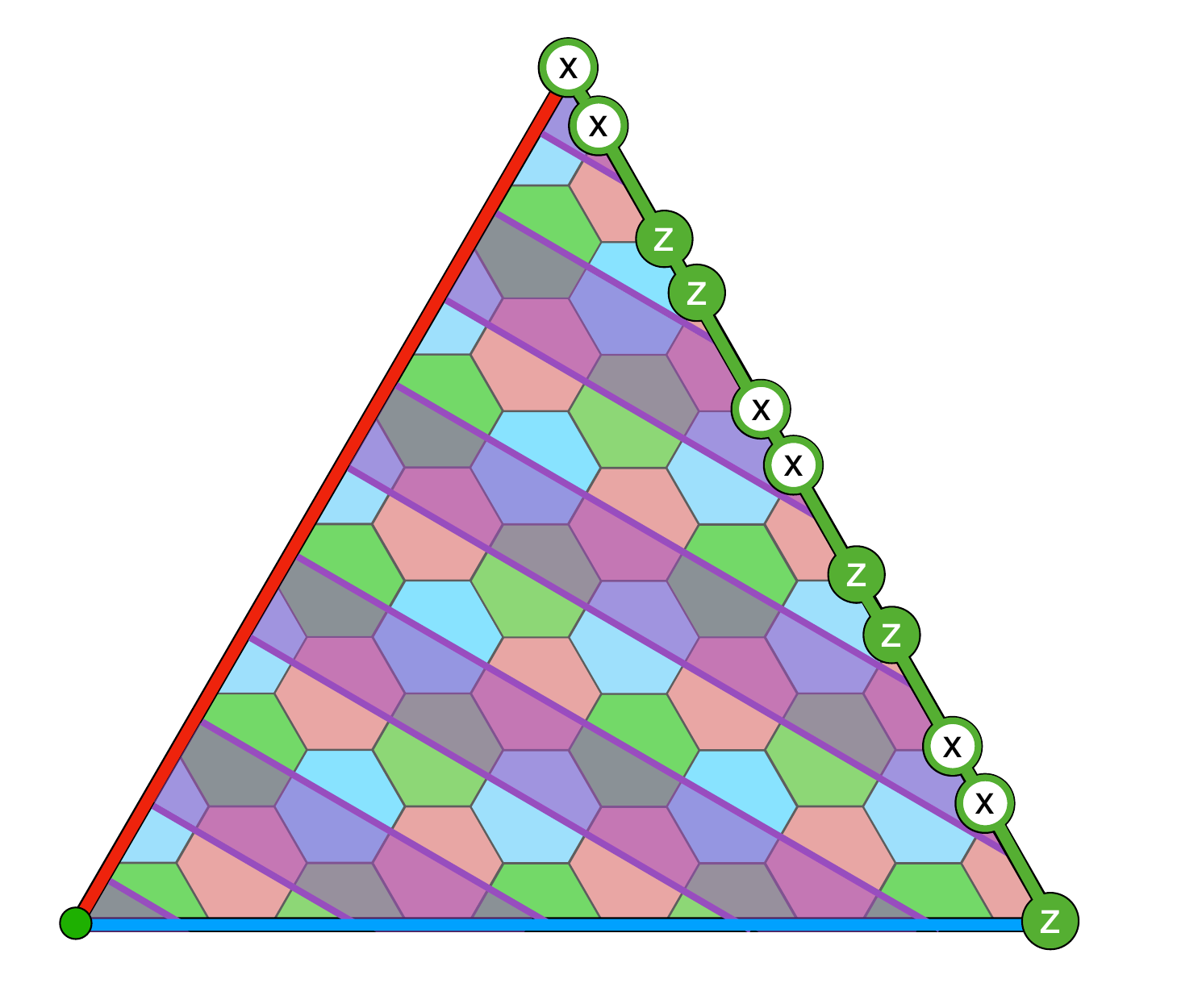}
        \label{fig:x3z3_1}
    }
    \subfloat[]{
        \includegraphics[width=0.33\columnwidth]{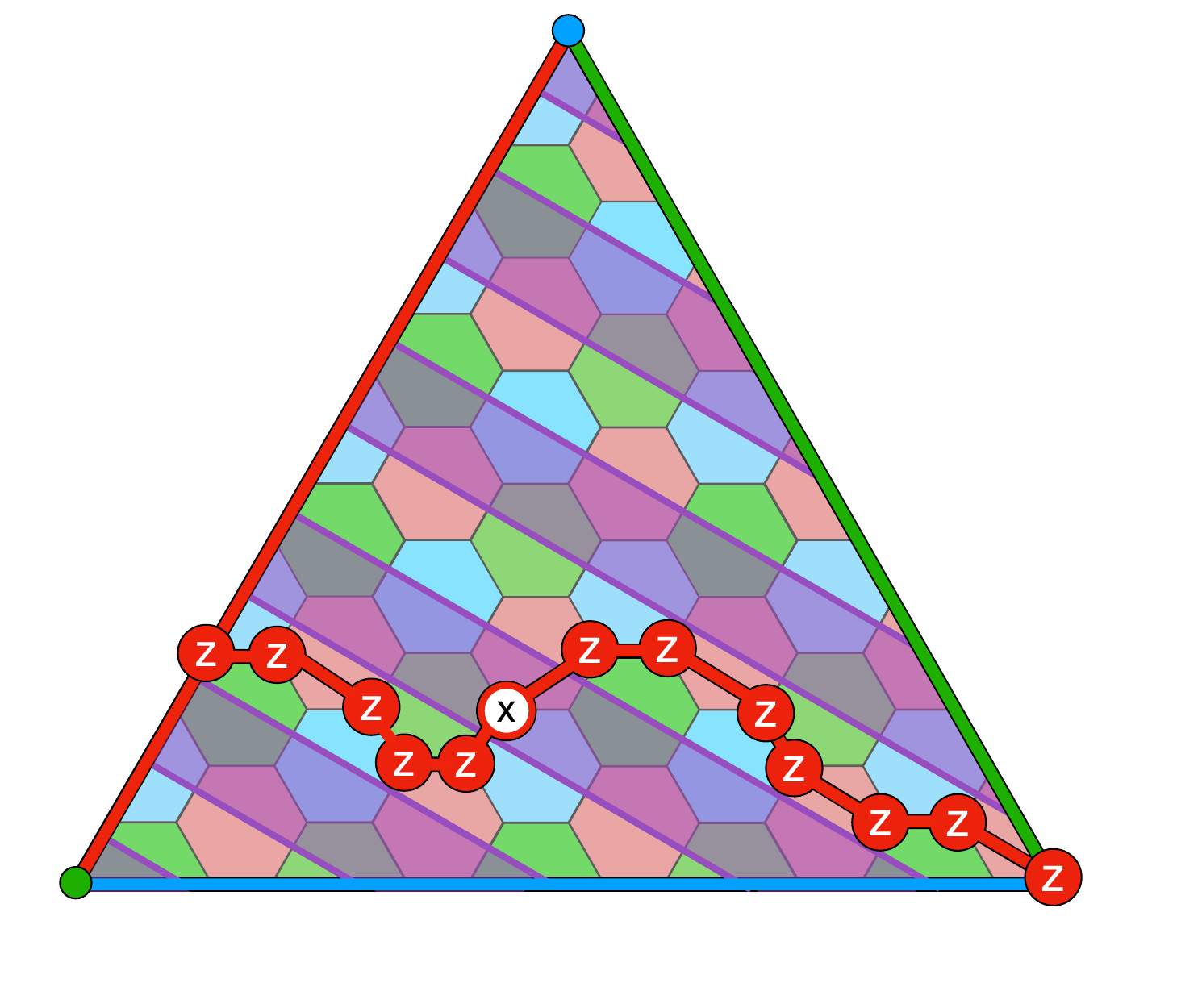}
        \label{fig:x3z3_2}
    }
    \subfloat[]{
        \includegraphics[width=0.33\columnwidth]{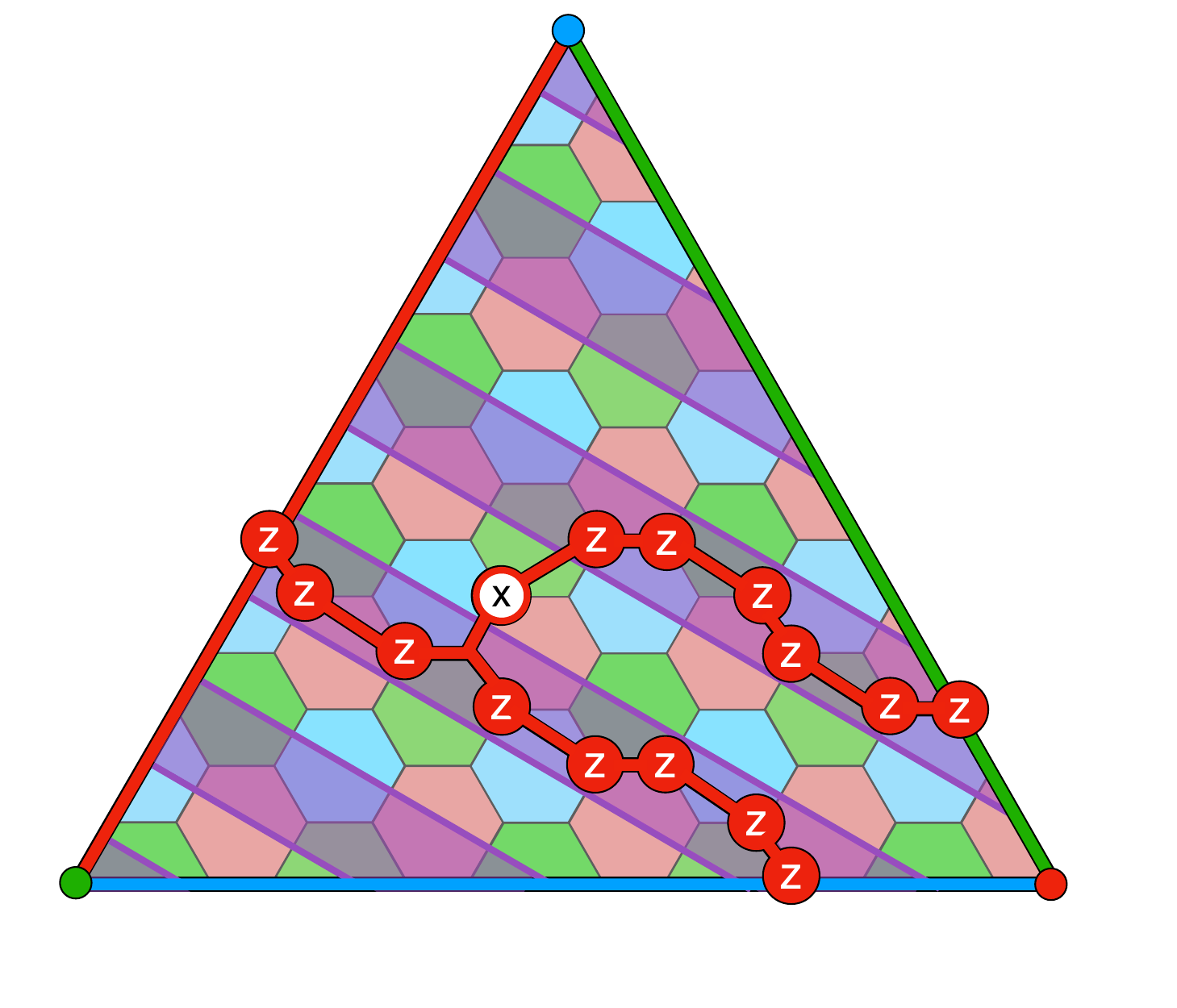}
        \label{fig:x3z3_3}
    }
    \caption{
    Distance-11 \xz{} code and logical operators of types 1, 2, and 3 described in the text.}
    \label{fig:x3z3} 
\end{figure}

Consider the \xz{} code of Fig.~\ref{fig:x3z3}. The only logical operator in the zeroth order in $p_X$ lies within the domain connecting the red corner to the red boundary, as explained in the main text. Below we will prove that a number of weight-$d$ logical operators in first order in $p_X$ is fixed and does not grow with $d$ for sufficiently large codes. 

\begin{proof} 
Recall that there are three types of logical operators in the color code:
\begin{enumerate}
    \item Strings of single-qubit Paulis that connect two corners of the lattice. For sufficiently large codes, logical operators of this type have to cross DWs many times, hence containing more than one operator of each type, see Fig.~\ref{fig:x3z3_1} for an example. 
    \item Strings of single-qubit Paulis that connect a corner to the opposite boundary of the code. Any logical operators connecting the blue or green corners to the respective boundaries will contain many low- and high-rate errors for sufficiently large $d$, similarly to as the case 1. Assume now a string extending from the red corner and crossing a DW in the bulk of the lattice, i.e., $w > 2$ stabilizers away from the code boundaries. Any string that crosses a DW only once contains at least $w > 2$ low-rate $X$ errors and is hence not first-order in $p_X$. Operators that are first-order in $p_X$ cross DWs twice in the direction orthogonal to the shortest path and are hence longer than $d$. An example of such a first-order logical operator is shown in Fig.~\ref{fig:x3z3_2}. On the other hand, o length-$d$ logical operators with $w \leq 1$ can be first-order in $p_X$. However, the low-rate error has to take place on a qubit either close to the red corner or close to the red boundary and the central domain, meaning that there is only a small fixed number of such paths for any $d$.  
    \item Strings of single-qubit Paulis that terminate at each of the three distinct boundaries of the lattice. Again, consider first a case where three strings branch in the bulk of the code, located $w > 2$ stabilizers away from the code boundaries. In order for the logical operator to be the first order in $p_X$, it has it cross DWs exactly twice, as in Fig.~\ref{fig:x3z3_3}. Any such logical operator will then be longer than $d$, since each of the three strings extending from the branching point has to connect to the code boundary by propagating within the respective domain, i.e., not along the shortest paths. As in the previous case, length-$d$ logical operators can be first order in $p_X$ when a branching is located close to the red corner or the red boundary; their number is hence fixed for any $d$ large enough. 
\end{enumerate}

Therefore, in the first order in $p_X$, there is a fixed number of shortest-path logical operators which does not grow with $d$ for $d \gg 1$. Finite-size effects hence become less noticeable for larger codes and, as a consequence, the threshold saturates for $d \gg 1$. 

\end{proof}

The same analysis can be directly applied to the XZZX code to explain the threshold saturation at large distances, which was observed in Ref.~\cite{XZZX}.

\end{document}